\documentclass[a4paper,12pt]{article}
\pdfoutput=1 
\usepackage{jheppub} 
\usepackage[utf8]{inputenc}
\hypersetup{
    bookmarks=true,         
    unicode=false,          
    pdftoolbar=true,        
    pdfmenubar=true,        
    pdffitwindow=false,     
    pdfstartview={FitH},    
    pdftitle={My title},    
    pdfauthor={Author},     
    pdfsubject={Subject},   
    pdfcreator={Creator},   
    pdfproducer={Producer}, 
    pdfkeywords={keyword1} {key2} {key3}, 
    pdfnewwindow=true,      
    colorlinks=true,       
    linkcolor=black,          
    citecolor=purple,        
    filecolor=magenta,      
    urlcolor=blue,           
    linktocpage=true
}
\usepackage{amsmath,amssymb,amsthm,graphicx,latexsym}
\usepackage{subfigure}
\usepackage{sparticles}

\usepackage{dcolumn}
\usepackage{bm}
\usepackage{textcomp}
\usepackage{pstricks}
\usepackage{color}
\usepackage{epsfig}
\usepackage{amsfonts}
\usepackage{rotating}
\usepackage{sparticles}
\usepackage{dcolumn}
\usepackage{bm}
\usepackage{textcomp}
\usepackage{subfigure}
\usepackage{pstricks}
\usepackage{amsfonts}
\usepackage{mathrsfs,amsmath}
\usepackage{epsfig}
\usepackage{amsmath,amssymb,amsthm,graphicx,latexsym}
\usepackage{float}
\usepackage{textcomp}
\usepackage{array}

\usepackage{subfigure}
\usepackage{sparticles}
\usepackage{dcolumn}
\usepackage{bm}
\usepackage{textcomp}
\usepackage{pstricks}
\usepackage{color}
\usepackage{epsfig}
\usepackage{amsfonts}
\usepackage{rotating}
\usepackage{sparticles}
\usepackage{dcolumn}
\usepackage{bm}
\usepackage{textcomp}
\usepackage{subfigure}
\usepackage{pstricks}
\usepackage{amsfonts}
\usepackage{mathrsfs,amsmath}
\usepackage{epsfig}
\usepackage{amsmath,amssymb,amsthm,graphicx,latexsym}
\usepackage{float}
\usepackage{textcomp}
\usepackage{array}

\usepackage{bigints}

\usepackage{mathrsfs}
\usepackage[toc,page]{appendix} 
\usepackage{amsmath}
\usepackage{mathrsfs}
\usepackage{hyperref}

\newcommand{\be}{\begin{equation}}
\newcommand{\ee}{\end{equation}}
\newcommand{\bea}{\begin{eqnarray}}
\newcommand{\eea}{\end{eqnarray}}
\newcommand{\bml}{\begin{subequations}}
\newcommand{\eml}{\end{subequations}}
\newcommand{\bfig}{\begin{figure}}
\newcommand{\efig}{\end{figure}}

\newcommand{\ag}{\alpha}
\newcommand{\bg}{\beta}

\newcommand{\del}{\delta}

\newcommand{\lb}{\lambda}

\newcommand{\og}{\omega}

\newcommand{\Del}{\Delta}

\newcommand{\slashed}{\hspace{-1.1ex}/}

\begin{document}
\title{ \textbf{Spectral Form Factor for Time-dependent Matrix model}}

 \author[a]{Arkaprava Mukherjee}	
\affiliation[a]{Department of Physical Sciences, Indian Institute of Science Education
and Research Kolkata, Mohanpur, West Bengal 741246, India.}		
\author[b]{, Shinobu Hikami}
	\affiliation[b]{Okinawa Institute of Science and Technology Graduate University, 1919-1
Tancha, Okinawa 904-0495, Japan.}

\emailAdd{ am16ms058@iiserkol.ac.in, hikami@oist.jp}
\abstract{The quantum chaos is related to  a Gaussian random matrix model, which shows a dip-ramp-plateau behavior in the
   spectral form factor for the  large size $N$. The spectral form factor of time dependent Gaussian random matrix model shows also dip-ramp-plateau
   behavior with a rounding behavior instead of a kink near Heisenberg time. This model is converted to two matrix model, made of  
   $M_1$ and $M_2$. The numerical evaluation for finite $N$ and analytic expression in the large $N$ are compared for the spectral form factor.}
\maketitle

 \section{Introduction}
 
 Recently the quantum chaos, which is  related to random matrix theory, has attracted interest from the view point of universality. The spectral form factor shows the dip-ramp-plateau behavior for various situations, and
 this behavior is considered as a universal signature of quantum chaos. This transition behavior has been studied and there are many discussions of the universality behaviors in wide area.  For instance, such behavior was noticed before in the level statistics of complex system \cite{Leviandier}.
 Recently, it was found that  black hole has also  dip-ramp-plateau transition in a late time \cite{Cotler,Cotler2,Balasubramanian}. 
 
 The level statistics of a random matrix has a universal behavior, known as Dyson sine kernel, and it coincides the the distribution of the zeros of Riemann zeta function.
 The proof of the universal behavior of the sine kernel, independent on  the external  deterministic term, is given in \cite{SFFOLD}.  The spectral form factor, the Fourier transform of the square of this sine kernel,  provides the ramp-plateau transition. The kink point is denoted here as Heisenberg time.
 
 These three phases, dip, ramp and plateau, show  the different behaviors of the spectral form factor.
 When the eigenvalue of the hermitian random matrix $M$ is denoted by $x_i$, we consider the variance of the quantity $\sum_j e^{i \lambda x_j}= {\rm tr}e^{i \lambda M}$. Two point
 correlation function $\rho^{(2)}(\lambda,\mu)$ is defined as \cite{SFFRMT},
 \be
 \rho^{(2)}(\lambda,\mu) = <\frac{1}{N} {\rm tr} \hskip 1mm \delta(\lambda- M) \frac{1}{N} {\rm tr} \hskip 1mm \delta(\mu - M)>
 \ee
 where the bracket means the average of the Gaussian weight $e^{- \frac{N}{2} {\rm tr} M^2}$.
 We define the connected part $\rho_c^{(2)}$ and disconnected part as covariance \cite{SFFRMT}.
 
 \be\label{dip}
 \rho^{(2)}_c (\lambda,\mu) = \rho^{(2)}(\lambda,\mu) - \rho(\lambda)\rho(\mu)
 \ee
 where $\rho(\lambda) = \left\langle\frac{1}{N}{\rm tr} \delta(\lambda- M)\right\rangle$. In the large $N$ limit with a fixed $N(\lambda-\mu)$ finite (Dyson limit), $\rho^{(2)}(\lambda,\mu)$ is expressed as \cite{SFFRMT},
 \be\label{delta}
 \rho_c^{(2)}(\lambda,\mu) =\frac{1}{N}\delta(\lambda-\mu)\rho(\lambda) - \rho(\lambda)\rho(\mu) \frac{{\rm sin^2} x}{x^2}
 \ee
 where $x= \pi N (\lambda-\mu)\rho(\frac{1}{2}(\lambda+\mu))$.
 The spectral form factor $S(t)$ is defined as Fourier transform of $\rho^{(2)}(\lambda,\mu)$, (we put $\lambda=0$),
 \bea\label{formfactor}
 S(t) &=& \int d\mu e^{i \mu t} \rho^{(2)}(0,\mu)\nonumber\\
 &=& \int e^{i \mu t} d\mu \biggl(\rho(0)\rho(\mu) - \rho(0)\rho(\mu) \frac{({\rm sin }x)^2}{x^2}  + \frac{1}{N} \delta (-\mu)\rho(0)\biggr)\nonumber\\
 &=& S^{(1)}+ S^{(2)} + S^{(3)}
 \eea
 The first term becomes for finite $N$ \cite{Brezin:2016eax}
 \be
 S^{(1)} (t) = \rho(0)(-\frac{1}{it}) e^{-\frac{t^2}{2N}}\oint \frac{du}{2i \pi} ( 1- \frac{it}{Nu})^N e^{- i t u}
 \ee
 This contour integral becomes in the large $N$ limit, by the exponentiating of the  integrand,
 \be
 -\frac{1}{it}\oint \frac{du}{2i\pi} e^{-i t(u+ \frac{1}{u})}= \frac{1}{t} J_1(2 t)
 \ee
 Fourier inverse transformation of above term is the density of state $\rho(x)$,
 \be\label{densityone}
\rho(x) =  \int_{-\infty}^\infty \frac{dt}{2\pi} \frac{1}{t} J_1(2 t) e^{- i x t} = \frac{1}{2\pi}\sqrt{4 - x^2}
 \ee
 which is normalized as
 \be
 \int_{-\infty}^\infty dx \rho(x) = 1
\ee
 Thus we obtain the first term of (\ref{formfactor}) as the Fourier transform of the density of state in the large $N$ limit,
 \be
 S^{(1)}(t) = \frac{1}{\pi t} J_1(2 t)
 \ee
 which gives a dip (decay) region of the spectral form factor in increasing time $t$.
 
 The second term of (\ref{formfactor}) gives the ramp region.
  We use Fourier transform for $|t| < 2$,
 \be
 \int_{-\infty}^\infty dx (\frac{{\rm sin} x}{x})^2 e^{i t x} = \pi ( 1- \frac{t}{2})
 \ee
and for $|t| > 2$, it vanishes. The second term $S^{(2)}(t)$ becomes
\bea\label{S2}
S^{(2)}(t) &=& \frac{1}{2 N^2 \pi} t - \frac{1}{N}\rho(0), \hskip 3mm( |t| < 2\pi N \rho(0))\nonumber\\
&=& 0,   \hskip 25mm (|t| > 2 \pi N\rho(0))
\eea
where we used the density of state $\rho(\mu)= \rho(0)$. This $S^{(2)}(t)$ has a kink at $t= 2 \pi N \rho(0)$, and beyond this kink, it becomes 0.

The last term of (\ref{formfactor}) gives a constant term, which cancels with the  constant term of (\ref{S2}),
\be
S^{(3)}(t) = \frac{1}{N} \rho(0)
\ee

Thus we find that the ramp region is order of $\frac{1}{N^2}$ and the plateau is order of $\frac{1}{N}$.
  The dip and plateau regions depend upon the density of state $\rho(E)$, thus this region is not universal, but the ramp region is  universal since it is universal Dyson kernel  \cite{SFFOLD,SFFRMT}.
 
In this paper, we consider the time dependent random matrix theory, which becomes equivalent two matrix model. The two point correlation functions of two matrix model was studied 
before in \cite{D'Anna}. For two matrix model, coupled matrices $M_1$ and $M_2$, two point function $\rho^{(2)}(\lambda,\mu)$ has two types;
\be
\rho^{(2)}(\lambda,\mu) = <\frac{1}{N}{\rm tr}\delta(\lambda - M_1)\frac{1}{N}{\rm tr}\delta(\mu - M_1)>
\ee
and 
\be
\rho^{(2)}(\lambda,\mu) = <\frac{1}{N}{\rm tr}\delta(\lambda - M_1)\frac{1}{N}{\rm tr}\delta(\mu - M_2)>
\ee
We call these two different correlation functions as $M_1-M_1$ and $M_1-M_2$ correlation functions.

The spectral form factor has two different types according to above difference. In the previous paper, the kink behavior of the spectral form factor
is found to be smeared out due to a factor $e^{-ct}$ in the ramp region \cite{SFFRMT}.
\be
S(t) = \int d\mu e^{-i \mu t} \rho_c (0,\mu) = t e^{-t \sqrt{1-c}/c }
\ee

We study further this rounding near Heisenberg time by the numerical analysis based on an exact formula of finite $N$. We also compare its result with the
saddle point analysis for the large $N$. These rounding behaviors are also observed in different ensembles \cite{Liu:2018hlr,Forrester,Okuyama:2018yep}.

 \vskip 3mm
 \section{Time dependent random matrix}
 \vskip 3mm

  We consider the Hamiltonian
  $H$ as 
  \be\label{td11}
  H = \frac{1}{2} {\rm tr} (p^2+ M^2)
  \ee
  where $p = d M(t)/dt$. The matrix $M$ depends on time, $M=M(t)$. The time dependent correlation function of the different time $t_1$ and $t_2$ is defined as
  \be
  \rho(\lambda,\mu) = \frac{1}{N^2} <{\rm tr}\delta(\lambda- M(t_1) {\rm tr} \delta(\mu - M(t_2)>
  \ee

  The Fourier transform of above correlation function is
  \be \label{Hikamicorr1}
  U(\alpha,\beta) = \frac{1}{N^2}<{\rm tr} e^{i\alpha M(t_1)}{\rm tr} e^{i \beta M(t_2)}>
  \ee
 
 It has been shown that the correlation function of the time dependent random matrix theory is equivalent to the correlation function of the 
 two matrix model by path integral in \cite{SFFRMT, D'Anna}.
 
  \be\label{corr2}
 U(\alpha,\beta) = \frac{1}{Z} \int dM_1 dM_2 {\rm tr} e^{i \alpha M_1} {\rm tr}e^{i \beta M_2} e^{-\frac{1}{2}{\rm tr} (M_1^2 + M_2^2 - 2 c M_1 M_2)}
 \ee
 where $c = e^{-t}$. The time $t$ is the difference of $t_1$ and $t_2$, $t= t_1-t_2$.
 Now the average $<\cdots>$ means the Gaussian distribution, $P(M_1,M_2) =  e^{- H_{1,2}}/Z$,
 \be\label{corr1}
 H_{1,2} = \frac{1}{2}{\rm tr} ( M_1^2+M_2^2 - 2 c M_1 M_2)
 \ee
 The density of state $\rho(\lambda)$ becomes in the large $N$ limit,
 \be
 \rho(\lambda) = \frac{\sqrt{1 - c^2}}{2\pi} \sqrt{4-(1- c^2) \lambda^2}
 \ee
 which is normalized to be 1 by the integration. When $c=0$, it reduces to the density of state $\rho(\lambda)$ in (\ref{densityone}).

Now we use the method of external matrix to compute the exact expression for two-point correlation function defined in Eq:-\ref{Hikamicorr1}.  By the integral expression in \cite{SFFRMT}, we have
\bea \label{compare123}
\begin{array}{lll}
\displaystyle
U_{A}(z_{1},z_{2})=\frac{1}{N^2}\sum_{\ag_{1},\ag_{2}}^{N} \int\int e^{iz_{1}r_{\ag_{1}}+ iz_{2}\xi_{\ag_{2}}} e^{-\frac{N}{2}\sum r_{i}^{2}-\frac{N}{2}\sum \xi_{i}^{2}+c N\sum r_{i}\xi_{i}-N\sum a_{i}r_{i}}\frac{\Del(\xi)dr d\xi}{\Del(A)}~~~~~~
\end{array}
\eea
In contour integral representation with $\ag_{1}=\ag_{2}$ and taking the external matrix at zero ($A\rightarrow 0$).
\bea
\begin{array}{lll}
\displaystyle
U_{0}^{I}(z_{1},z_{2})=-\frac{1}{iN(z_{1}+\frac{z_{2}}{c})}\oint \frac{du}{2\pi i}\bigg[1-\frac{i}{Nu}(z_{1}+\frac{z_{2}}{c})\bigg]^{N} \mathrm{Exp}\bigg[-\frac{z_{1}^{2}}{2N(1-c^{2})}~~
\\
\displaystyle
~~~~~~~~~~~~~~~~~~~~~~~~~~~~~~~-\frac{z_{2}^{2}}{2N(1-c^{2})}-(\frac{iz_{1}}{1-c^2}-\frac{icz_{2}}{1-c^{2}})u-\frac{c z_{1}z_{2}}{N(1-c^{2})}\bigg]~~
\end{array}
\eea
With a scaling, $u\rightarrow\bar  u (z_{1}+\frac{z_{2}}{c})$ using the transformation $z_{1}\rightarrow\frac{1}{\sqrt{1-c^2}}(z'_{1}-cz'_{2})$ and  $z_{2}\rightarrow\frac{1}{\sqrt{1-c^2}}(z'_{2}-cz'_{1})$ and after integrating over $z_{1}$ following \cite{SFFRMT}
\bea\label{rho0}
\begin{array}{lll}
\displaystyle
\rho^{I}_{c}(\lambda,\mu)=\frac{i \mathrm{Exp}\{-\frac{N}{2}(\lambda-c\mu)^2\}}{N(1-c^2)^{2}}\int \frac{dz'_{2}}{2\pi} \oint \frac{du}{2\pi i}(1-\frac{i}{Nu})^{N}
\\
\displaystyle
~~~~~~~~~~~~~~~~\mathrm{Exp}\bigg[-i\mu z'_{2}+Nu z'_{2}(\mu-\frac{\lambda}{c})-\frac{i u (z'_{2})^2}{1-c^2}-\frac{N u^2 (z'_{2})^2}{2c^2}-\frac{(z'_{2})^{2}}{2N(1-c^2)}\bigg]~~~~~~

\end{array}
\eea
For $\ag_{1}\neq\ag_{2}$
\bea
\begin{array}{lll}\label{tot}
\displaystyle
U_{A}(z_{1},z_{2})=-\frac{c}{z_{1}z_{2}}\oint \frac{dudv}{(2\pi i)^{2}}\bigg(1-\frac{iz_{1}}{Nu}\bigg)^{N}\bigg[1-\frac{z_{1}z_{2}}{c N^{2}(u-v-\frac{iz_{1}}{N})(u-v+\frac{iz_{2}}{cN})}\bigg]~~~~
\\
\\
\displaystyle
~~~~~~~~\times\bigg(1-\frac{iz_{2}}{cNv}\bigg)^{N} \mathrm{Exp}\bigg\{-\frac{iz_{1}u}{1-c^{2}}-\frac{iz_{2}c v}{1-c^2}-\frac{z_{1}^{2}}{2N(1-c^2)}-\frac{z_{2}^{2}}{2N(1-c^2)}\bigg\}~~~~
\end{array}
\eea
First term in parenthesis of Eq:-(\ref{tot}) on Fourier transform gives the disconnected part of two point correlation function:-
\bea
\begin{array}{lll}\label{aa} 
\displaystyle
\rho^{d}(\lambda,\mu)=-c \bigg\{\int \frac{d z_1}{2\pi z_{1}}\oint \frac{du}{(2\pi i)}\bigg(1-\frac{iz_{1}}{Nu}\bigg)^{N}e^{-\frac{iz_{1}u}{1-c^{2}}-\frac{z_{1}^{2}}{2N(1-c^2)}-iz_{1}\lambda}
\\
\displaystyle
~~~~~~~~~~~~~~~~~~~~~~~~\times\int \frac{dz_2}{2\pi z_{2}} \oint \frac{dv}{2\pi i}(1-\frac{iz_{2}}{cNv})^{N}e^{-\frac{iz_{2}c v}{1-c^2}-\frac{z_{2}^{2}}{2N(1-c^2)}-iz_{2}\mu}\bigg\}
\end{array}
\eea
Therefore disconnected two point correlation function Eq:- (\ref{aa}) is very similar to one matrix model density of states. 
\be
\rho^{d}(\lb,\mu)=\rho(\lb) \rho(\mu)
\ee
After a Fourier transform and setting the values $\lb=0$ and $\mu=\omega$ we get the spectral form factor of the disconnected part.
\be \label{teddc_1}
S_{d}(\tau)=\int \frac{1}{2\pi}\rho^{d}(0,\og)e^{i\og \tau} d\og
\ee
 We have averaged this dynamical form factor over an interval [0,t] and plot this average value:-
\be \label{td411}
\langle S_{d}\rangle_{Average}=\int_{0}^{t} S(\tau)d\tau
\ee
\begin{figure}[H]
       \centering
         \subfigure[Disconnected part of SFF $S_{d}$ for c=0.9, N=10 from \ref{teddc_1}]{
      \includegraphics[width=7cm,height=5.5cm] {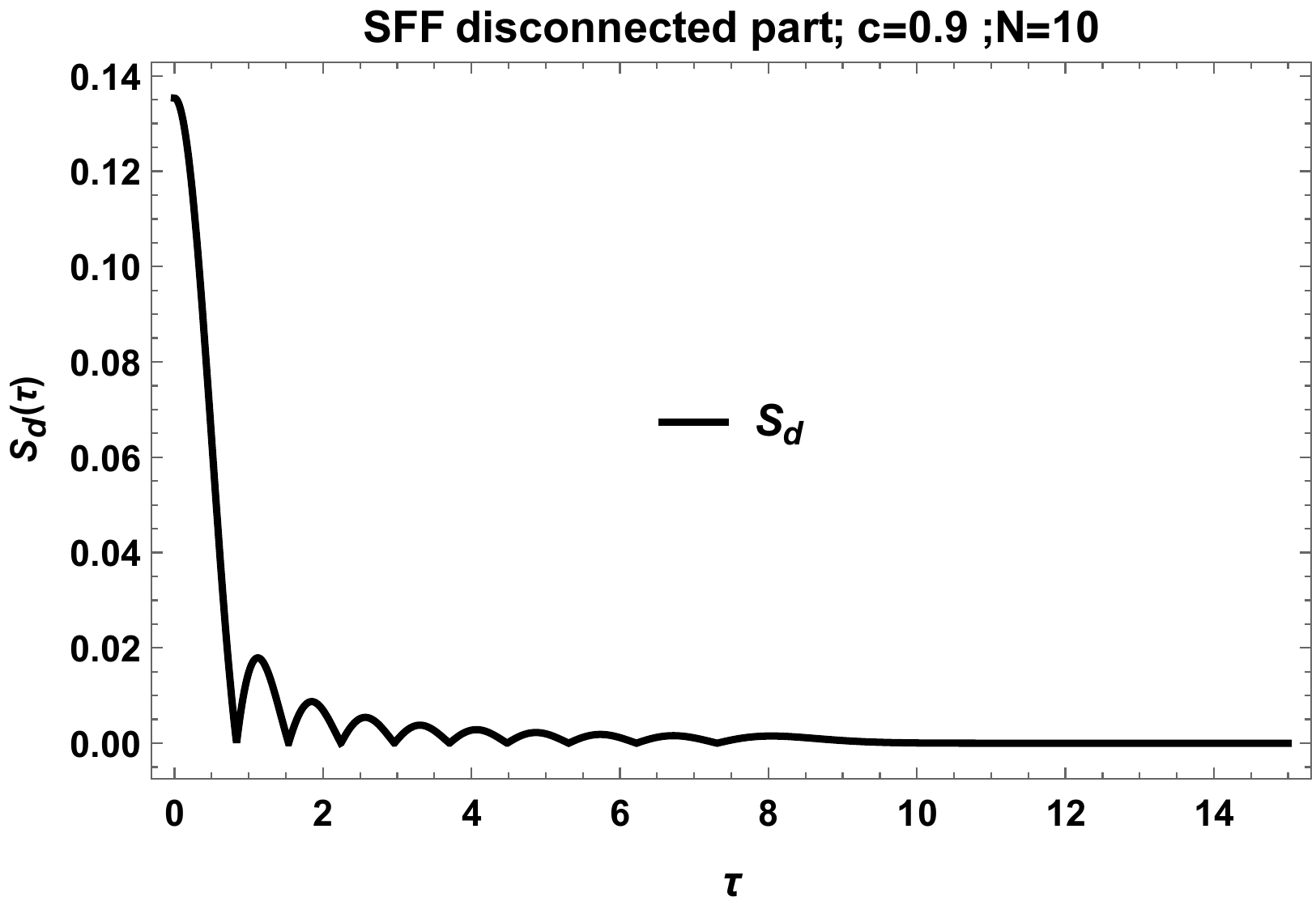}
     \label{}
         } 
        \subfigure[Averaging SFF disconnected part $S_{d}$ over (0,t) from \ref{td411}]{
	  \includegraphics[width=7cm,height=5.5cm] {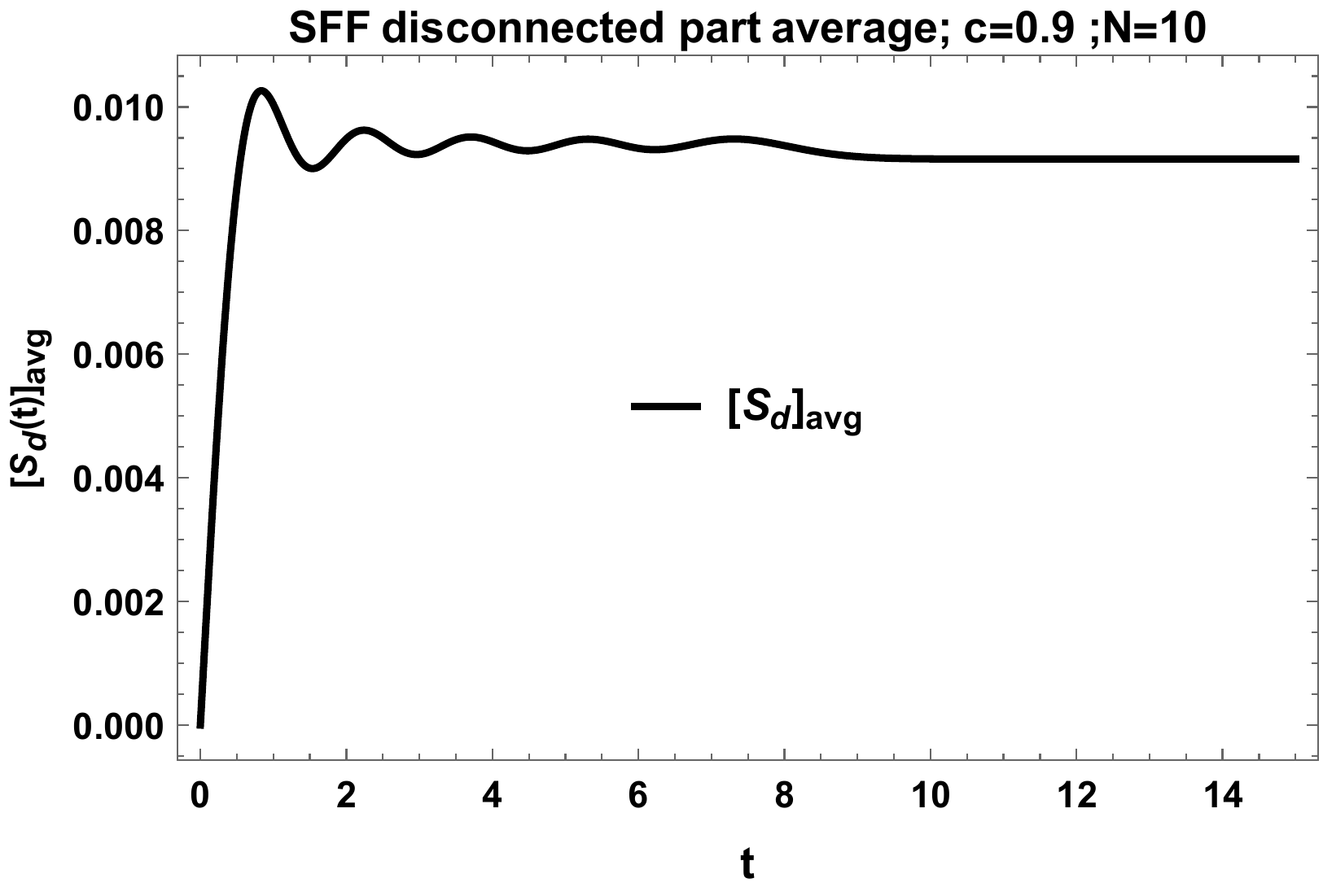}
         } 
          \caption[Optional caption for list of figures]{Spectral Form Factor disconnected part $S_{d}$ for two matrix model  from Eq:-(\ref{aa})} 
           \label{g1}
          \end{figure}  
Now second term in parenthesis of Eq:- \ref{tot} gives the connected part of two point correlation function:-
\bea
\begin{array}{lll}
\displaystyle
U_{A}(z_{1},z_{2})=-\frac{1}{N^2}\oint \frac{dudv}{(2\pi i)^{2}}(1-\frac{iz_{1}}{Nu})^{N}(1-\frac{iz_{2}}{cNv})^{N}\frac{1}{(u-v-\frac{iz_{1}}{N})}\frac{1}{(u-v+\frac{iz_{2}}{cN})}~~~~~~~\\
\displaystyle
~~~~~~~~~~~~~~~~~~~~~~\times\mathrm{Exp}\bigg[-\frac{iz_{1}u}{1-c^{2}}-\frac{iz_{2}c v}{1-c^2}-\frac{z_{1}^{2}}{2N(1-c^2)}-\frac{z_{2}^{2}}{2N(1-c^2)}\bigg]
\end{array}
\eea
If $z_{1}\rightarrow z'_{1}-iuN$,$z_{2}\rightarrow z'_{2}-ivcN$, it is written as a product form.
\bea
\begin{array}{lll}
\displaystyle
\rho_{c}^{II}(\lambda,\mu)=-\frac{1}{N^2}\oint\frac{du}{2\pi i}\int \frac{dz_{2}}{2\pi}\bigg[\frac{z_{2}}{cuN}\bigg]^{N}\frac{\mathrm{Exp}\big[-\frac{Nu^{2}}{2(1-c^2)}-\frac{z_{2}^{2}}{2N(1-c^2)}-iz_{2}\mu-uN\lambda\big]}{u+\frac{iz_{2}}{cN}}~~
\\
\displaystyle
~~~~~~~~~~~~\times \oint\frac{dv}{2\pi i}\int \frac{dz_{1}}{2\pi}\bigg[\frac{z_{1}}{vN}\bigg]^{N}\frac{\mathrm{Exp}\big[-\frac{Nv^{2}c^{2}}{2(1-c^2)}-\frac{z_{1}^{2}}{2N(1-c^2)}- iz_{1}\lb- vcN\mu\big]}{v+\frac{iz_{1}}{N}}
\end{array}
\eea
where the contours of $u$ and $v$ are around the poles of $u=0$ and $v=0$. If we take the pole of $v= -iz_1/N$, we obtain $\rho^{(I)}(\lambda,\mu)$ in (\ref{rho0}). In kernel form this can be written as -
 \be \label{hik12}
 \rho_c^{(II)}(\lambda,\mu) = -\frac{1}{N^2} K_N(\lambda,\mu) \bar K_N(\lambda,\mu)
 \ee
From $\ag_{1}=\ag_{2}$ contribution we have other part of connected correlation function in kernel form as-
\be\label{rho1}
 \rho_c^{(I)}(\lambda,\mu) = \frac{1}{N^2} \sqrt{\frac{N}{2\pi}} K_N(\lambda,\mu) e^{-\frac{N}{2}(\lambda- c \mu)^2}
 \ee
 Two point function $\rho^{(2)}(\lambda,\mu)$ becomes the sum of the product of the density of state $\rho(\lambda)\rho(\mu)$ and the connected part
 $\rho_c^{(2)}(\lambda,\mu)$.
 \be
 \rho^{(2)}(\lambda,\mu) = \rho(\lambda)\rho(\mu) + \rho_c^{(2)}(\lambda,\mu)
 \ee
 The Fourier transform of the first term for $\lambda=0$ in the large $N$ limit becomes
 \be
 S^{(1)}(t) = \frac{\sqrt{1-c^2}}{\pi t}\rho(0) J_1( \frac{2t}{\sqrt{1-c^2}})
 \ee
 where $\rho(0) = \sqrt{1- c^2}/\pi$. This term gives the dip (decay) behavior.
 
 The connected correlation function $\rho_c^{(2)}(\lambda,\mu)$ is expressed as \cite{SFFRMT},
 \be\label{connectedcorrel}
 \rho_c^{(2)}(\lambda,\mu) = \rho_c^{(I)}(\lambda,\mu) + \rho_c^{(II)}(\lambda,\mu)
 \ee
 where these two terms are given by Eq:-\ref{hik12},Eq:-\ref{rho1}
These two terms correspond to two terms of (\ref{delta}) in the large $N$ limit. Note that the term of (\ref{rho1}) becomes delta-function
in the Dyson limit for large $N$ since we have,
\be
\lim_{N\to \infty} \sqrt{\frac{N}{2\pi}} e^{-\frac{N}{2}(\lambda- c\mu)^2}= \delta(\lambda - c \mu)
\ee
Previously we have represented connected correlation as a product of kernels $K_N(\lambda,\mu)$,
$ \bar{K}_N(\lambda,\mu)$,
 \bea\label{kernel}
\begin{array}{lll}
\displaystyle
 \rho_c^{(II)}(\lambda,\mu)= - \frac{1}{N^2} K_N(\lambda,\mu) \bar K_N(\lambda,\mu) \\
\displaystyle
\rho_c^{(I)}(\lambda,\mu) = \frac{1}{N^2} \sqrt{\frac{N}{2\pi}} K_N(\lambda,\mu) e^{-\frac{N}{2}(\lambda- c \mu)^2}
 \end{array}
\eea
 where $K_N(\lambda,\mu)$ is written as
 \bea\label{K}
\begin{array}{lll}
\displaystyle
 K_N (\lambda,\mu) = (-i)^N \oint \frac{du}{2 i \pi}\int \frac{dz}{2\pi}\biggl(\frac{[1- (\frac{-iz}{N c u})^N ]-1}{(1 - \frac{iz}{N c u})}\biggr) \frac{1}{u}\\
\displaystyle
~~~~~~~~~~\times e^{-[N u^2/2(1-c^2)] - [ z^2/2N(1-c^2)] - i z \mu - u N \lambda}\\
 \displaystyle
=\sqrt{\frac{N(1-c^2)}{2\pi}} e^{-\frac{N \mu^2(1-c^2)}{2}}   \sum_{l=0}^{N-1} \frac{1}{l! c^l} H_l(\mu \sqrt{N(1-c^2)} ) H_l(\lambda \sqrt{N(1-c^2)}) \\
\displaystyle
 - (-i)^N \oint \frac{du}{2i\pi}\int \frac{dz}{2\pi} \frac{1}{u - \frac{iz}{Nc}}e^{-[N u^2/2(1-c^2)] - [ z^2/2N(1-c^2)] - i z \mu - u N \lambda}
\end{array}
 \eea
 where  we used a formula $(1- a^N)/(1-a) = \sum a^l$ and $H_l(x)$ is Hermite polynomial, which is represented as \cite{SFFRMT}
 \bea\label{Hermite}
 H_l(x) &=& \oint \frac{du}{2i\pi} \frac{l!}{u^{l+1}}e^{x u - (1/2) u^2}\nonumber\\
 &=& \int_{-\infty}^\infty \frac{dt}{\sqrt{2\pi}}(it)^l e^{-t^2/2 - i t x + x^2/2}
 \eea
 where  $H_0(x)= 1$, $H_1(x)= x, H_2(x)= x^2-1$. The second term of (\ref{K}) is vanishing.

 Another kernel $\bar K_n(\lambda,\mu)$ in (\ref{kernel}) is
 \bea\label{Kbar}
 &&\bar K_N(\lambda,\mu) = \sqrt{\frac{N(1-c^2)}{2\pi}} e^{-\frac{N\lambda^2(1-c^2)}{2}}  \sum_{l=0}^{N-1} \frac{c^l}{l!} H_l(\lambda \sqrt{N(1-c^2)}) H_l (\mu \sqrt{N(1-c^2)})\nonumber\\
&+ & i^N \oint\frac{dv}{2i\pi} \int \frac{dz}{2\pi} \frac{1}{v- \frac{iz}{N}} e^{-[v^2 c^2 N/2(1-c^2)]-[z^2/2N(1-c^2)]-i z\lambda - v c \mu N}
 \eea
 The second term is vanishing.
 
 In the limit $c \to 0$, $\bar K_N(\lambda,\mu) = \sqrt{N/(2\pi)}e^{-(N/2)\lambda^2}$ and $\rho_c^{(2)}(\lambda,\mu) = \rho_c^{(I)}(\lambda,\mu) + \rho_c^{(II)}(\lambda,\mu)$ becomes
 \bea
 \rho_c^{(2)}(\lambda,\mu) &=& \sqrt{\frac{N}{2\pi}}K_N(\lambda,\mu) e^{-\frac{N}{2}\lambda^2} - \sqrt{\frac{N}{2\pi}} K_N(\lambda,\mu) e^{-\frac{N}{2}\lambda^2}\nonumber\\
 &=&0
 \eea
 which means that there is no correlation for $t\to \infty$ ($c= e^{-t}$) as expected.

 For the later use, we evaluate $\rho^{(2)}(\lambda,\mu)$ for small $N$ with the expression of $K_N(\lambda,\mu)$ and $\bar K_N(\lambda,\mu)$. For $N=2$,
 
 \be
 K_N(\lambda,\mu) = \sqrt{\frac{1-c^2}{\pi}} ( 1+ \frac{2\lambda \mu}{c}(1-c^2))e^{-\mu^2(1-c^2)}
 \ee
 and 
 \be
 \bar K_N(\lambda,\mu) = \sqrt{\frac{1- c^2}{\pi}}(1 + 2 c \lambda \mu (1 - c^2)) e^{-\lambda^2(1- c^2)}
 \ee
 
 For $\lambda=0$, the product of $K_N(0,\mu)\bar K_N(0,\mu)$ becomes for $N=2$,
 \be
 K_N(0,\mu)\bar K_N(0,\mu) = \frac{1-c^2}{\pi}e^{-\mu^2(1-c^2)}
 \ee

 and its Fourier transform $S^{(II)}(t)$ is
 \bea
 S^{(II)}(t) &=& - \int_{-\infty}^\infty d\mu e^{ i \mu t} K_N(0,\mu)\bar K_N(0,\mu)\nonumber\\
 &=& - \frac{\sqrt{1-c^2}}{\sqrt{\pi}} e^{- \frac{t^2}{4(1-c^2)}}
 \eea
 This corresponds to the ramp behavior for $N=2$. 
 For the plateau region, $S^{(I)}(t)$ of $N=2$, we obtain Fourier transform of  (\ref{rho1}),
 \be
 S^{(I)}(t)=\sqrt{\frac{1-c^2}{\pi}} e^{-\frac{t^2}{4}}
 \ee
 
\begin{figure}[H]
       \centering
         \subfigure[SFF $\langle S_c(t)\rangle_{average}$ for c=0.9, N=30]{
      \includegraphics[width=7cm,height=5.5cm] {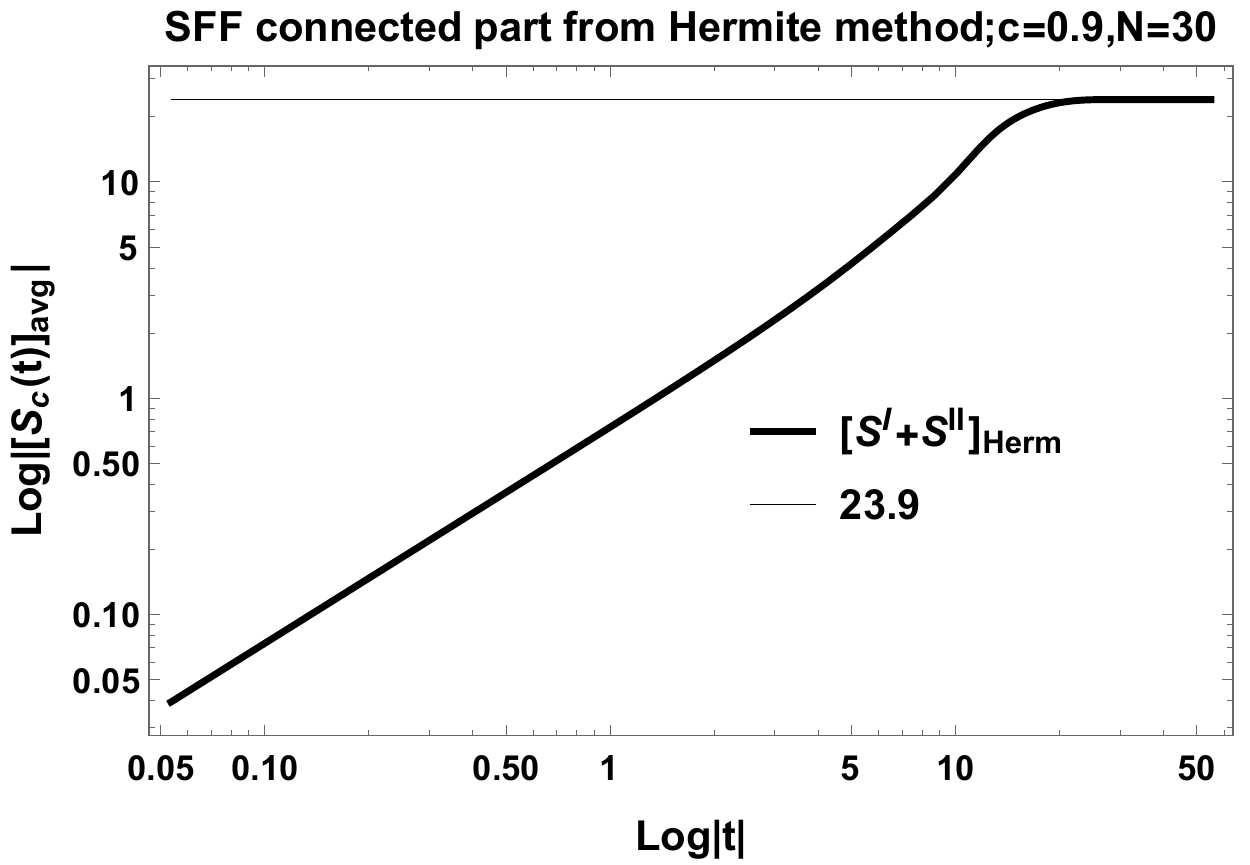}
     \label{}
         }
    \subfigure[SFF  $\langle S_c(t)\rangle_{average}$  c=0.9, N=10]{
         \includegraphics[width=7cm,height=5.5cm] {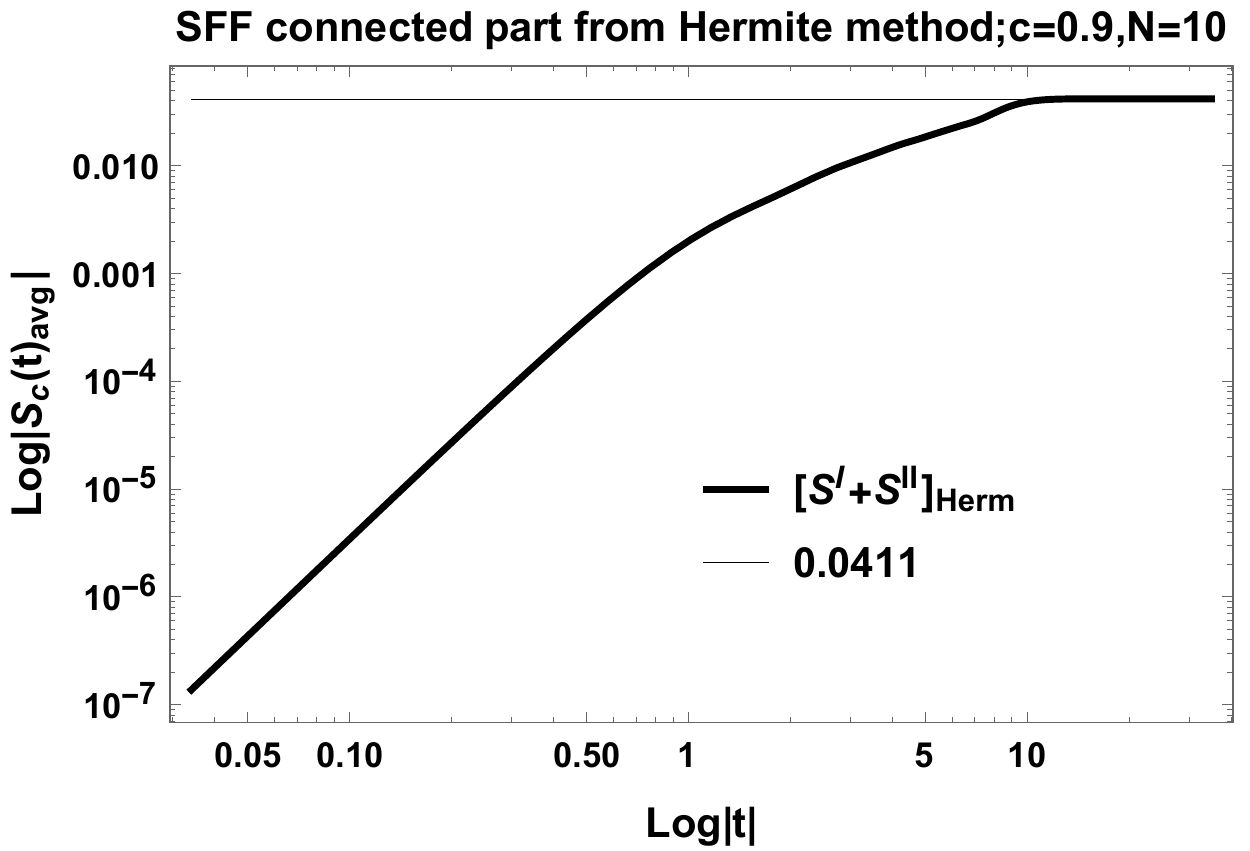}
     \label{}
            }   
                  \caption[Optional caption for list of figures]{Spectral Form Factor average $S_c=(S^{I}+S^{II})_{Hermite}+S^{d}$ where the connected part is evaluated using the hermite polynomial representation of the kernel } 
                                                                    \label{hermite123}
                                                                    \end{figure}

Then we write a connected part as $S_c(\tau)$ and its average as $\langle S_c\rangle_{average}$,
   \bea \label{art12}
   \begin{array}{lll}
   \displaystyle
S_{c}(\tau)=\int \frac{1}{2\pi}\rho_{c}^{2}(0,\og)e^{i\og \tau} d\og \\
\displaystyle
\langle S_c(t)\rangle_{average}=\int_{0}^{t} S(\tau)d\tau
\end{array}
\eea
And after averaging this over an interval $[0,t]$ we plot it in Fig:-\ref{hermite123}

In Fig:-\ref{hermite123} we have computed the the connected part of the correlation function from Eq:-\ref{kernel} using the definition in Eq:-\ref{K},\ref{Kbar}
\subsection*{Saddle point analysis for large $N$}
 For the numerical analysis of finite $N$, we here represent the detail analysis of the large $N$  by the saddle point method. The saddle point analysis is same as \cite{SFFRMT}.\\
First kernel for this case:-
\be 
K_{N}(\lb,\mu)=\oint\frac{du}{2\pi i}\int \frac{dz_{2}}{2\pi}(\frac{z_{2}}{cuN})^{N}\frac{1}{u+\frac{iz_{2}}{cN}}e^{-(\frac{Nu^{2}}{2(1-c^2)}+\frac{z_{2}^{2}}{2N(1-c^2)}+iz_{2}u+uN\lambda)}
\ee
By a scaling $z\rightarrow zNc$
\bea
\begin{array}{lll}\label{1stsaddle}
\displaystyle
K_{N}(\lb,\mu)=cN \oint\frac{du}{2\pi i}\int \frac{dz'_{2}}{2\pi}\frac{1}{u+ i z'_{2}}e^{-N f(z'_{2},u)}
\end{array}
\eea 
Saddle point equation for first kernel:-
\be
f(z'_{2},u)=\frac{u^2}{2 (1-c^2)}+\frac{c^2 (z'_{2})^2}{2 (1-c^2)}+i c \mu  z'_{2}+\lambda  u+\log (u)-\log (z'_{2})
\ee
here $\lb=\frac{2}{\sqrt{1-c^2}}\sin(\theta)$, $\mu=\frac{2}{\sqrt{1-c^2}}\sin(\phi)$.
Solving the equations saddle points are 
\[
\begin{array}{lll}
\displaystyle
(z'_{2},u) :\rightarrow\left( \frac{\sqrt{1-c^2} e^{-i \phi }}{c},-i \sqrt{1-c^2} e^{-i \theta }\right) ;\left(-\frac{\sqrt{1-c^2} e^{i \phi }}{c} , -i \sqrt{1-c^2} e^{-i \theta } \right);
\\
\displaystyle
~~~~~~~~~~~~~\left(\frac{\sqrt{1-c^2} e^{-i \phi }}{c}, i \sqrt{1-c^2} e^{i \theta }\right) ;\left(-\frac{\sqrt{1-c^2} e^{i \phi }}{c}, i \sqrt{1-c^2} e^{i \theta }\right)
\end{array}
\]
Considering fluctuation around saddle points the solution of kernel is:-
\bea
\begin{array}{lll}
\displaystyle
K_{N}(\lb,\mu)=-\frac{c N}{2\pi N}\sum \frac{e^{-N(f(z'_{2},u))}}{u+i z'_{2}}\frac{1}{\sqrt{\frac{\partial^2f}{\partial u^2}\frac{\partial^2 f}{\partial {z'_{2}}^2}}}
\end{array}
\eea

\bea\label{KKTRY1}
\begin{array}{lll}
\displaystyle
K_{N}(\theta,\phi)=-\frac{i \sqrt{1-c^2}c e^{\frac{1}{2} i (\theta +\phi )}}{4 \sqrt{\cos (\theta ) \cos (\phi )}} \bigg(\frac{\mathrm{Exp}[\frac{N}{2}(G[\theta,\phi]-e^{2i\phi}+1)]}{(-ic)^{N}\left(c+e^{i (\theta +\phi )}\right)}
\\
\\
\displaystyle
~~~~~~~~-\frac{\mathrm{Exp}[\frac{N}{2}(G[\theta,-\phi])]}{(ic)^{N}\left(e^{i \theta }-c e^{i \phi }\right)}-\frac{\mathrm{Exp}[\frac{N}{2} (G[-\theta,-\phi])]}{(-ic)^{N}\left(1+c e^{i (\theta +\phi )}\right)}+\frac{\mathrm{Exp}[\frac{N}{2}(G[-\theta,\phi])]}{(ic)^{N}\left(e^{i \phi }-c e^{i \theta }\right)}\bigg)
\end{array}
\eea
where we define $G[\theta,\phi]=2i(\theta+\phi)-e^{-2i\theta}+e^{2i\phi}$\\
The second kernel is similarly given as
\be 
\displaystyle
\bar K_{N}(\lb,\mu)=\oint\frac{dv}{2\pi i}\int \frac{dz_{1}}{2\pi}(\frac{z_{1}}{vN})^{N}\frac{1}{v+\frac{iz_{1}}{N}}e^{-(\frac{Nc^{2}v^{2}}{2(1-c^2)}+\frac{z_{1}^{2}}{2N(1-c^2)}+iz_{1}\lambda+cv\mu N)}
\ee
With a scaling $z\rightarrow zN$
\bea
\begin{array}{lll}\label{2ndsaddle}
\displaystyle
\bar{K}_{N}(\lb,\mu)=N \oint\frac{dv}{2\pi i}\int \frac{dz'_{1}}{2\pi}\frac{1}{v+ i z'_{1}}e^{-N\bar{f}(z'_{1},v)}
\end{array}
\eea 
here also $\lb=\frac{2}{\sqrt{1-c^2}}\sin(\theta)$, $\mu=\frac{2}{\sqrt{1-c^2}}\sin(\phi)$ and saddle point equation for the second kernel:-
\be
\bar f(z'_{1},v)=(\frac{c^2v^{2}}{2(1-c^2)}+\frac{(z'_{1})^{2}}{2(1-c^2)}+iz'_{1}\lambda+cv\mu+ln(v)-ln(z'_{1}))
\ee
This gives the saddle points:-
\[
\begin{array}{lll}
\displaystyle
\left(z'_{1} , v\right):\rightarrow\left( -\sqrt{1-c^2} e^{i \theta },-\frac{i \sqrt{1-c^2} e^{-i \phi }}{c} \right);\left( -\sqrt{1-c^2} e^{i \theta} ,\frac{i \sqrt{1-c^2} e^{i \phi }}{c}\right); 
\\
\displaystyle
~~~~~~~~~~~~~\left( \sqrt{1-c^2} e^{-i \theta },-\frac{i \sqrt{1-c^2} e^{-i \phi }}{c}\right);\left( \sqrt{1-c^2} e^{-i \theta }, \frac{i \sqrt{1-c^2} e^{i \phi }}{c}\right)
\end{array}
\]
With the fluctuation around the saddle points, one get the solution of kernel:-
\bea
\begin{array}{lll}
\displaystyle
\bar{K}_{N}(\lb,\mu)=-\frac{N}{2\pi N}\sum \frac{e^{-N(\bar f(z'_{1},v))}}{v+i z'_{1}}\frac{1}{\sqrt{\frac{\partial^2\bar f}{\partial v^2}\frac{\partial^2\bar f}{\partial {z'_{1}}^2}}}
\end{array}
\eea
\bea
\begin{array}{lll}
\displaystyle
\bar{K}_{N}(\theta,\phi)\text{:=}\frac{i \sqrt{1-c^2}c e^{\frac{1}{2} i (\theta +\phi )}}{4 \sqrt{\cos (\theta ) \cos (\phi )}} \Bigg(-\frac{\mathrm{Exp} (\frac{N}{2} (F[\theta,\phi]))}{(-ic)^{N}\left(e^{i \theta }-c e^{i \phi }\right)}+\frac{\mathrm{Exp} (\frac{N}{2} (F[-\theta,\phi]))}{(-ic)^{N}\left(e^{i \phi }-c e^{i \theta }\right)}\\
\\
\displaystyle
~~~~~~~~~~~~~~~~~~~~~~~~~~~~~~~~~~~~~~~+\frac{\mathrm{Exp} (\frac{N}{2} (F[\theta,-\phi]))}{(ic)^{N}\left(c+e^{i (\theta +\phi )}\right)}-\frac{\mathrm{Exp} (\frac{N}{2} (F[-\theta,\phi]))}{(ic)^{N}\left(1+c e^{i (\theta +\phi )}\right)}\Bigg)
\end{array}
\eea
Where $F[\theta,\phi]=2 i (\theta -\phi )-e^{-2 i \theta }+e^{-2 i \phi }$.
Two point Correlation Function then represented as Eq:-\ref{kernel}.
One get by putting $\theta=0$ and by the change $\phi$ with 
\[
\omega=\frac{2\sin{\phi}}{\sqrt{1-c^2}},
\]

\bea \label{bessel}
\begin{array}{lll}
\displaystyle
\rho^{II}_c[0,\omega]=\frac{(1-c^2)A\times\mathrm{Exp}[-N]}{4 N^{2} \sqrt{4-(1-c^2) \omega ^2}}
\Bigg[\frac{2(2-A) }{A (c A-2)(-ic)^{N}}\mathrm{Exp} \left\{\frac{ N A^2}{8}\right\}
\\
\displaystyle
~~~~~~~~~~~~~~~~~~~~~~+\frac{2 A^2}{(A^2-4 c^2)(ic)^{N}}\mathrm{Exp} \left\{\frac{ N A^2}{8}\right\}\Bigg]\times
\\
\displaystyle
\Bigg[\frac{2 A^2 }{( A^2-4 c^2)(-ic)^{N}}\mathrm{Exp} \left\{\frac{N}{8A^2}\right\}
+\frac{16}{A(c^2 A^2-4)(ic)^{N}}\mathrm{Exp} \left\{\frac{2N}{A^2}\right\}\Bigg]
\end{array}
\eea
Where $A= (\sqrt{4-(1-c^2) \omega ^2}+i \sqrt{1-c^2} \omega )$. For $\rho^{I}(0,\og)$ we simply used the definition in Eq:-\ref{kernel} and exact expression for first kernel from Eq:-\ref{KKTRY1}
\begin{figure}[H]
       \centering
         \subfigure[Two-point correlation function connected part $\rho^{I},\rho^{II}$ for c=0.9, N=20]{
      \includegraphics[width=7cm,height=5.5cm] {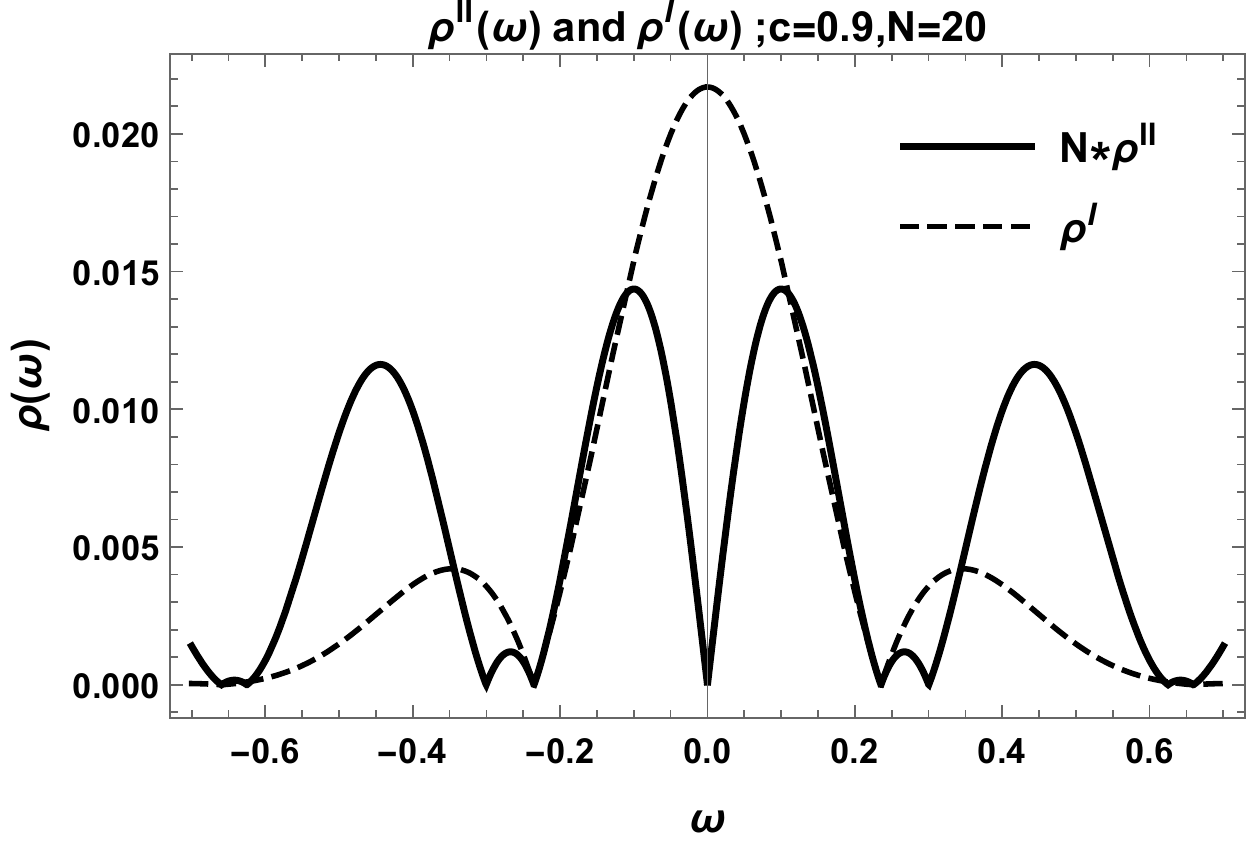}
     \label{}
         }
                  \subfigure[Two-point correlation function  connected part and Bessel Function of first kind comparison for c=0.9, N=50]{
      \includegraphics[width=7cm,height=5.5cm] {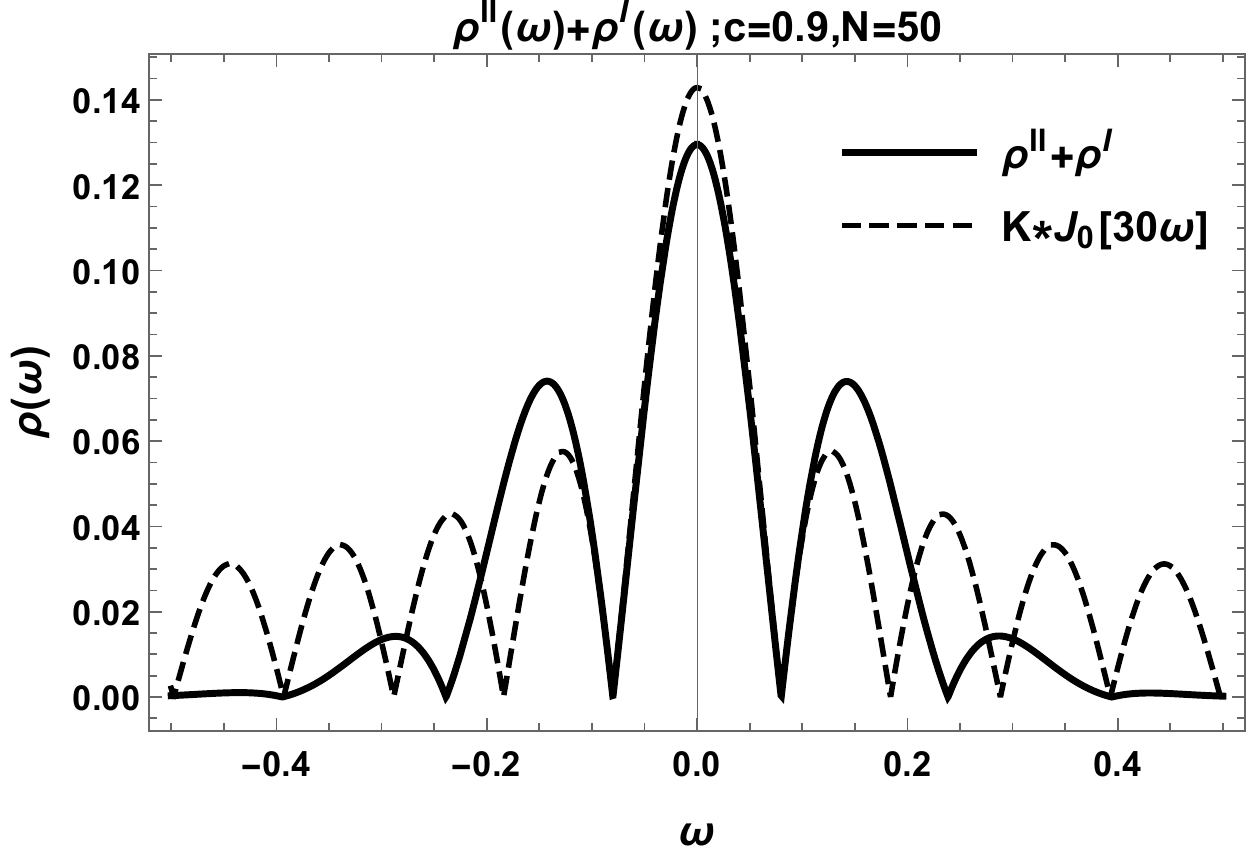}
     \label{}
         }

                  \caption[Optional caption for list of figures]{Two-point correlation function connected part  behavior using Eq:-\ref{bessel} and Eq:-\ref{connectedcorrel}.This is compared with Bessel function of first kind.} 
                                                                    \label{gw}
                                                                    \end{figure}   
Now we need to compute Fourier transform of this two point correlation function to get the dynamical form factor.
\be \label{S1}
\displaystyle
S_{c}(\tau)=\int \frac{d\omega}{2\pi}e^{i\omega\tau}\rho^{(2)}_c(E,\omega)
\ee
We choose the singularities of the above equation to evaluate the contour integral. Poles of the equation are
\bea\label{singular}
\begin{array}{lll}
\displaystyle
\omega= -\frac{i\sqrt{c^2-1}}{c}~~~~\omega= \frac{i\sqrt{c^2-1}}{c}
\end{array}
\eea
Evaluating the integral w.r.t the second saddle point $\frac{i\sqrt{c^2-1}}{c}$ suggests that
\bea \label{heisencontrol}
\begin{array}{lll}
\rho(\tau)\simeq \tau e^{-\frac{\sqrt{1-c}}{\epsilon} \tau}
\end{array}
\eea
 Fig:-\ref{g1} support our argument.
\begin{figure}[H]
       \centering
         \subfigure[SFF for c=0.9, N=5, showing behavior as predicted in \ref{S1}]{
      \includegraphics[width=7cm,height=5.5cm] {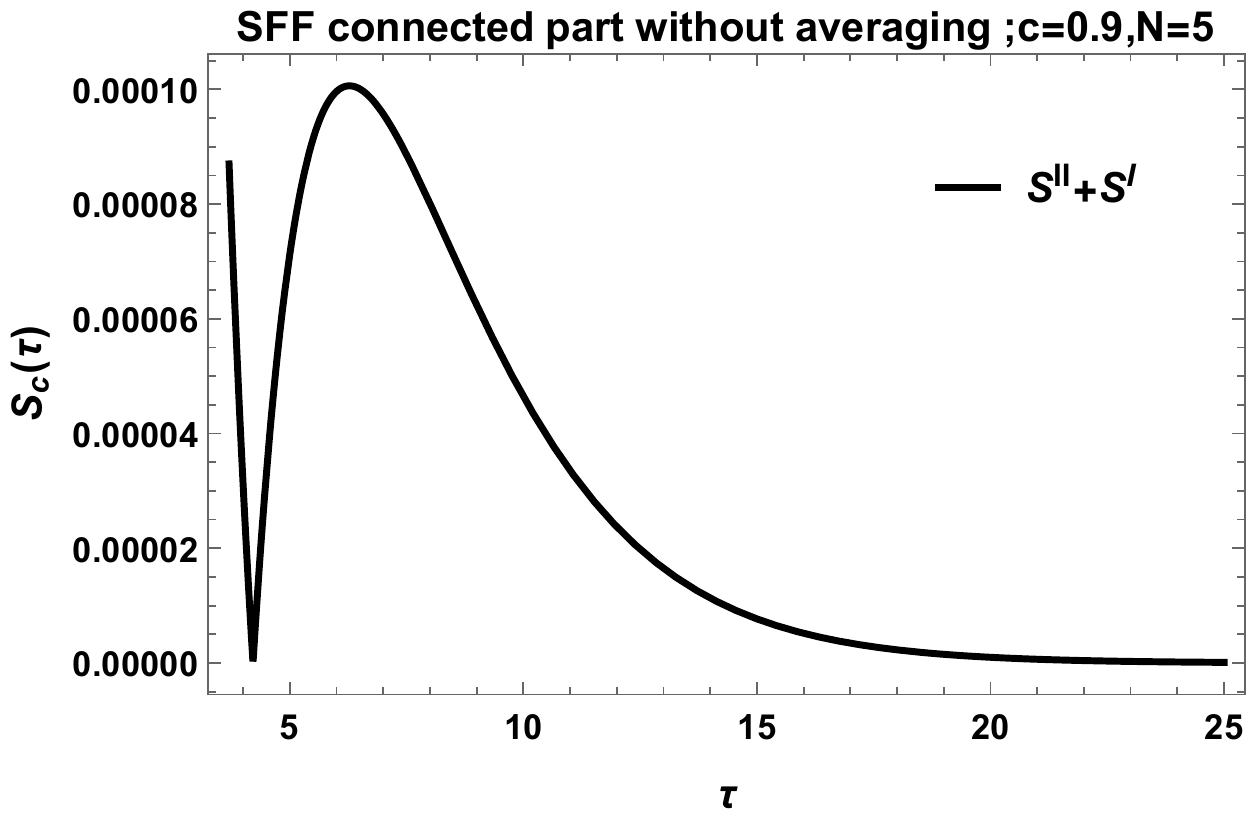}
     \label{}
         }   
             \subfigure[SFF for c=0.9, N=7, showing behavior as predicted in \ref{S1}]{
      \includegraphics[width=7cm,height=5.5cm] {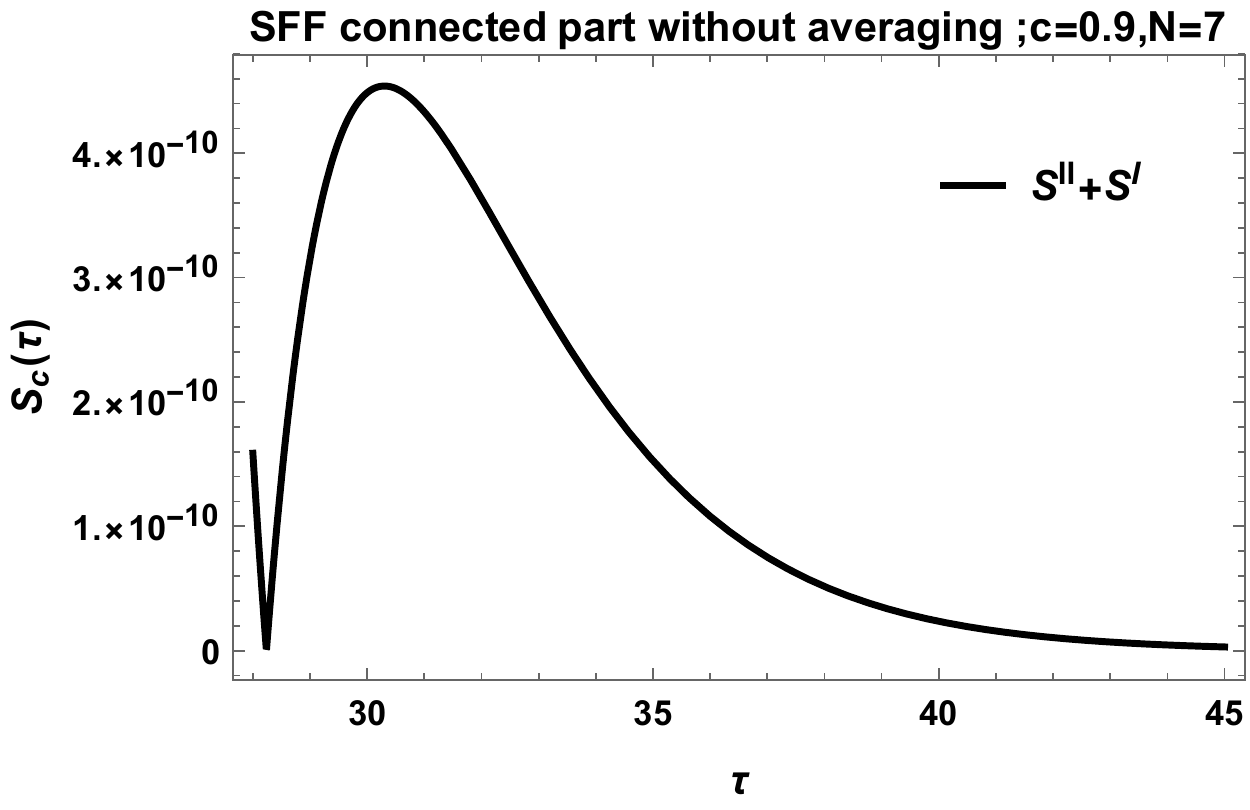}
     \label{}
         }
                  \caption[Optional caption for list of figures]{SFF of connected part $S_c$ ($S^{I}+S^{II}$) from Eq:-(\ref{S1})} 
                                                                    \label{g1}
                                                                    \end{figure}                       
                                                                                                  
\subsection{Average of SFF}
\begin{figure}[H]
       \centering
        \subfigure[SFF average c=0.9 for N=7 in LogLog scale, with saturation value at $0.028$]{
      \includegraphics[width=7cm,height=5cm] {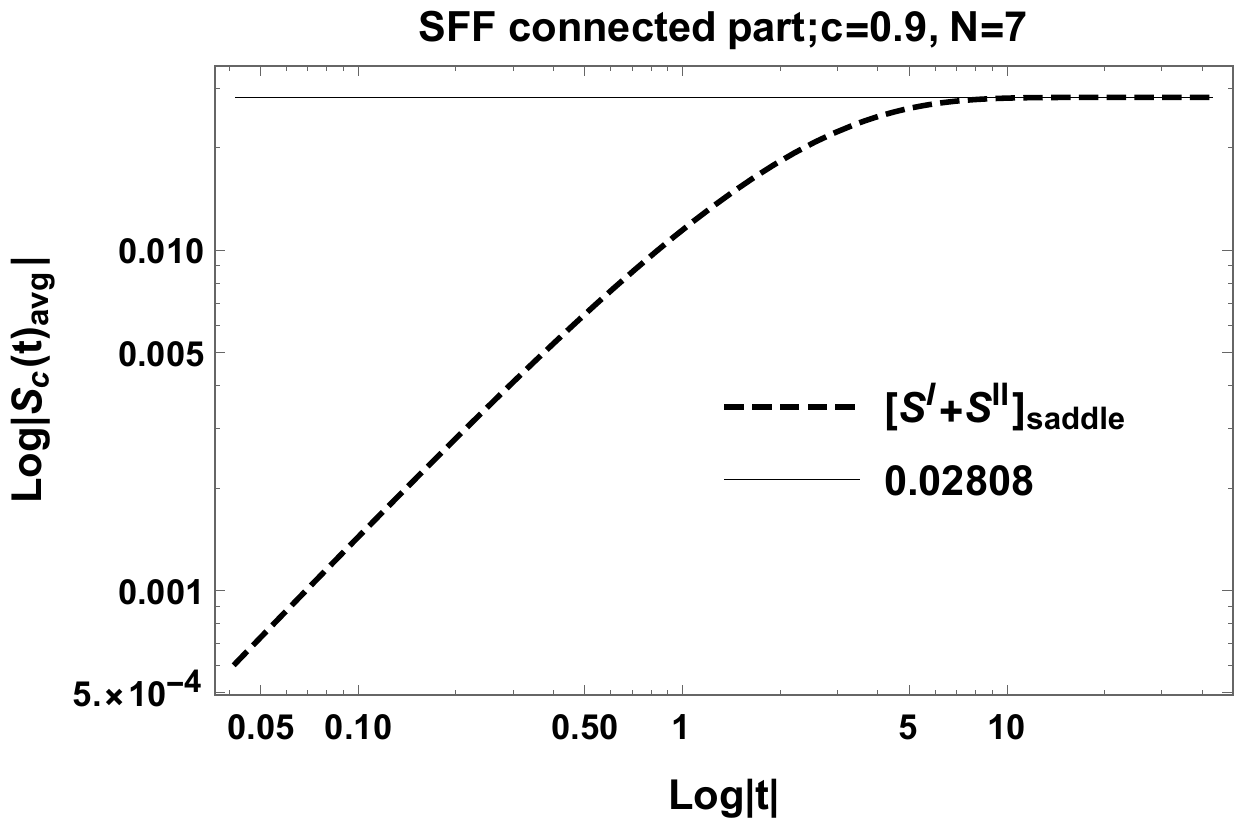}
     \label{}
         }
       \subfigure[SFF average c=0.9 for N=10 in LogLog scale, with saturation value 0.0321]{
      \includegraphics[width=7cm,height=5cm] {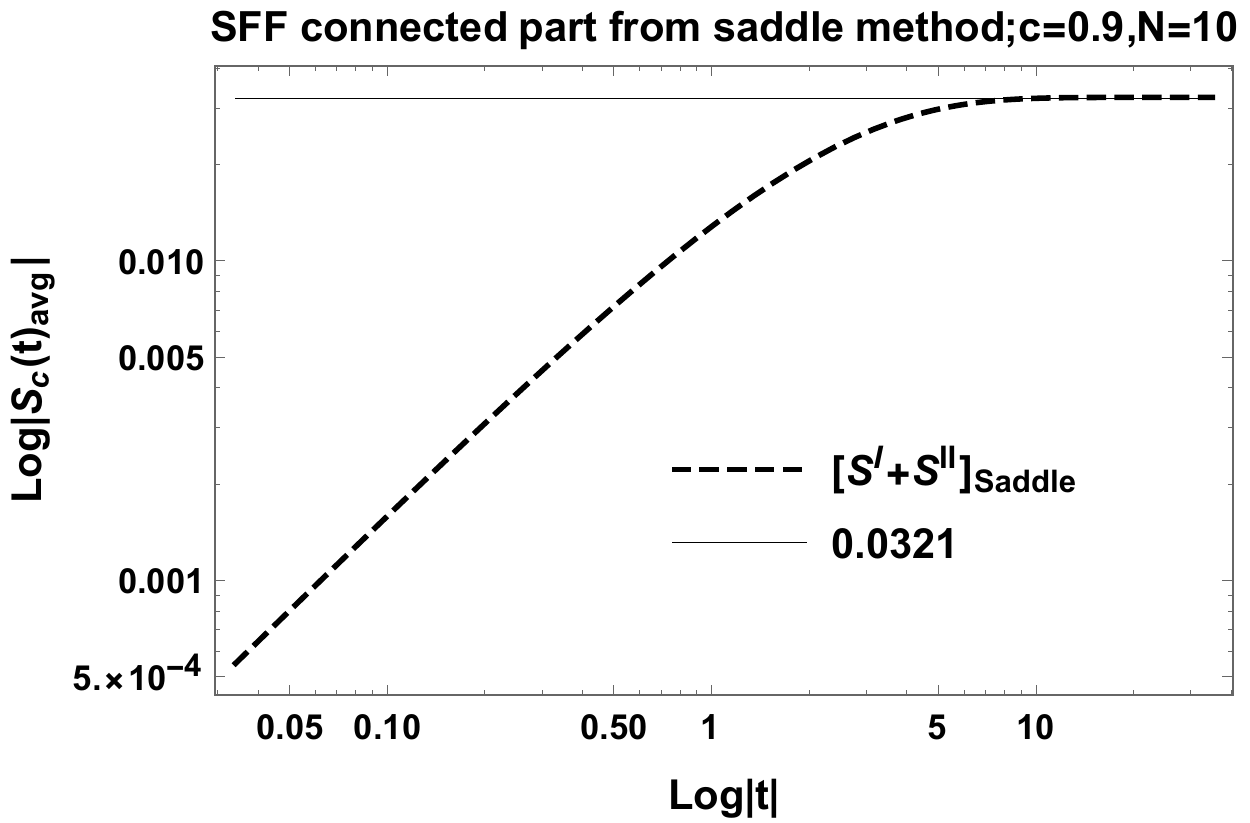}
     \label{}
         }
        \subfigure[SFF average c=0.9 for N=20 in LogLog scale,with saturation value around 0.57]{
      \includegraphics[width=7cm,height=5cm] {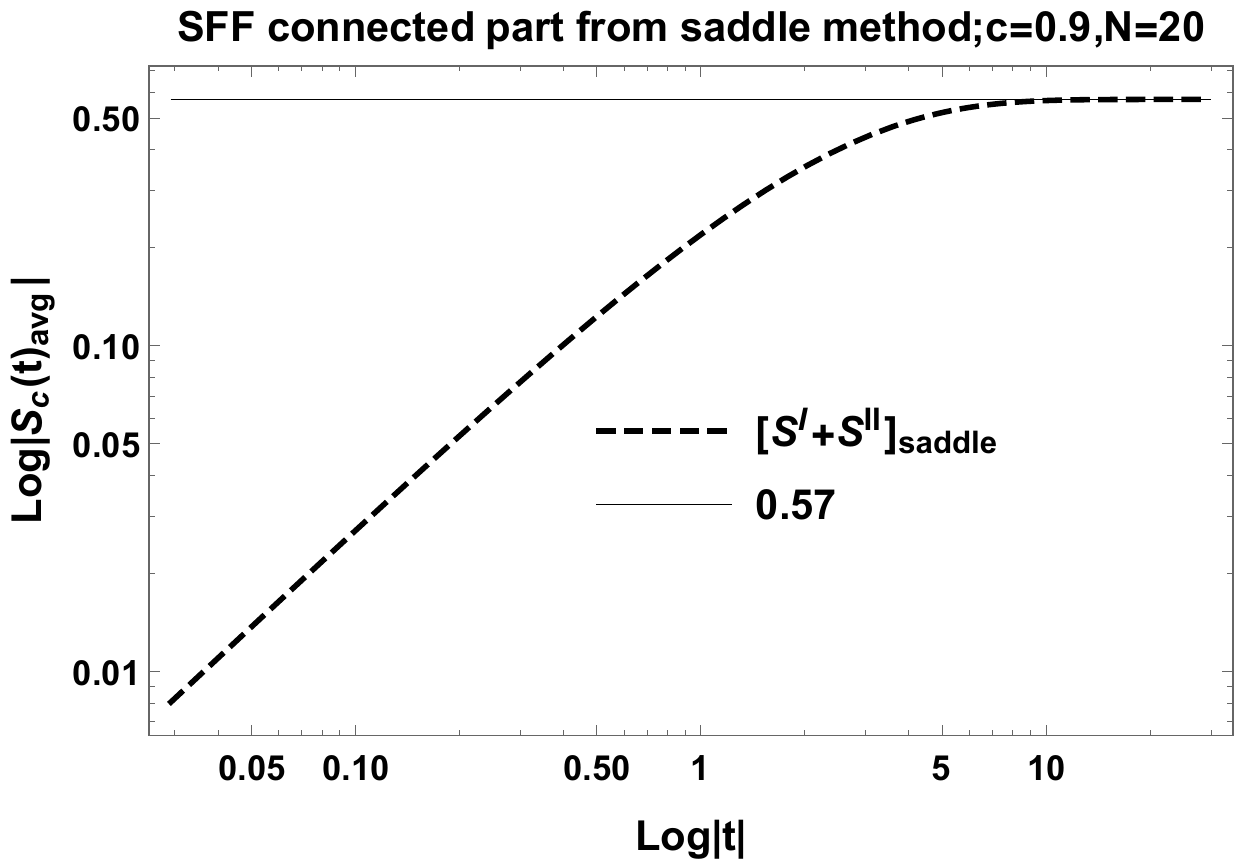}
     \label{}
         }
          \subfigure[SFF average c=0.9 for N=30 in LogLog scale, with saturation near 17.36]{
      \includegraphics[width=7cm,height=5cm] {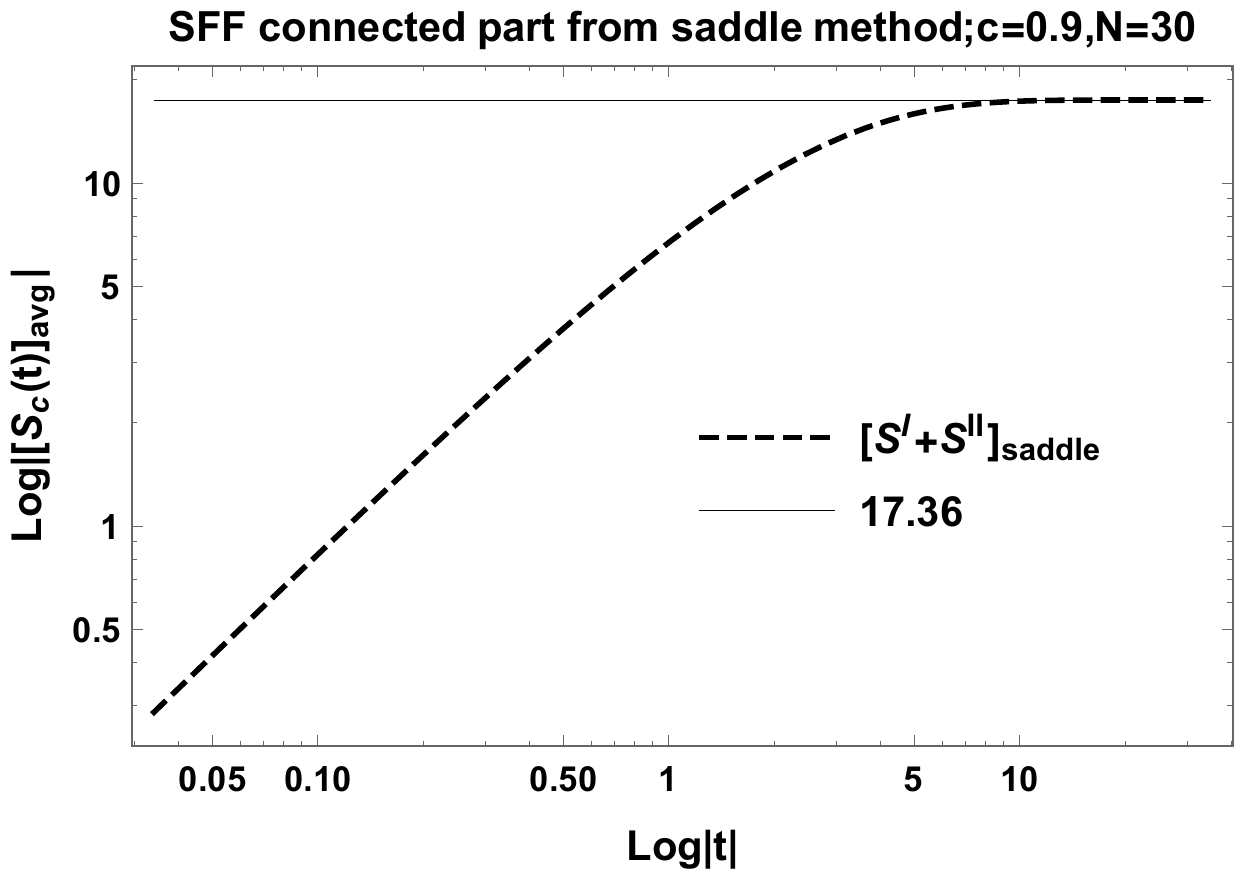}
     \label{}
         }
                  \caption[Optional caption for list of figures]{Spectral Form Factor averaged over an interval $[0,t]$  from  Eq:-(\ref{S2}) is plotted in Log Log scale. We have considered both connected and disconnected part in this case.} 
                                                                    \label{g278}
                                                                    \end{figure}

SFF can be averaged over an interval (0,t) and plotting that shows a continuous behavior instead of kink at Heisenberg time
\be \label{S2}
\langle S(\tau)\rangle=\int_{0}^{t} d\tau S(\tau)=[S_{c}(t)]_{avg}
\ee
From Fig:-\ref{hermite123} and Fig:-\ref{g278} SFF averaged over interval $[0,t]$.  $\left[ S_{c}(t)\right]_{avg}$  from Hermite polynomial method and saddle point analysis shows same kind of continuous transition behavior near Heisenberg time. We have compared solution of SFF average from both Hermite method \ref{art12} and saddle point analysis \ref{S2}. In Section :-\ref{sec4} we have compared the change in saturation value of SFF and nature of this rounding off (Heisenberg Time).

\section{$\frac{1}{N}$ expansion for correlation function}
\label{sec4}
Now we repeat well know saddle point method of one variable for two variable saddle point approximation
\be
\int_{-\infty}^{\infty} dx dy f(x,y)e^{AH(x,y)}
\ee
For saddle point approximation we choose the main contributing points of the integral and this set of point is given by:-
\be
\frac{\partial H(x,y)}{\partial x}|_{(x_{0},y_{0})}=0, \frac{\partial H(x,y)}{\partial y}|_{(x_{0},y_{0})}=0, \frac{\partial^{2} H(x,y)}{\partial y \partial x}|_{(x_{0},y_{0})}=0
\ee
 Now we change the integration variable to 
\be 
x=x_{0}+\frac{w}{\sqrt{A}} ~~~~~~~y=y_{0}+\frac{z}{\sqrt{A}}
\ee

Now we make Taylor expansions of $AH(x,y)$  and $F(x,y)$ around $x_{0}$  and $y_{0}$ and choose up to certain terms to get the $\frac{1}{A^{n}}$ terms completely for $n=\frac{1}{2},1$ in  $e^{AH(w,z)}F(w,z)$ and then perform Gaussian Integration to get the 1st and second term of saddle point approximation
\bea
\begin{array}{lll}\label{saddle}
\displaystyle
\int_{-\infty}^{\infty} dx dy F(x,y)e^{AH(x,y)}=\Bigg\{\frac{2 \pi  F\left(x_0,y_0\right) e^{A H\left(x_0,y_0\right)}}{A \sqrt{H^{(2,0)}\left(x_0,y_0\right)} \sqrt{H^{(0,2)}\left(x_0,y_0\right)}}\Bigg\}+
\\
\displaystyle
\Bigg(\frac{\pi  e^{A H\left(x_0,y_0\right)}}{4 A^2 H^{(0,2)}\left(x_0,y_0\right){}^{5/2} H^{(2,0)}\left(x_0,y_0\right){}^{5/2}}\Bigg)\times
\\
\displaystyle
\Bigg\{H^{(0,2)}\left(x_0,y_0\right))^{2}\bigg[4 F^{(2,0)}\left(x_0,y_0\right) H^{(2,0)}\left(x_0,y_0\right)+4 F^{(1,0)}\left(x_0,y_0\right) H^{(3,0)}\left(x_0,y_0\right)
\\
\displaystyle
~~~+F\left(x_0,y_0\right) H^{(4,0)}\left(x_0,y_0\right)\bigg]+H^{(0,2)}\left(x_0,y_0\right) \bigg[4 F^{(0,2)}\left(x_0,y_0\right) H^{(2,0)}\left(x_0,y_0\right){}^2
\\
\displaystyle
+4 F^{(1,0)}\left(x_0,y_0\right) H^{(1,2)}\left(x_0,y_0\right) H^{(2,0)}\left(x_0,y_0\right)+4 F^{(0,1)}\left(x_0,y_0\right) H^{(2,1)}\left(x_0,y_0\right) H^{(2,0)}\left(x_0,y_0\right)
\\
\displaystyle
+2 F\left(x_0,y_0\right) H^{(2,2)}\left(x_0,y_0\right) H^{(2,0)}\left(x_0,y_0\right)+3 F\left(x_0,y_0\right) H^{(2,1)}\left(x_0,y_0\right){}^2
\\
\displaystyle
+2 F\left(x_0,y_0\right) H^{(1,2)}\left(x_0,y_0\right) H^{(3,0)}\left(x_0,y_0\right)\bigg]+H^{(2,0)}\left(x_0,y_0\right)\bigg[4 F^{(0,1)}\left(x_0,y_0\right) H^{(0,3)}\left(x_0,y_0\right) 
\\
\displaystyle
H^{(2,0)}\left(x_0,y_0\right)+F\left(x_0,y_0\right) (3 H^{(1,2)}\left(x_0,y_0\right){}^2+H^{(0,4)}\left(x_0,y_0\right) H^{(2,0)}\left(x_0,y_0\right)
\\
\displaystyle
~~~~~~~~~~~~~~~~~~~~~~~~~~~~~+2 H^{(0,3)}\left(x_0,y_0\right) H^{(2,1)}\left(x_0,y_0\right))\bigg]\Bigg\}
\end{array}
\eea
\subsection{Second order contribution of SFF}
Now we evaluate next order contribution for correlation function using second term of the Eq:-\ref{saddle}.
We use this relation for expression of both the kernels in Eq:-(\ref{1stsaddle}) and Eq:-(\ref{2ndsaddle}). We follow the exactly same procedure thereafter and at first evaluate the two-point correlation function.
 \begin{figure}[H]
       \centering
         \subfigure[1st order correction for c=0.9, N=100]{
      \includegraphics[width=7cm,height=5cm] {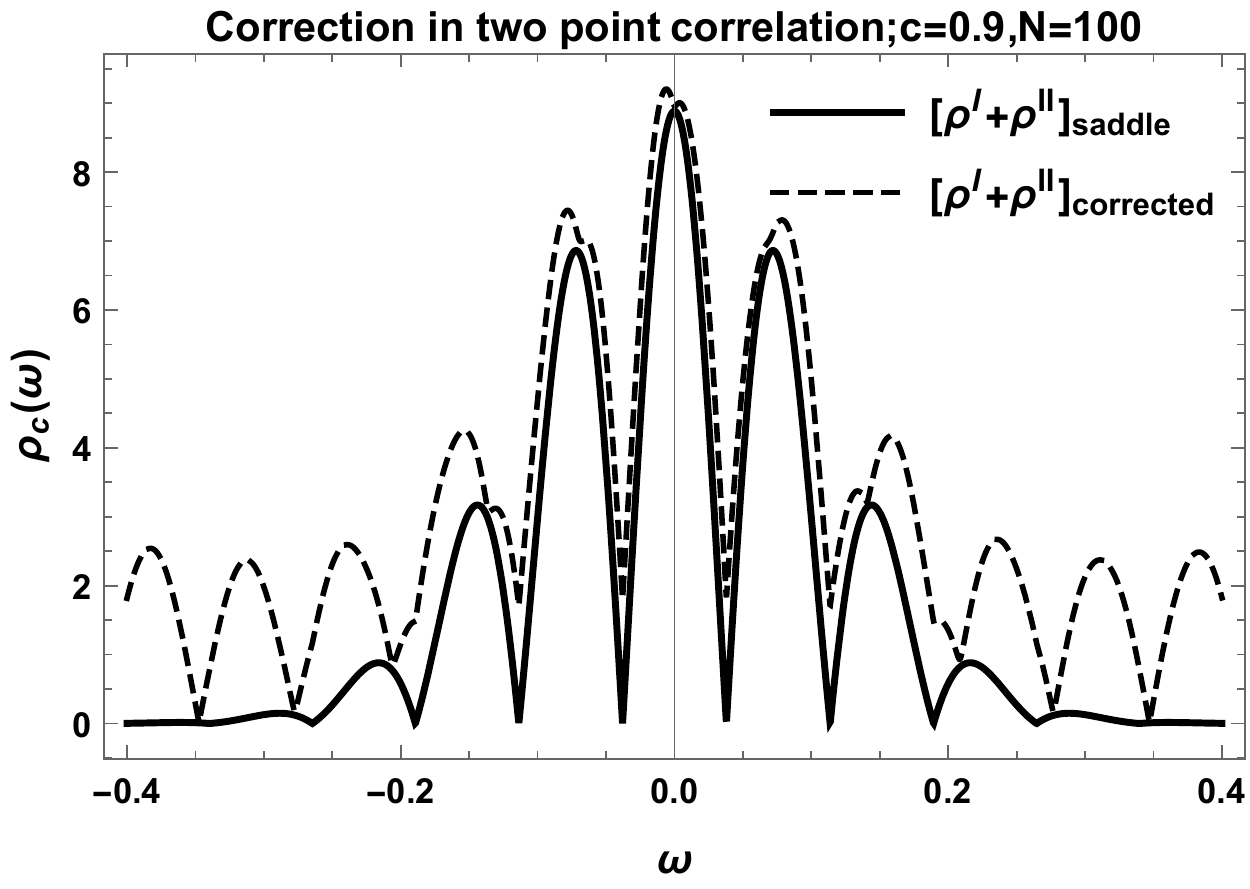}
     \label{}
         }
          \subfigure[1st order correction for c=0.9, N=20]{
      \includegraphics[width=7cm,height=5cm] {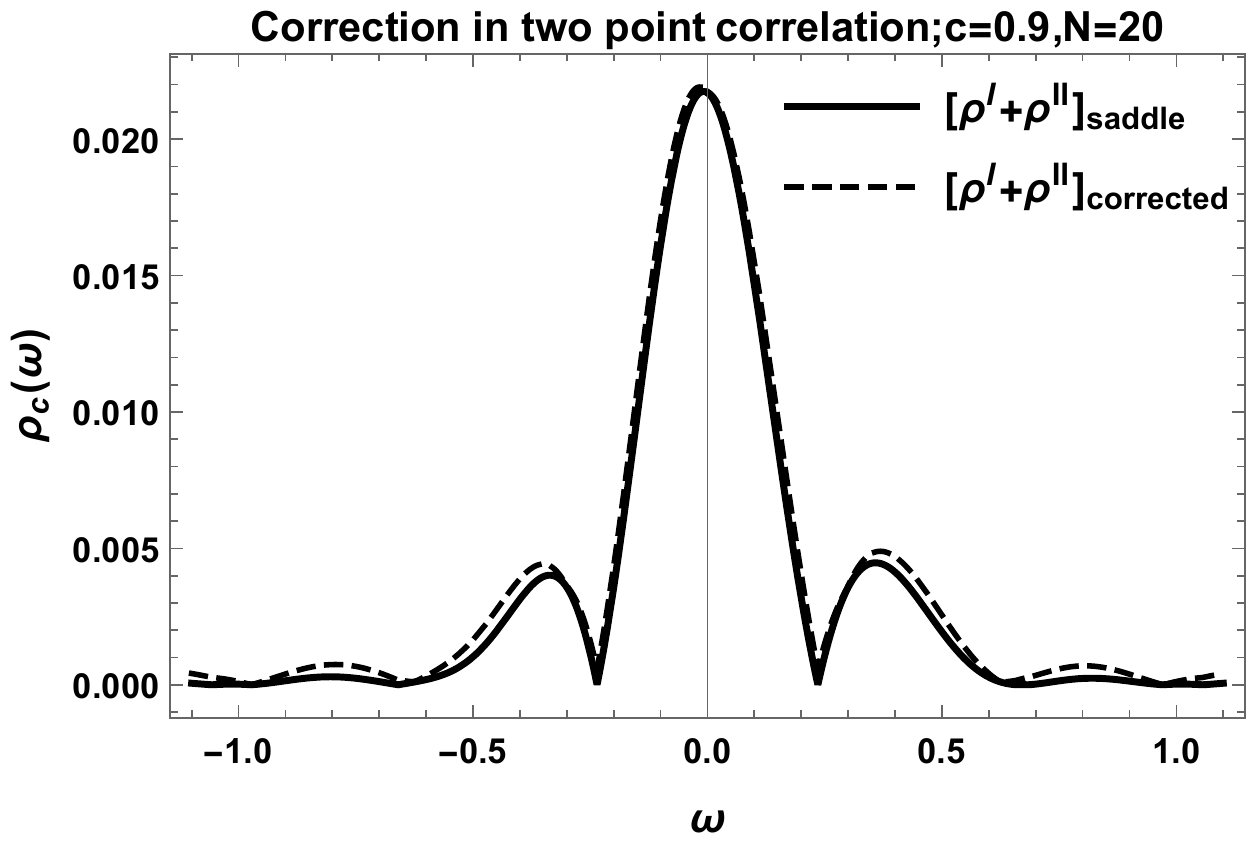}
     \label{}
         }
                  \caption[Optional caption for list of figures]{Correlation function w.r.t $\og$ for different dimension of Matrix(N) and effect of correction from second term of Eq:-\ref{saddle}} 
                                                                    \label{g4}
                                                                    \end{figure} 
                                                                   
\begin{figure}[H]
       \centering
       \subfigure[SFF average from different method for c=0.9 for N=30]{
      \includegraphics[width=7cm,height=5.8cm] {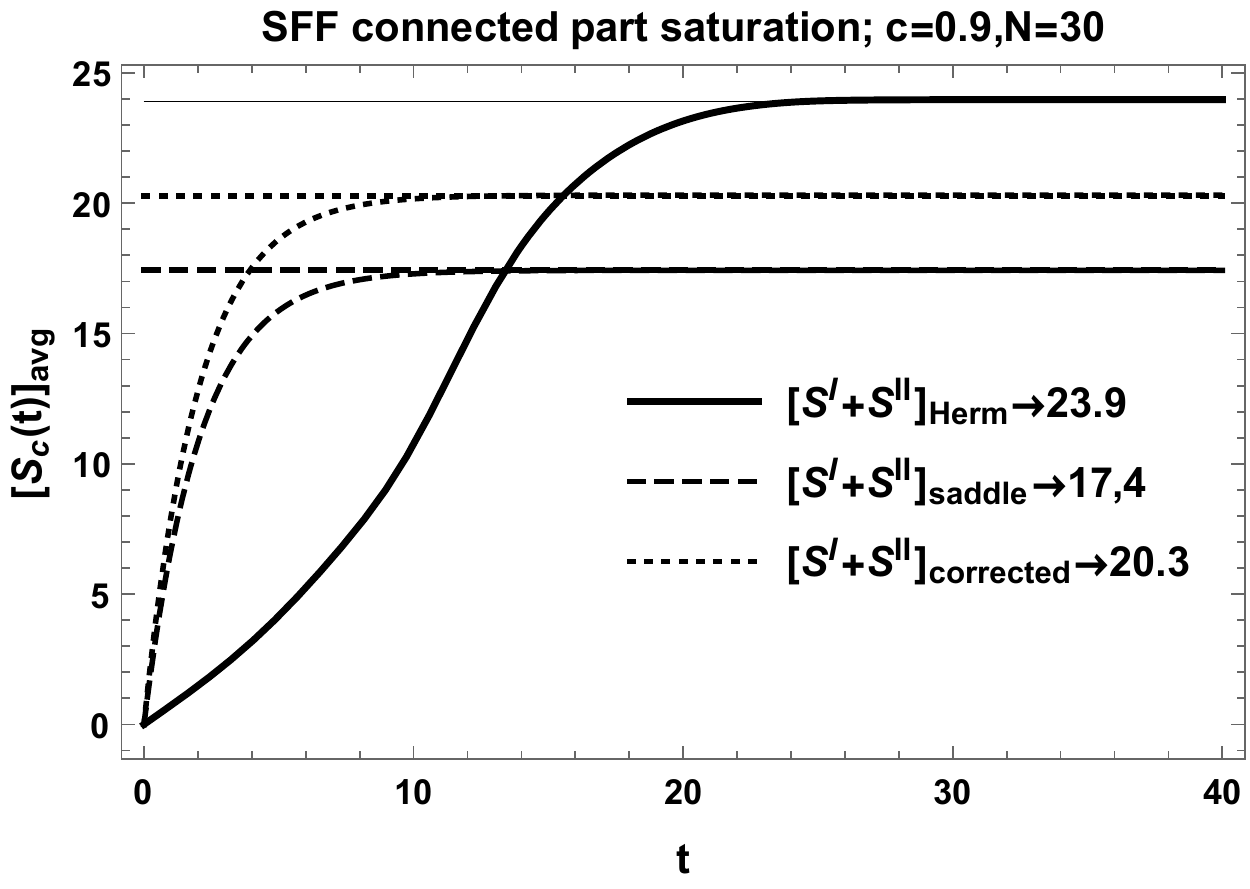}
     \label{}
         }
        \subfigure[Comparison of SFF average from various method and correction for c=0.9 for N=15]{
      \includegraphics[width=7cm,height=5.8cm] {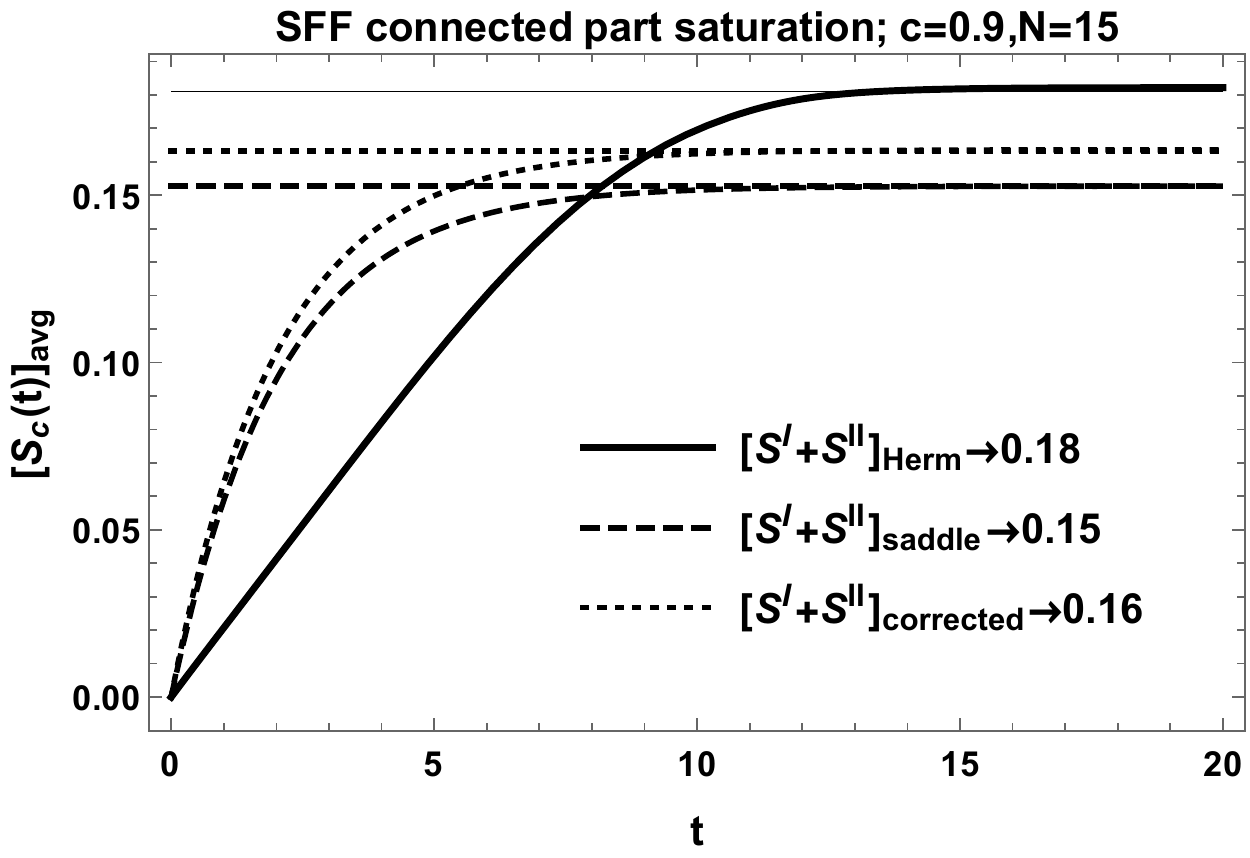}
     \label{}
         }
                \subfigure[Comparison of SFF average with , without correction, hermite method solution for c=0.9 for N=10]{
      \includegraphics[width=6.8cm,height=5.8cm] {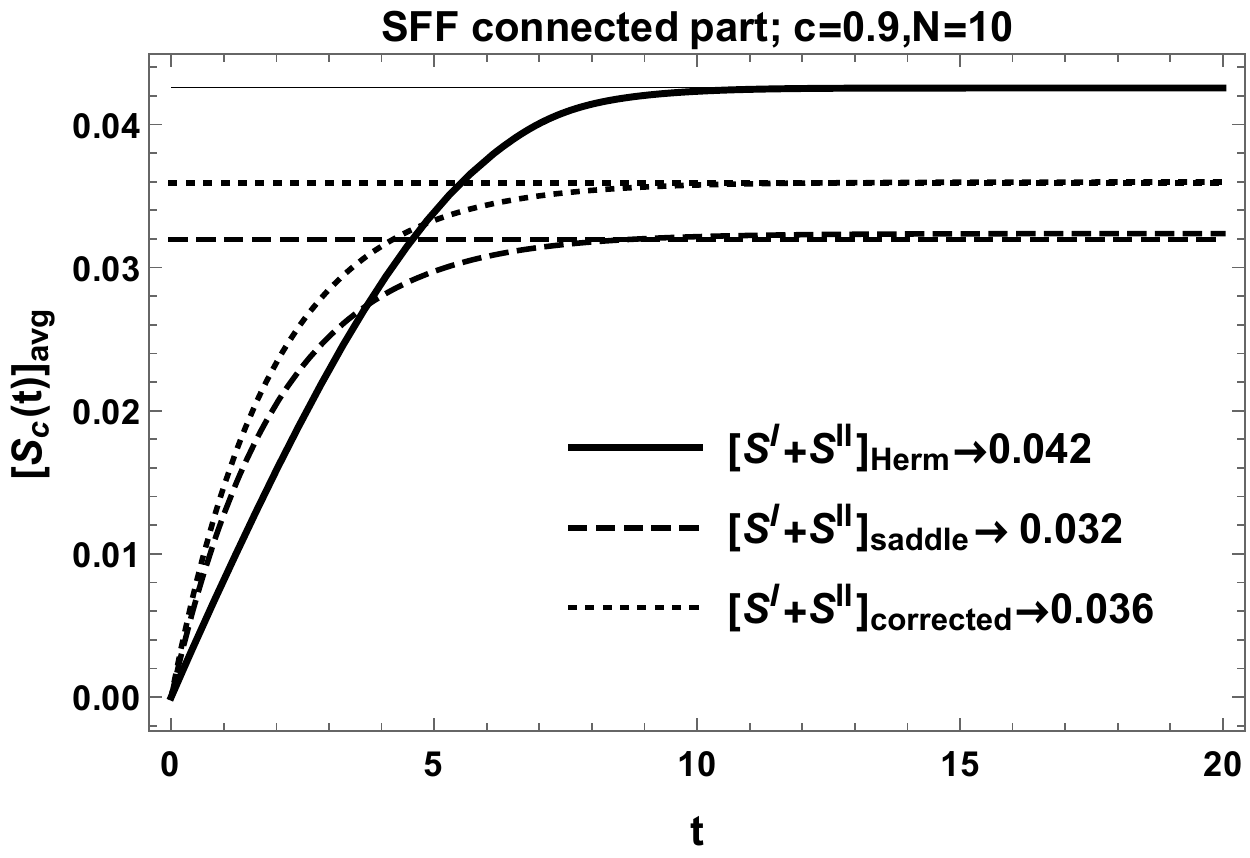}
    \label{}
         }   
                   \subfigure[SFF average from zeroth order,with 1st order saddle approximation, hermite method for c=0.9 for N=7]{
      \includegraphics[width=7cm,height=5.8cm] {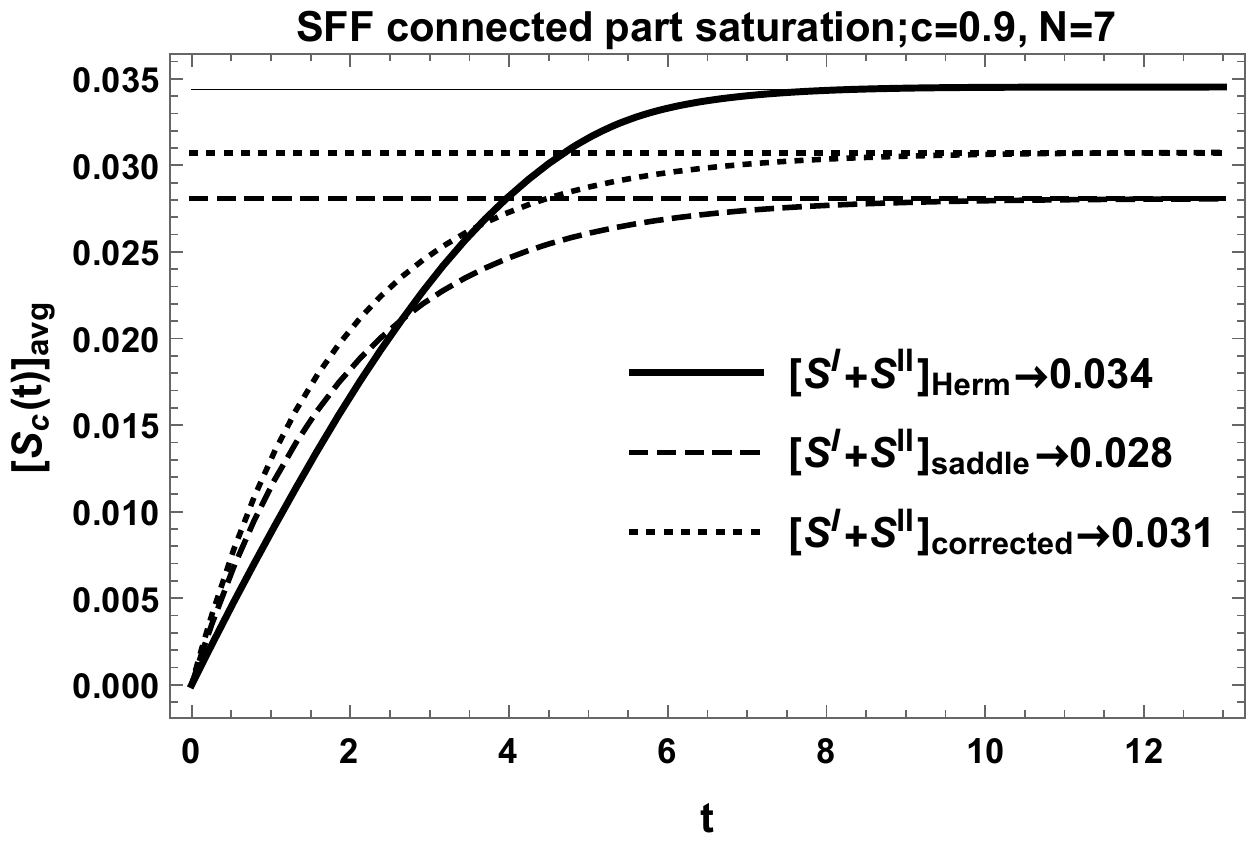}
     \label{}
         }
                  \caption[Optional caption for list of figures]{ SFF average has shift in saturation values. Here different solution from Hermite method, zeroth order saddle method and 1st order corrected saddle approximation is compared.} 
                                                                    \label{g8}
                                                                    \end{figure}

Now we evaluate spectral form factor by Fourier transform exactly as Eq:-(\ref{S1}). We choose the singularities of this equation to find the integral by residue theorem. The singularities are same as previous case (Eq:-\ref{singular}). Then we compute residue w.r.t these  points.
 \begin{figure}[H]
       \centering
          \subfigure[Effect of 1st order correction in SFF connected part for c=0.9, N=5]{
      \includegraphics[width=7cm,height=5cm] {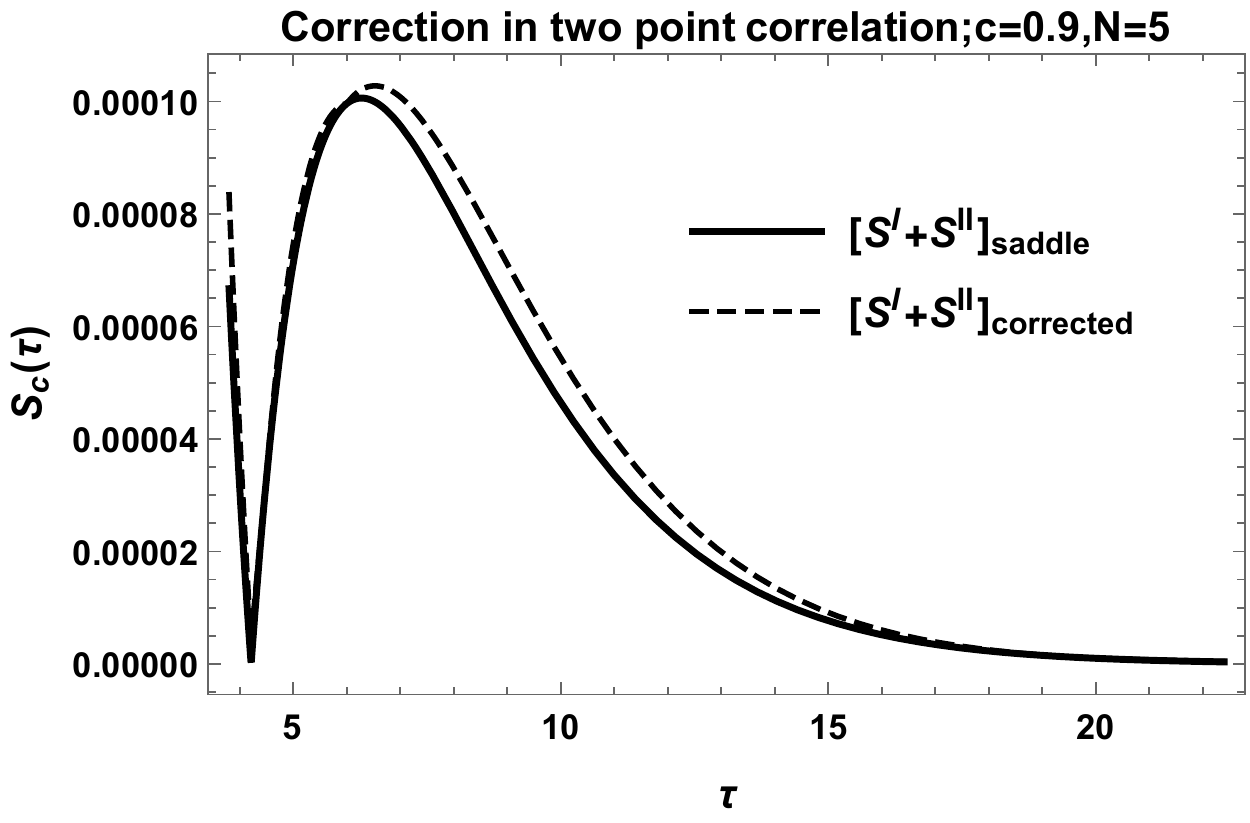}
     \label{}
         }
         \subfigure[Effect of 1st order correction in SFF connected part for c=0.9,N=7]{
      \includegraphics[width=7cm,height=5cm] {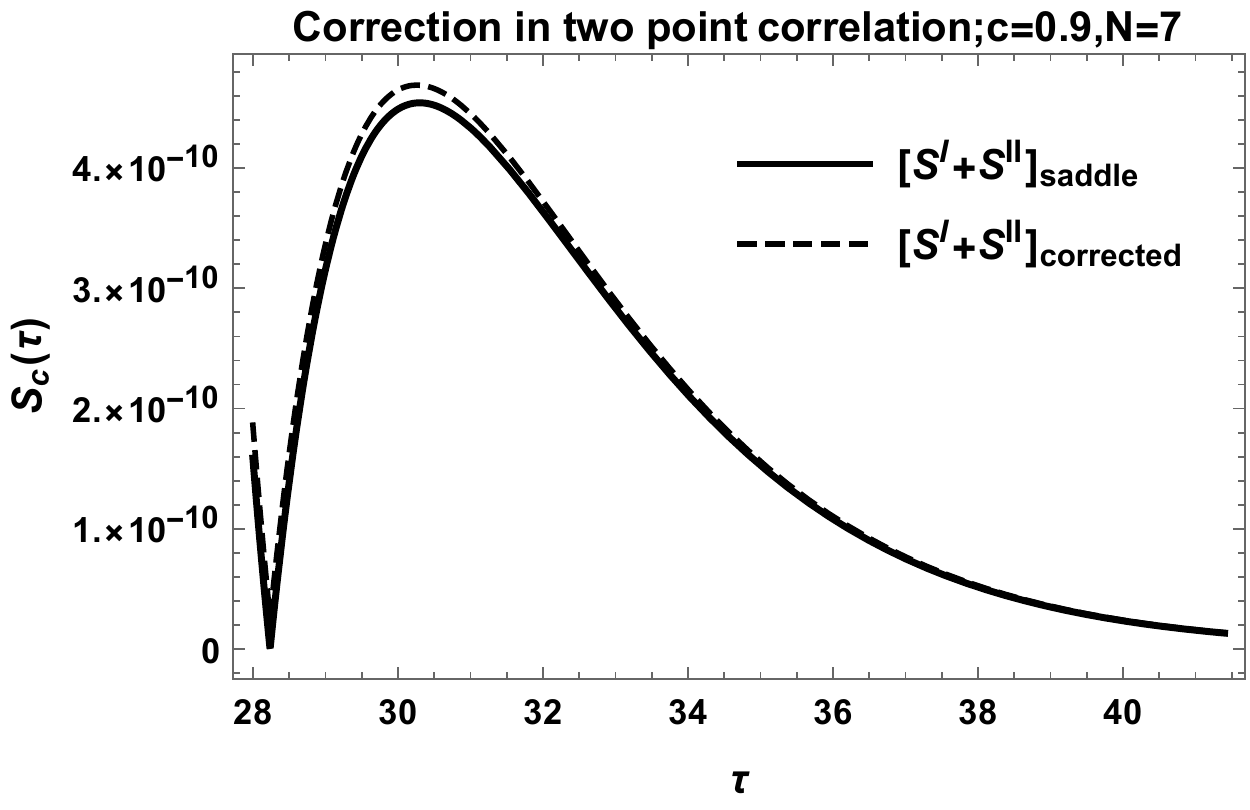}
     \label{sfcomp}
         }
                  \caption[Optional caption for list of figures]{1st order correction of Spectral Form Factor w.r.t $\tau$ for $N$=20 behave in the same way as the zeroth order solution. } 
                                                                    \label{g6}
                                                                    \end{figure} 
Considering the second term of Eq:-\ref{saddle} we have added the correction term in SFF and plotted it in Fig:-\ref{sfcomp}
SFF can be averaged over an interval [0,t] and plotting that shows a continuous behavior instead of kink at Heisenberg time
\be \label{S21}
\langle S(\tau)\rangle=\int_{0}^{t} d\tau S(\tau)=S_{avg}(t)
\ee
\begin{figure}[H]
       \centering
        \subfigure[Correction in SFF connected part average c=0.9 for N=10]{
      \includegraphics[width=7cm,height=4.5cm] {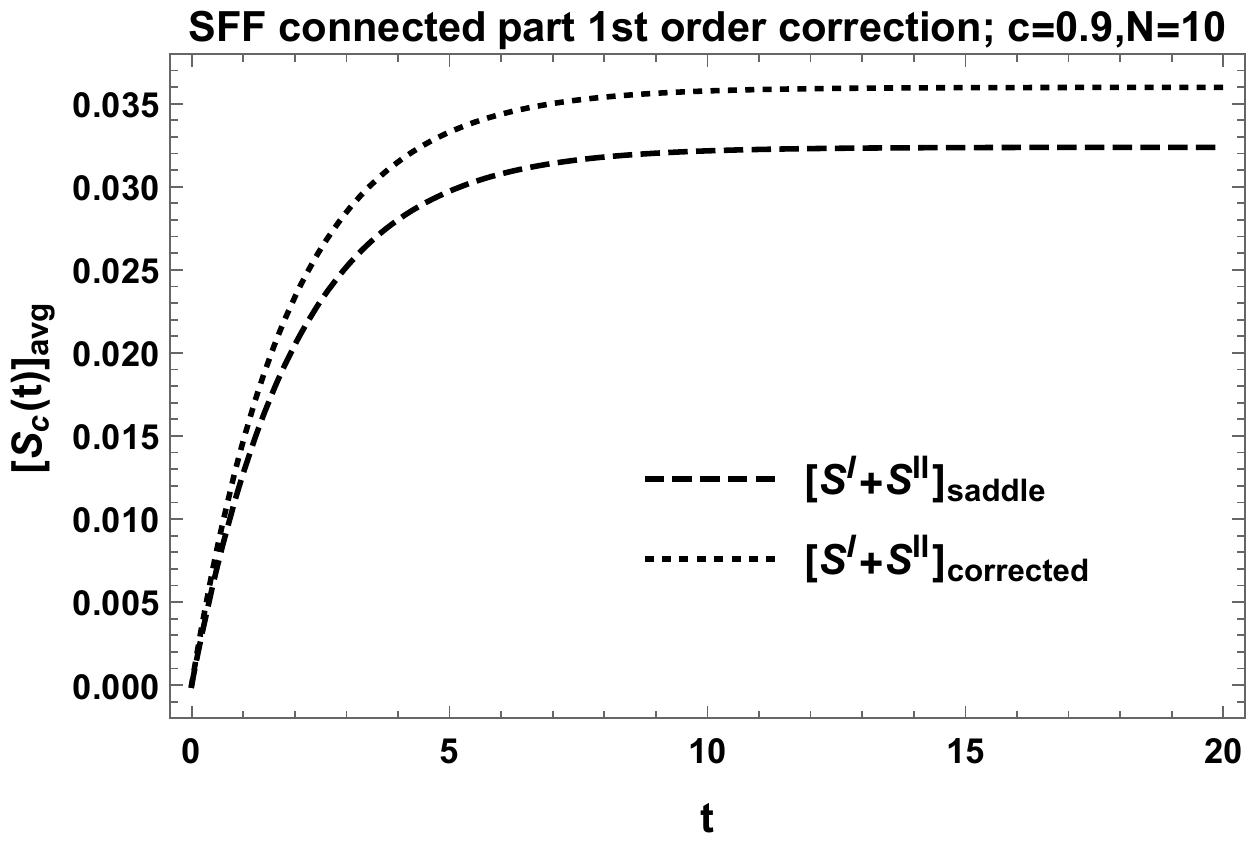}
     \label{}
         }
           \subfigure[Correction in SFF connected part average c=0.9 for N=30]{
      \includegraphics[width=7cm,height=4.5cm] {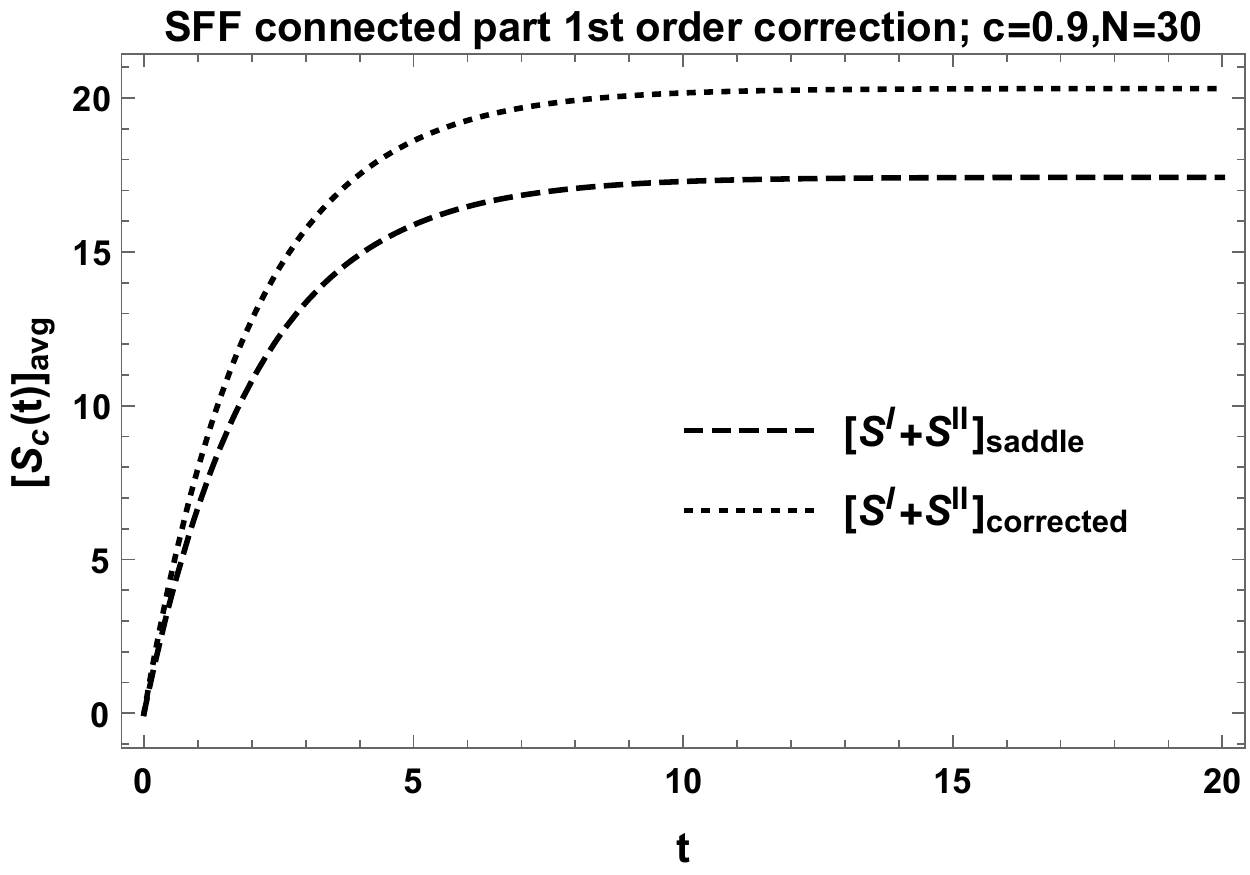}
     \label{}
         }
                  \caption[Optional caption for list of figures]{1st order of correction in SFF connected part average over interval [0,t] from Eq:-\ref{saddle}} 
                                                                    \label{g7}
                                                                    \end{figure}                                   
We can compare solution of SFF with first order correction term to our previously evaluated SFF with only zeroth order saddle approximation term and hermite polynomial solution for kernels 
\begin{figure}[H]
       \centering
        \subfigure[Comparison of SFF average for Hermite method solution and saddle method with and without correction for c=0.9, N=7]{
      \includegraphics[width=6.5cm,height=5cm] {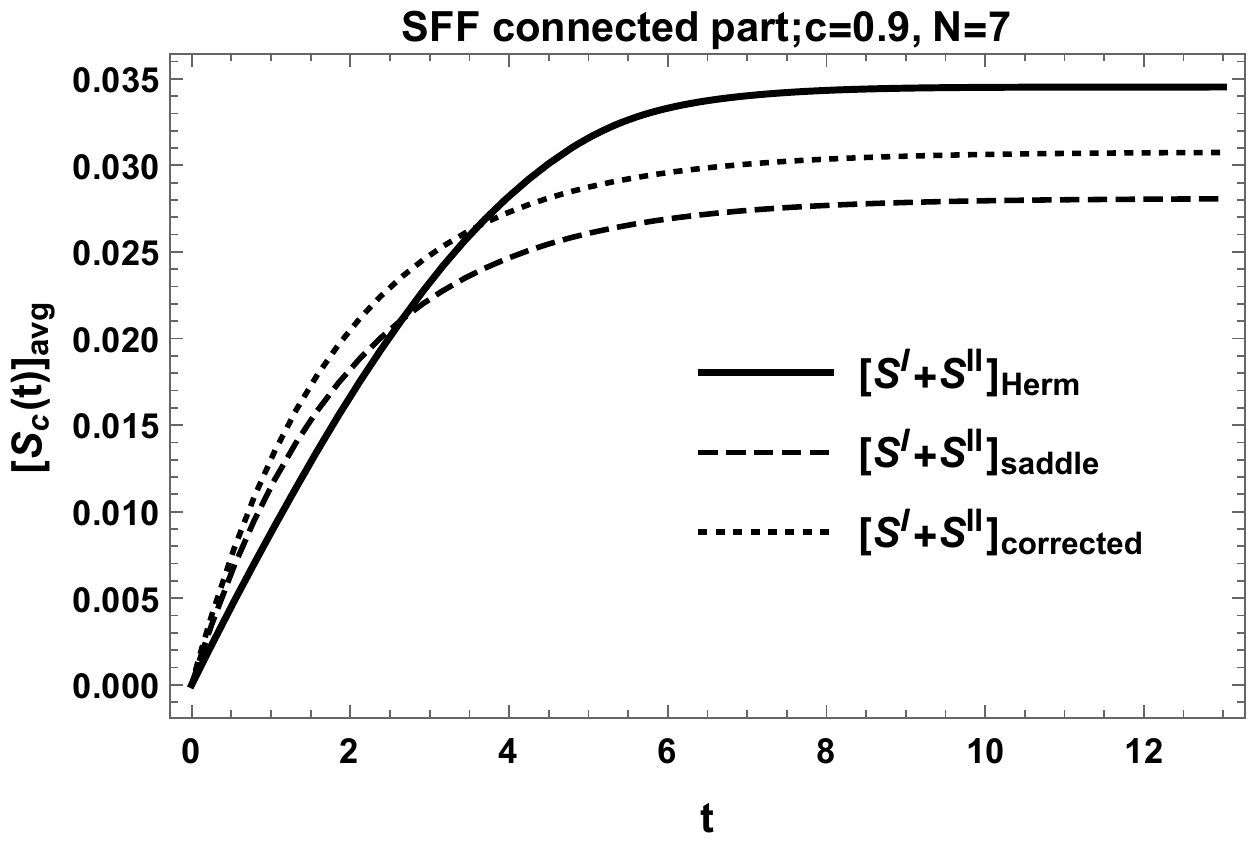}
     \label{}
         }
                   \subfigure[Comparison of SFF average from different method for c=0.9 for N=10]{
      \includegraphics[width=6.5cm,height=5cm] {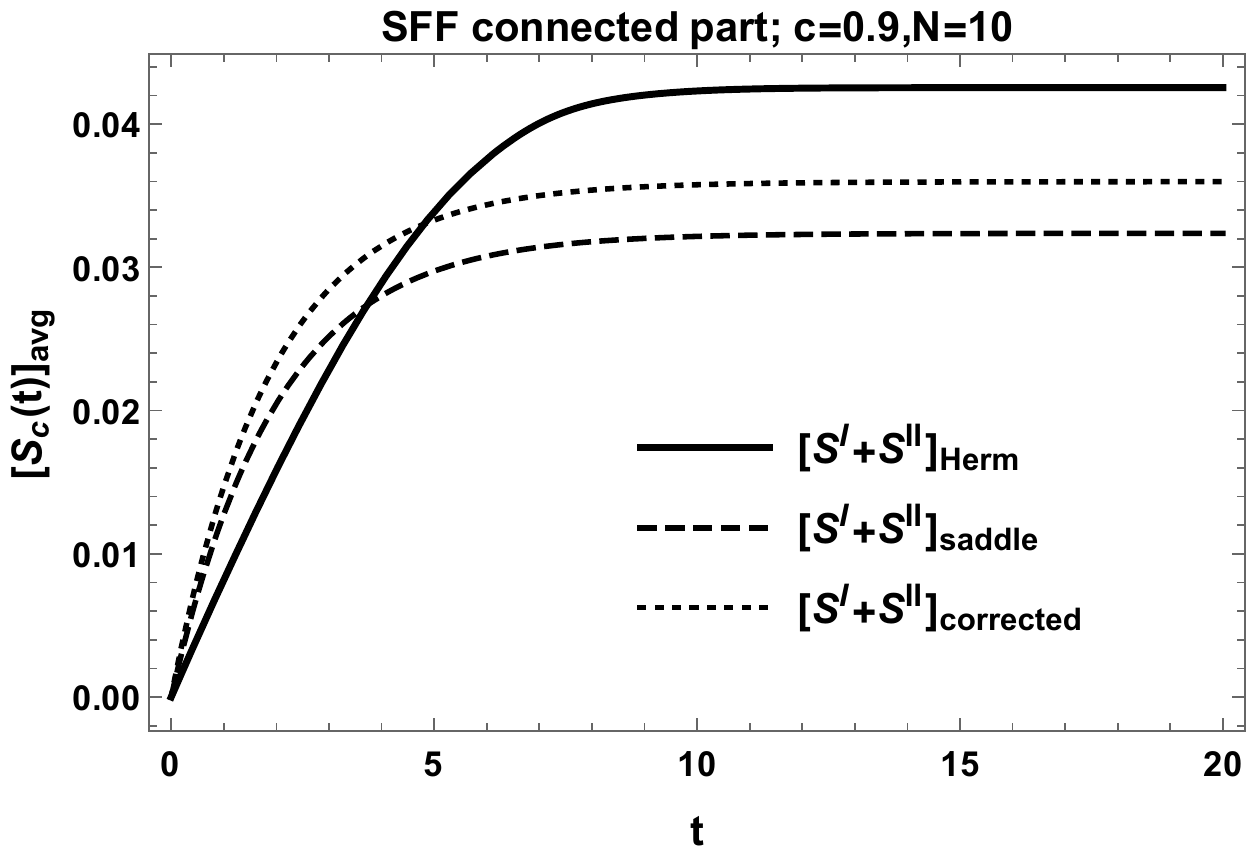}
     \label{}
         }   
          \subfigure[Comparison of SFF average from different method for c=0.9 for N=15]{
      \includegraphics[width=6.5cm,height=5cm] {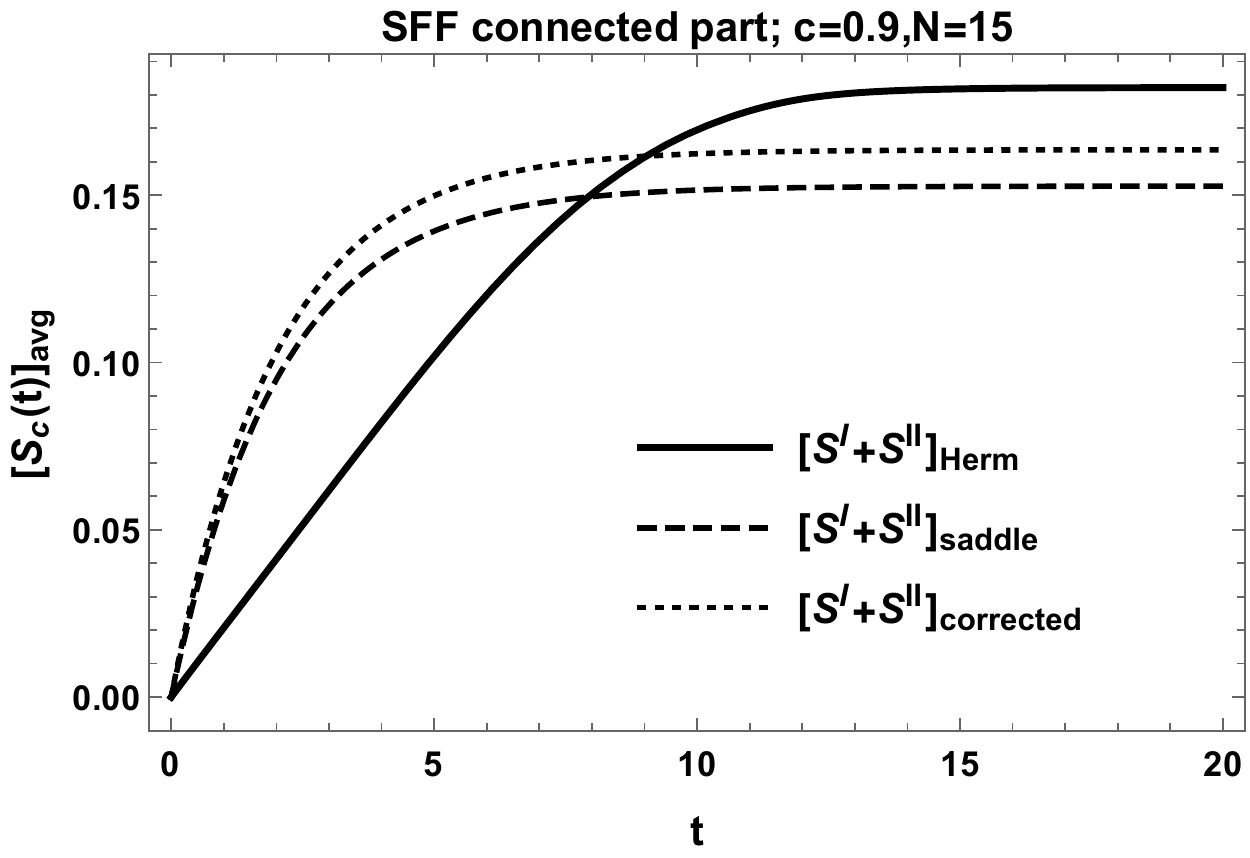}
     \label{}
         }
           \subfigure[Comparison of SFF average from different method for c=0.9 for N=30]{
      \includegraphics[width=6.5cm,height=5cm] {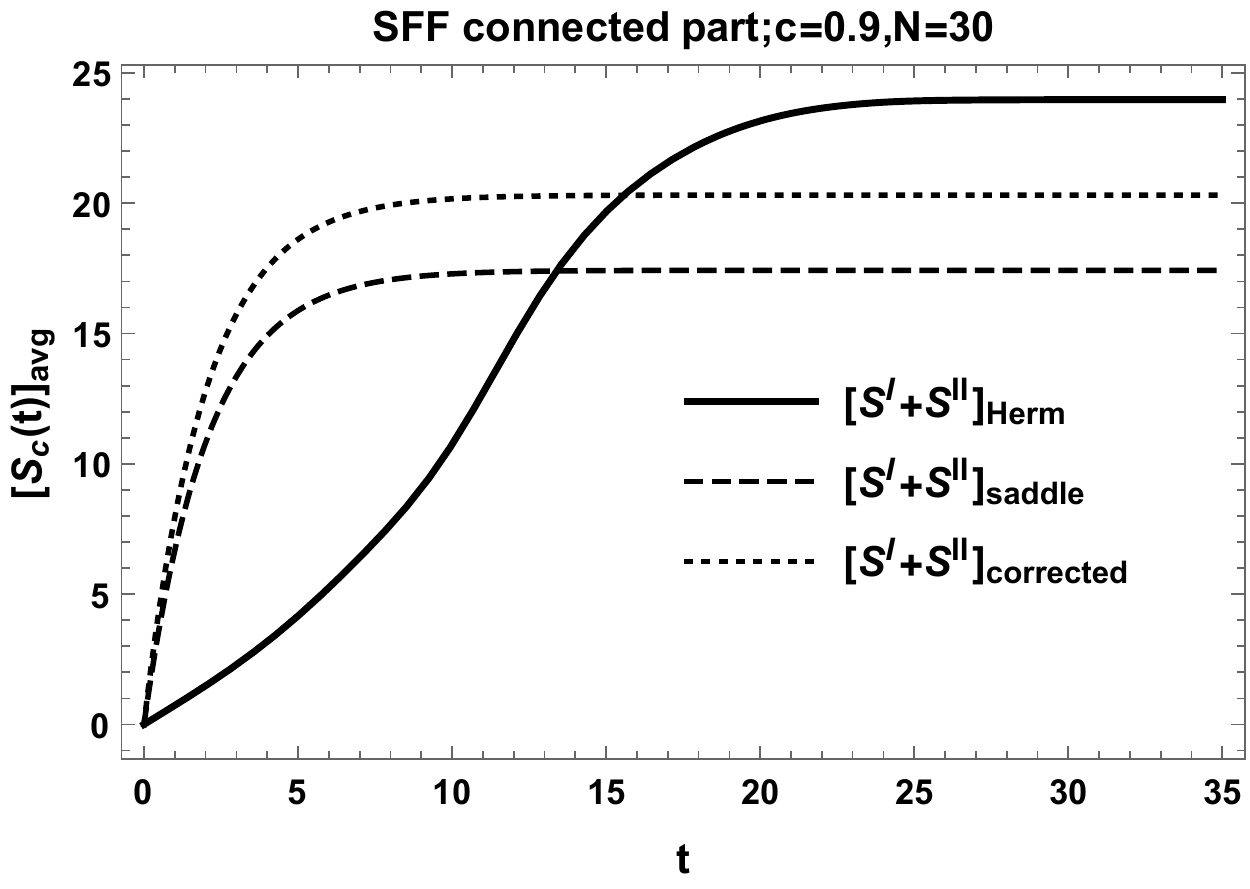}
     \label{}
         }
                  \caption[Optional caption for list of figures]{ SFF from saddle point approximation at zeroth order is compared with 1st order correction from saddle point method. Change in saturation value has been specified.} 
                                                                    \label{g18}
                                                                    \end{figure}   
Fig:-\ref{g18} shows that 1st order correction introduces extra shift in saturation value of SFF average. Different $N$ cases has been obtained in Fig:-\ref{g8} with saturation values explicitly mentioned. For $N$=7 zeroth order saddle point method solution has saturation at 0.028 which is shifted to 0.031 for solution with 1st order term in saddle point approximation. For $N=10,15,30$ cases zeroth order solution has saturation at $0.032, 0.15, 17.4$ which is shifted to $0.036, 0.16, 20.3$ with 1st order correction term. The first order correction shifted the saturation values closer to hermite polynomial representation solution. Hermite polynomial representation of kernels give saturation values for $N=4,10,15,30$ at around $0.034, 0.042, 0.18, 23.9$. In Fig:-\ref{g8} this comparison is shown in the plots explicitly.
\\
\subsection{Comparing Different solutions}
\subsection{Shift in Heisenberg Time}
\label{SHT12}
Here we compare our different solution to find what happens to Heisenberg time. Heisenberg time is the time scale after which SFF average saturates. To find this we have computed the relative fluctuation in SFF average. After certain time it is decayed to a value less than $10^{-3}$. We have considered that time as Heisenberg time.
Relative fluctuation in SFF average is defined as:-
\be \label{relfluc}
\frac{\Del S_{c}(t)}{S_{c}(t)}=\frac{S_{c}(t+\Del t)-S_{c}(t)}{S_{c}(t)}
\ee
\begin{figure}[H] 
       \centering
   \subfigure[Change in Heisenberg time for  N=20]{
      \includegraphics[width=7cm,height=5.5cm] {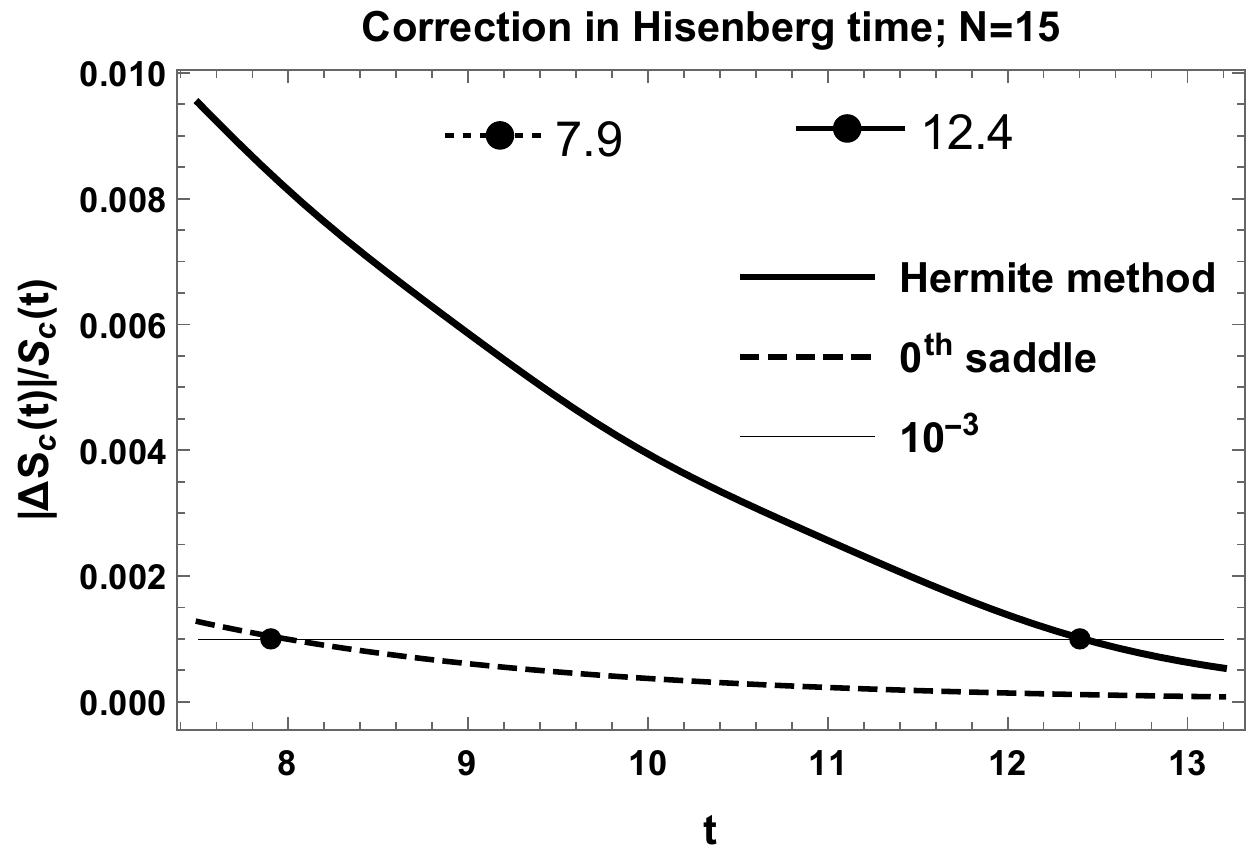}
     \label{g108}
         }
\subfigure[Change in Heisenberg time for  N=10]{
      \includegraphics[width=7cm,height=5.5cm] {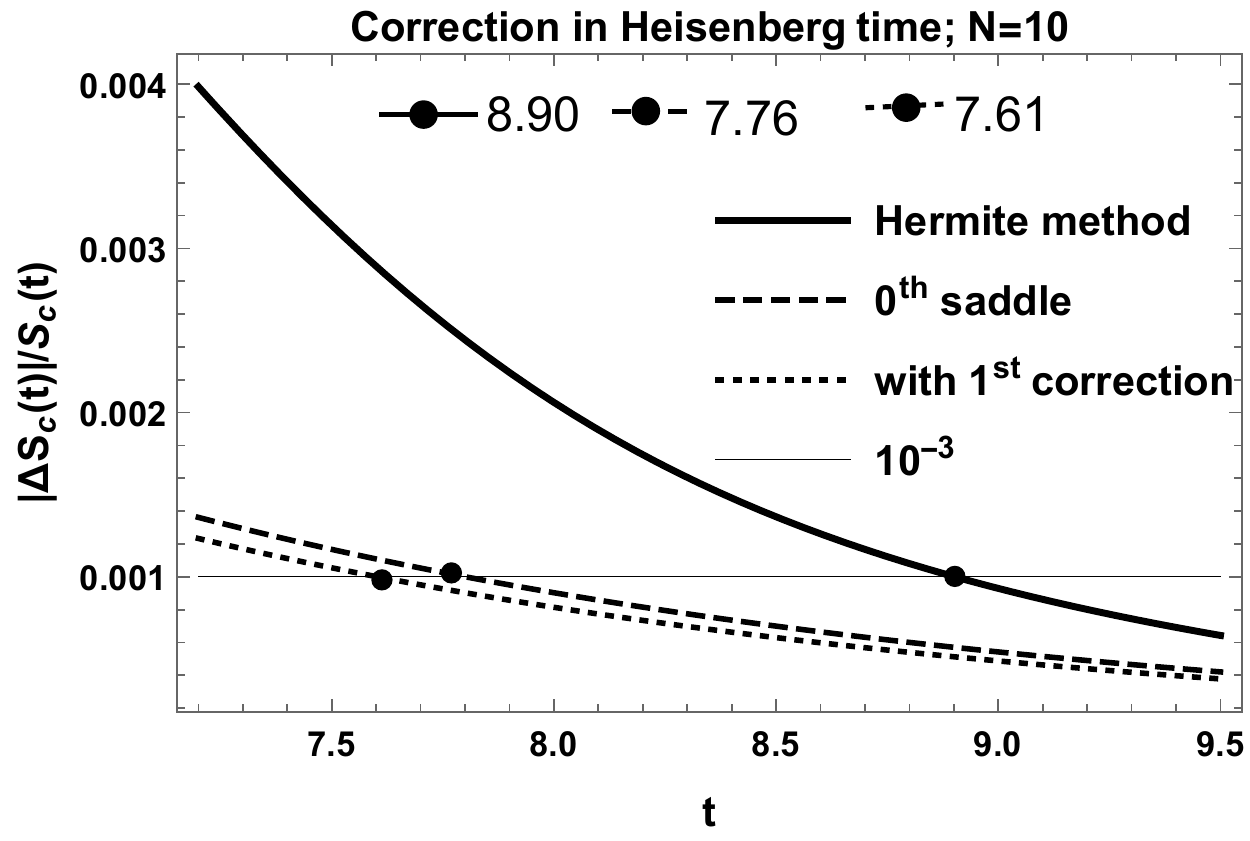}
     \label{g308}
         }
\caption[Optional caption for list of figures]{ Relative fluctuation defined in Eq:-\ref{relfluc} for SFF average at $N=15, 10$,$c=0.9$ w.r.t   $t$ is obtained using Eq:-\ref{relfluc} with $\Del t=0.1$.} 
                                                                    \label{g58}
                                                                    \end{figure}  
 The zeroth order saddle point method, saddle point method with 1st order correction and Hermite method solution all has Heisenberg time ($t_{H}$) at a point when this relative fluctuation decays to less than $10^{-3}$ at some fixed $t_{H}$. From Fig:-\ref{g108} for $N=15, c=0.9$ Heisenberg time for zeroth order saddle point is at $t_{H}=7.9$,and for Hermite method solution $t_{H}=12.4$. For Fig:-\ref{g308} Heisenberg time for $N=10,c=0.9$ is at $t_{H}=7.61,7.76,8.90$ for Saddle point solution with 1st order correction, zeroth order solution and from hermite polynomial solution.
\begin{figure}[H]       
\centering
      \includegraphics[width=9cm,height=5.5cm] {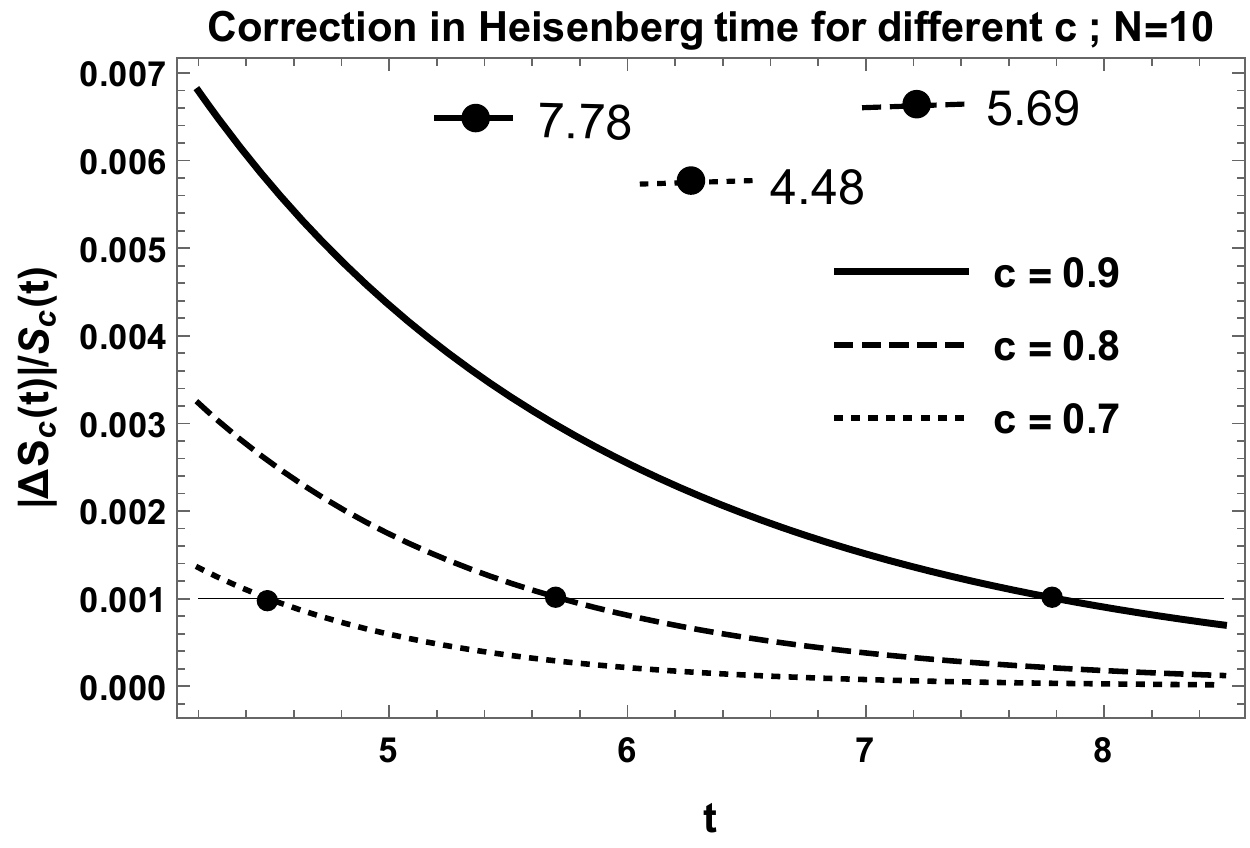}
       \label{g208}
\caption[Optional caption for list of figures]{ Heisenberg time from SFF average computed for zeroth order saddle point method solution for c=0.9,0.8,0.7 for N=10} 
                                                                  
                                                                    \end{figure}   
In Fig:-\ref{g208} shift in Heisenberg time form zeroth order saddle point solution for different $c$ values is obtained. For $c=0.9$ Heisenberg time is $t_{H}=7.78$ which is shifted to $t_{H}=5.69$ for $c=0.8$. For $c=0.7$ Heisenberg time is $t_{H}=4.48$. As the exponential decat behavior of SFF connected part dependes on $c$ explicitly (Eq:-\ref{heisencontrol}), chnaging $c$ values heavily shift Heisenberg time.

\section{Two point correlation function between same matrices}
In this section we will compute the two-point correlation function between same matrices. 
Two point correlation function for same matrices can be written as:- 
\be
\label{}
\displaystyle
U^{M_{1}}_{0}(z_{1},z_{2})=\langle\frac{1}{N}\mathrm{Tr} e^{iz_{1}M_{1}}\frac{1}{N}\mathrm{Tr} e^{iz_{2}M_{1}}\rangle
\ee
Following the same change of measure and distribution function as in Eq:-\ref{compare123}, we solve this by external source matrix A. The method here is same as in \cite{SFFRMT}. Using the modification of previous representation of $U(z_1,z_2)$ we have for the same matrix.
\[
\begin{array}{lll}
\displaystyle
U^{M_{1}}_A(z_{1},z_{2})=K\sum_{\ag_{1},\ag_{2}}^{N} \int \int e^{iz_{1}r_{\ag_{1}}}e^{iz_{2}r_{\ag_{2}}}e^{-\frac{N}{2}\sum r_{i}^{2}+\frac{N}{2}\sum \xi_{j}^{2} +cN\sum \xi_{i}r_{i}-N\sum a_{i}r_{i}}\Del(\xi)d r~d\xi
\end{array}
\]
Here $\prod_{i=1}^{N} d\xi_{i}=d\xi$ and $\prod_{i=1}^{N} d r_{i}=d r$ and $\frac{1}{N^{2}Z_{A}\Del(A)}=K$.
\\
Solving the integral in contour representation we decompose it in three parts.\\
For $\ag_{1}=\ag_{2}$ 
\bea
\begin{array}{lll}
\displaystyle
U_{0}^{M_{1}}\left(z_{1}, z_{2}\right)=\frac{c \sqrt{1-c^{2}}}{i N Z_{A}\left(z_{1}+z_{2}\right)} \oint\left(1-\frac{i\left(z_{1}+z_{2}\right)}{N u\left(1-c^{2}\right)}\right)^{N} \mathrm{Exp}\Bigg[-\frac{z_{1}^{2}}{2 N\left(1-c^{2}\right)}
\\
\displaystyle
~~~~~~~-\frac{z_{2}^{2}}{2 N\left(1-c^{2}\right)}-\frac{z_{1} z_{2}}{N\left(1-c^{2}\right)}-\frac{i\left(z_{1}+z_{2}\right) u}{1-c^{2}}\Bigg]
\end{array}
\eea
\bea\label{samecorel11}
\begin{array}{lll}
\displaystyle
\rho^{I}_{M_{1}}(\lb,\mu)=\frac{c \sqrt{1-c^{2}}}{i N Z_{A}}\int\frac{dz_{1}dz_{2}}{4\pi^{2}\left(z_{1}+z_{2}\right)}\mathrm{Exp}[-i z_{1}\lb-i z_{2}\mu]\oint\left(1-\frac{i\left(z_{1}+z_{2}\right)}{N u\left(1-c^{2}\right)}\right)^{N} 
\\
\displaystyle
~~~~~~~~~~~~~\mathrm{Exp}\Bigg[-\frac{z_{1}^{2}}{2 N\left(1-c^{2}\right)}-\frac{z_{2}^{2}}{2 N\left(1-c^{2}\right)}-\frac{z_{1} z_{2}}{N\left(1-c^{2}\right)}-\frac{i\left(z_{1}+z_{2}\right) u}{1-c^{2}}\Bigg]
\end{array}
\eea
Then we have done $z_{1},z_{2}$ integrals and  the contour integration over $u$ for the pole at $u=o$. 
\bea
\begin{array}{lll}
\displaystyle
S^{I}(\tau)=\int \frac{d\og}{2\pi} e^{i \og \tau}\rho^{I}(0,\og)
\end{array}
\eea
$S^{I}(\tau)$ is independent of $\tau$.
The disconnected part of correlation function has the form:-
\bea
\begin{array}{lll}
\displaystyle
U_{d}^{M_{1}}(z_{1},z_{2})=-\frac{c^2(1-c^2)}{z_{1}z_{2}}\oint \frac{dudv}{(2\pi i)^{2}}(1-\frac{iz_{1}}{N(1-c^2)u})^{N}(1-\frac{iz_{2}}{N(1-c^2)v})^{N}
\\
\displaystyle
~~~~~~~~~~~~~~~\mathrm{Exp}\bigg\{-\frac{z^{2}_{1}}{2N(1-c^2)}-\frac{z^{2}_{2}}{2N(1-c^2)}-\frac{iz_{1}u}{1-c^2}-\frac{iz_{2}v}{1-c^2}\bigg\}

\end{array}
\eea

\begin{figure}[H]
       \centering
         \subfigure[Correlation function from in Log scale for c=0.9, N=5]{
      \includegraphics[width=7cm,height=5.5cm] {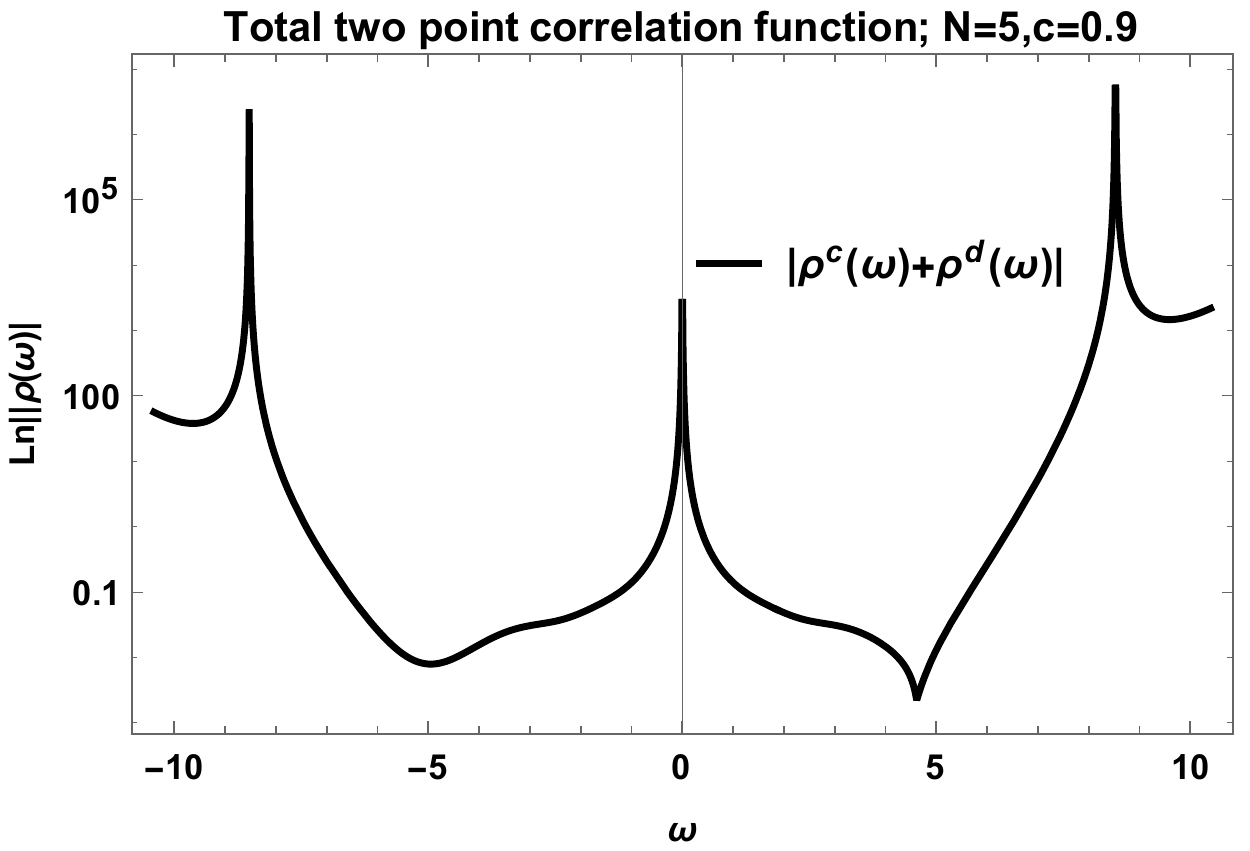}
     \label{}
         }   
        \subfigure[Correlation function in Log scale for c=0.9, N=20]{
         \includegraphics[width=7cm,height=5.5cm] {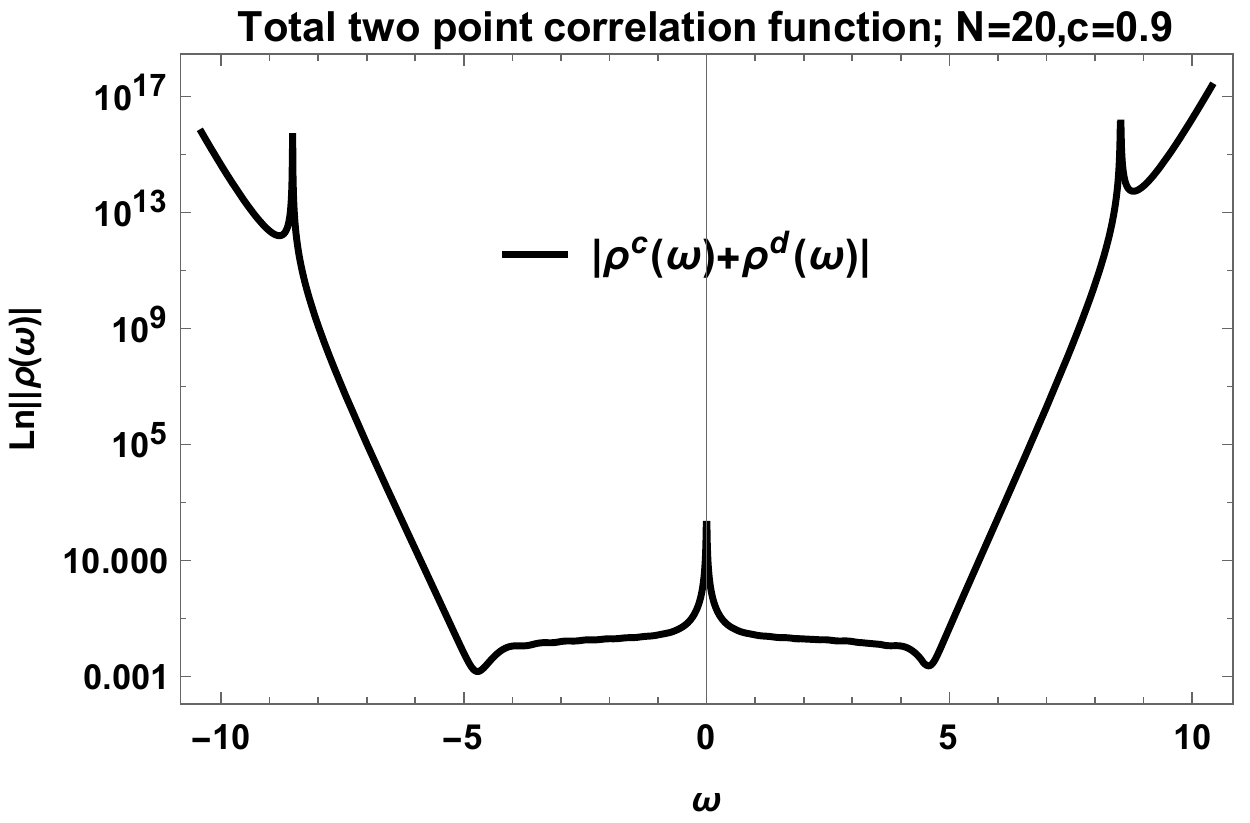}
     \label{}
             }   
                  \caption[Optional caption for list of figures]{Correlation function for same matrix interaction w.r.t $\og$ in Log Log scale using Eq:-\ref{samecorel33}} 
                                                                    \label{samemat2}
                                                                    \end{figure}

Then the disconnected part of two point correlation function for $M_{1}-M_{1}$ interaction is simply
\be\label{samecorel22}
\rho^{d}_{M_{1}}(\lb,\mu)=-\rho(\lb)\rho(\mu)
\ee
Where $\rho(\lb)$ is the level density for two matrix model.
Now we use one transformation for connected part :-
$z_{1}=z'_{1}-iN(1-c^2)u$ , $z_{2}=z'_{2}-iN(1-c^2)v$
\\
\[
\begin{array}{lll}
\displaystyle
U^{II}_{M_{1}}(z_{1},z_{2})=\frac{c}{N^{2}}\oint \frac{dudv}{(2\pi i)^{2}}\bigg[\frac{z'_{1}}{iN(1-c^2)u}\times\frac{z'_{2}}{iN(1-c^2)v}\bigg]^{N}\frac{1}{(v+\frac{ i z_{1}}{N(1-c^2)})(u+\frac{iz_{2}}{N(1-c^2)})} 
\\
\\
\displaystyle
\times \mathrm{Exp}\bigg[-\frac{z^{2}_{1}}{2N(1-c^2)}-\frac{z^{2}_{2}}{2N(1-c^2)}-\frac{iuz'_{1}c^{2}}{1-c^{2}}-\frac{ivz_{2}c^{2}}{1-c^2}-\frac{N(c^2+1)u^{2}}{2}-\frac{N(c^2+1)v^{2}}{2}\bigg]
\end{array}
\]
now we do the Fourier transform two get the two point correlation function:-
\[
\begin{array}{lll}
\displaystyle
\rho^{II}_{M_{1}}(\lb,\mu)=\frac{c}{N^{2}}\int \frac{dz_{1}dz_{2}}{(2\pi)^{2}}\oint \frac{dudv}{(2\pi i)^{2}}\bigg[\frac{z'_{1}}{iN(1-c^2)u}\times\frac{z'_{2}}{iN(1-c^2)v}\bigg]^{N}\frac{1}{(v+\frac{ i z_{1}}{N(1-c^2)})}
\\
\displaystyle
~~~~\frac{1}{(u+\frac{iz_{2}}{N(1-c^2)})}\mathrm{Exp}\bigg[-\frac{z'^{2}_{1}}{2N(1-c^2)}-\frac{z'^{2}_{2}}{2N(1-c^2)}-\frac{iuz'_{1}c^{2}}{1-c^{2}}-\frac{ivz'_{2}c^{2}}{1-c^2}-\frac{N(c^2+1)u^{2}}{2}~~~
\\
\displaystyle
~~~~~~~~~~~~~~~~~~~~~~~~~~~~~~~~~-\frac{N(c^2+1)v^{2}}{2}+iz_{1}\lb+N(1-c^2)u\lb+iz_{2}\mu+N(1-c^2)v\mu\bigg]
\end{array}
\]
replacing $z_{1}= \frac{z'_{1}}{N(1-c^2)}$ and $z_{2}=\frac{z'_{2}}{N(1-c^2)}$
\bea\label{410}
\begin{array}{lll}
\displaystyle
\rho_{M_{1}}^{II}(\lb,\mu)=c(1-c^2)^{2}\int \frac{dz_{1}dz_{2}}{(2\pi)^{2}}\oint \frac{dudv}{(2\pi i)^{2}} \bigg[\frac{z_{1}}{iu}\bigg]^{N}\bigg[\frac{z_{2}}{iv}\bigg]^{N}\frac{1}{(v+ iz_{1})(u+iz_{2})}
\\
\\
\displaystyle
~~\mathrm{Exp}\bigg[-N\Big\{\frac{(1-c^2)z_{1}^{2}}{2}+\frac{(1-c^{2})z_{2}^{2}}{2}+ic^{2} u z_{1}+ic^{2}vz_{2}+\frac{c^{2}+1}{2}u^{2}+\frac{c^{2}+1}{2}v^{2}
\\
\displaystyle
~~~~~~~~~~~~~~~~~~~~~~~~~~+i (1-c^2)z_{1}\lb+ i (1-c^2)z_{2}\mu+(1-c^2)u\lb+(1-c^2)v\mu\Big\}\bigg]

\end{array}
\eea
\textbf{Solving the Integral by four-variable saddle point method.}\\
Here we consider four variable saddle point solution discussed in \cite{bleistein2012saddle,AoA657}. Eq:- \ref{410} is characterized by the following form of four variables.
\bea
\begin{array}{lll}
\displaystyle
F(z_{1},z_{2},u,v)=\frac{1}{2} \left(c^2+1\right) \left(u^2+v^2\right)+\left(1-c^2\right) (\lambda  u+\mu  v)+i c^2 (u z_{1}+v z_{2})
\\
\displaystyle
~~~~~~~~~~~~~~~~~~~~~+\frac{1}{2} \left(1-c^2\right) z_{1}^2+\frac{1}{2} \left(1-c^2\right) z_{2}^2+i \left(1-c^2\right) (\lambda  z_{1}+\mu  z_{2})\\
\displaystyle
~~~~~~~~~~~~~~~~~~~~~~~~~-\log (u) -\log (v)+\log (x)+\log (y)~~~~~
\end{array}
\eea
So our saddle points are the simultaneous solution of four equations.
\bea
\begin{array}{lll}
\displaystyle
\frac{\partial F(z_{1},z_{2},u,v)}{\partial z_{1}}=0,~~~~~~~~~~~~~\frac{\partial F(z_{1},z_{2},u,v)}{\partial z_{2}}=0
\\
\displaystyle
\frac{\partial F(z_{1},z_{2},u,v)}{\partial u}=0,~~~~~~~~~~~~~~~\frac{\partial F(z_{1},z_{2},u,v)}{\partial v}=0\end{array}
\eea
Solving this equation gives sixteen set of solution as the saddle points.
Now using saddle point method for four variables with the transformation:-
\[
\begin{array}{lll}
\displaystyle
\lambda =\frac{\left(2 \sqrt{c^2+1}\right) \sin (\theta )}{\sqrt{c^2-1}}
,~~~~\mu =\frac{\left(2 \sqrt{c^2+1}\right) \sin (\phi )}{\sqrt{c^2-1}}
\end{array}
\]
\bea\label{pyaar}
\begin{array}{lll}
\displaystyle
\rho^{II}_{M_{1}}(\theta,\phi)=\frac{c(-1)^{(N+1)}}{(2\pi)^{4}}\Bigg\{\frac{(1-c^2)^{2}~e^{-N~F(z_{1},z_{2},u,v)}}{(v+iz_{1})(u+iz_{2})}
\\
\displaystyle
\bigg(1\bigg/\bigg[\frac{\partial^{2} F(z_{1},z_{2},u,v)}{\partial^{2} z_{1}}\frac{\partial^{2} F(z_{1},z_{2},u,v)}{\partial^{2} z_{2}}\frac{\partial^{2} F(z_{1},z_{2},u,v)}{\partial^{2} u}\frac{\partial^{2} F(z_{1},z_{2},u,v)}{\partial^{2} v}\bigg]\bigg)\Bigg\}

\end{array}
\eea
Now set $\lb=0$ and apply the reverse transformation to obtain its previous form  by
\be
\omega=\frac{\left(2 \sqrt{c^2+1}\right) \sin (\phi )}{\sqrt{c^2-1}}
\ee
Then this gives us the correlation function for same matrix model. We have plotted the same matrix correlation function for different $N$ in Fig:-
\ref{samemat2}.
So the total correlation function from Eq:-\ref{samecorel11},\ref{samecorel22},\ref{410}
\be \label{samecorel33}
\rho_{M_{1}}(\og)=\rho^{I}_{M_{1}}(\og)+\rho^{II}_{M_{1}}(\og)+\rho^{d}_{M_{1}}(\og)
\ee

Fourier transforming the correlation function generates the spectral form factor.
\be
S(\tau)=\int \frac{1}{2\pi}e^{i \og \tau}\rho_{M_{1}}(\og) d\og
\ee
We do this by contour integral over the poles. Poles of this function are:-
\be
\og\rightarrow-\frac{2 c^2}{c^2-1},\og\rightarrow \frac{\sqrt{-c^6+3 c^4-4}}{c^2-1}, \og\rightarrow \pm\frac{2 i c^2 \sqrt{c^4-1}}{\left(c^2-1\right)^2 \left(c^2+1\right)},\og\rightarrow 0
\ee
 Finding residue w.r.t this poles gives us the Spectral Form Factor. We plotted its time average defined as:-
\be
S_{c}(t)_{avg}=\int_{0}^{t}S_{c}(\tau) d\tau
\ee
\begin{figure}[H]
       \centering
         \subfigure[ (SFF average)/t  for c=0.9, N=5]{
      \includegraphics[width=7cm,height=5.5cm] {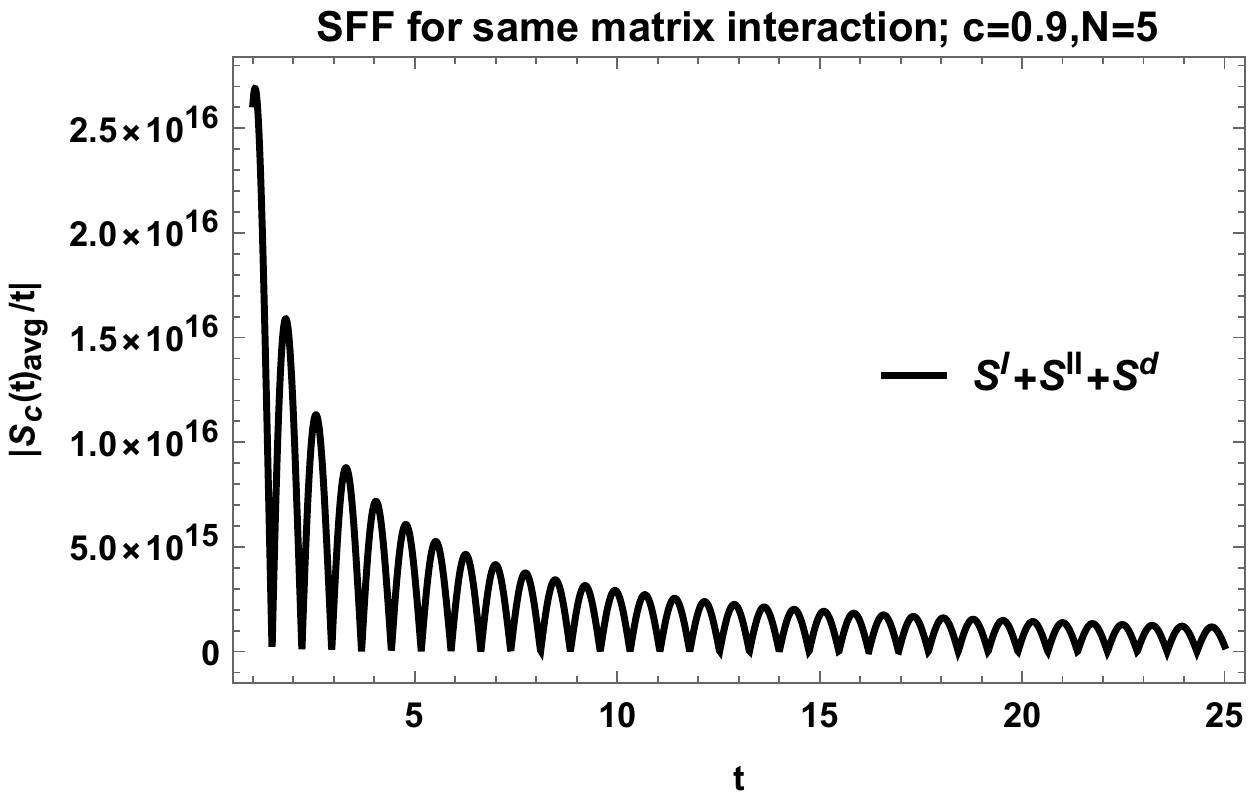}
     \label{}
         }
        \subfigure[SFF average for c=0.9, N=10 in LogLog scale]{
         \includegraphics[width=7cm,height=5.5cm] {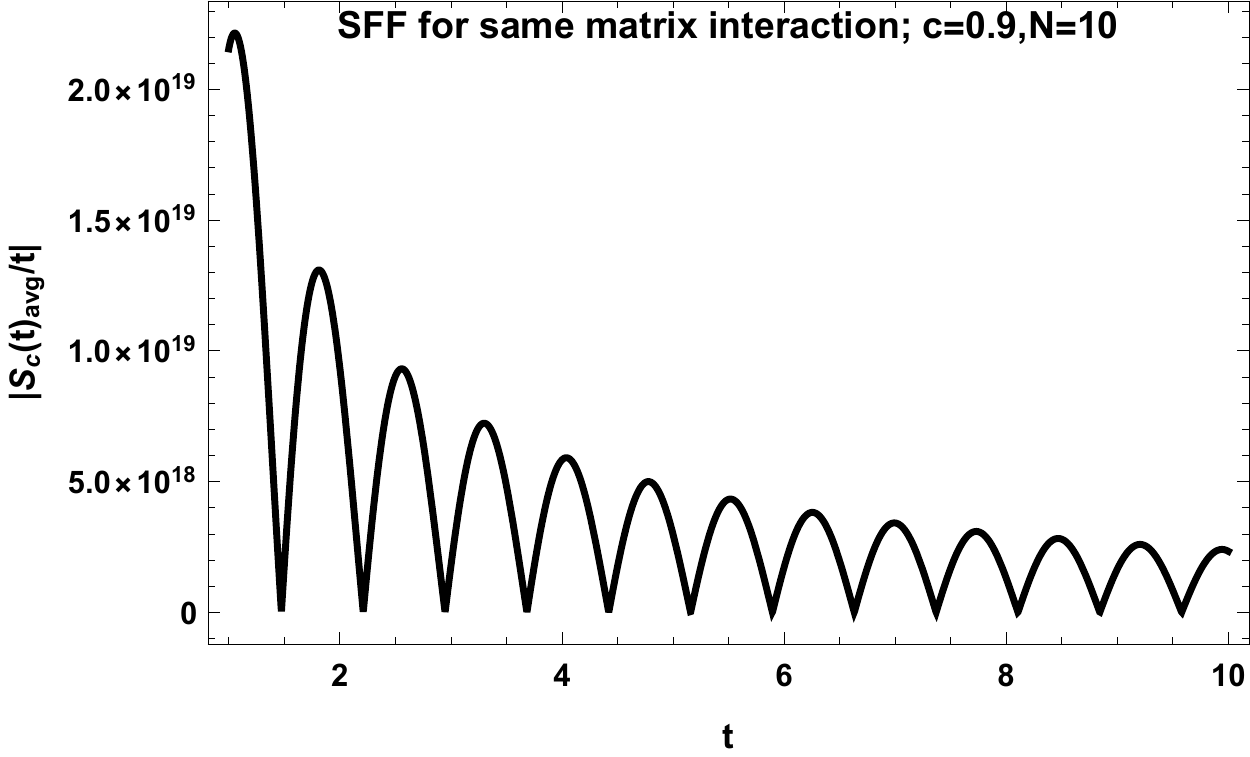}
     \label{}
             }  
        \subfigure[(SFF average)/t for c=0.9, N=20]{
         \includegraphics[width=7cm,height=5.5cm] {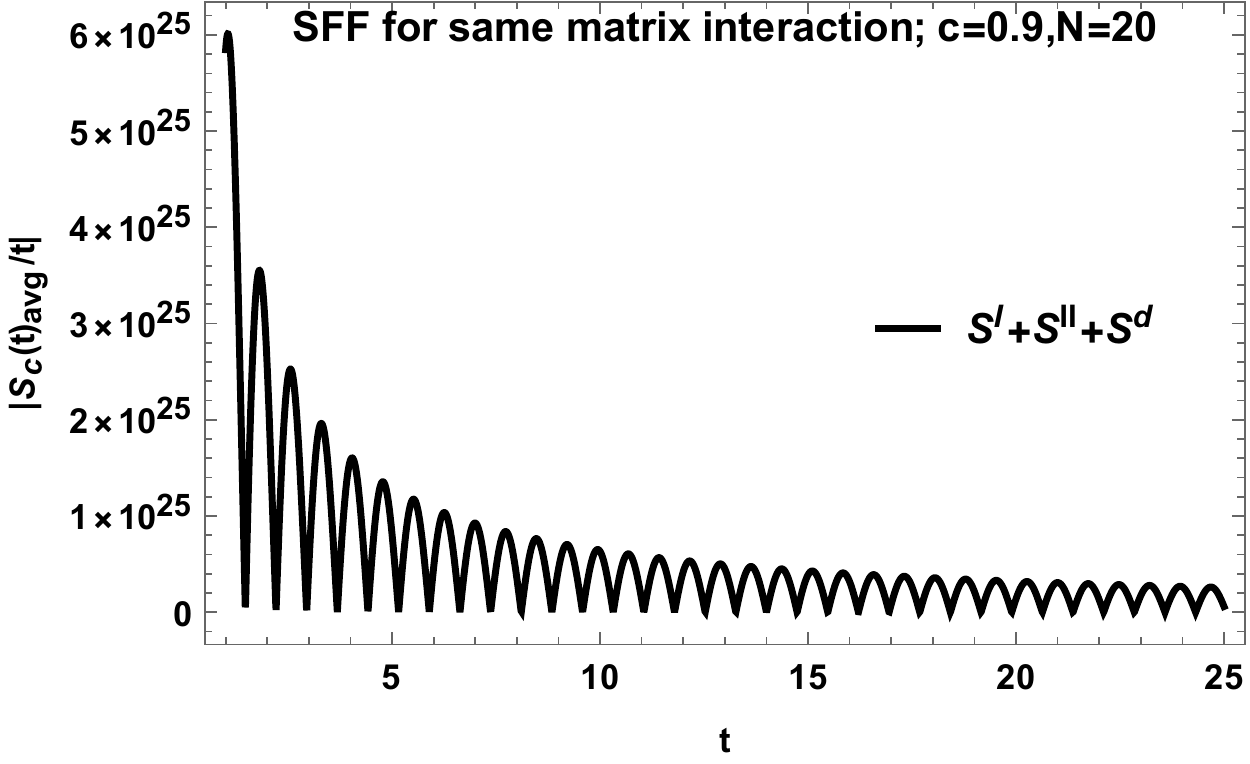}
     \label{}
             }  
           \subfigure[(SFF average) for c=0.1, N=10]{
         \includegraphics[width=7cm,height=5.5cm] {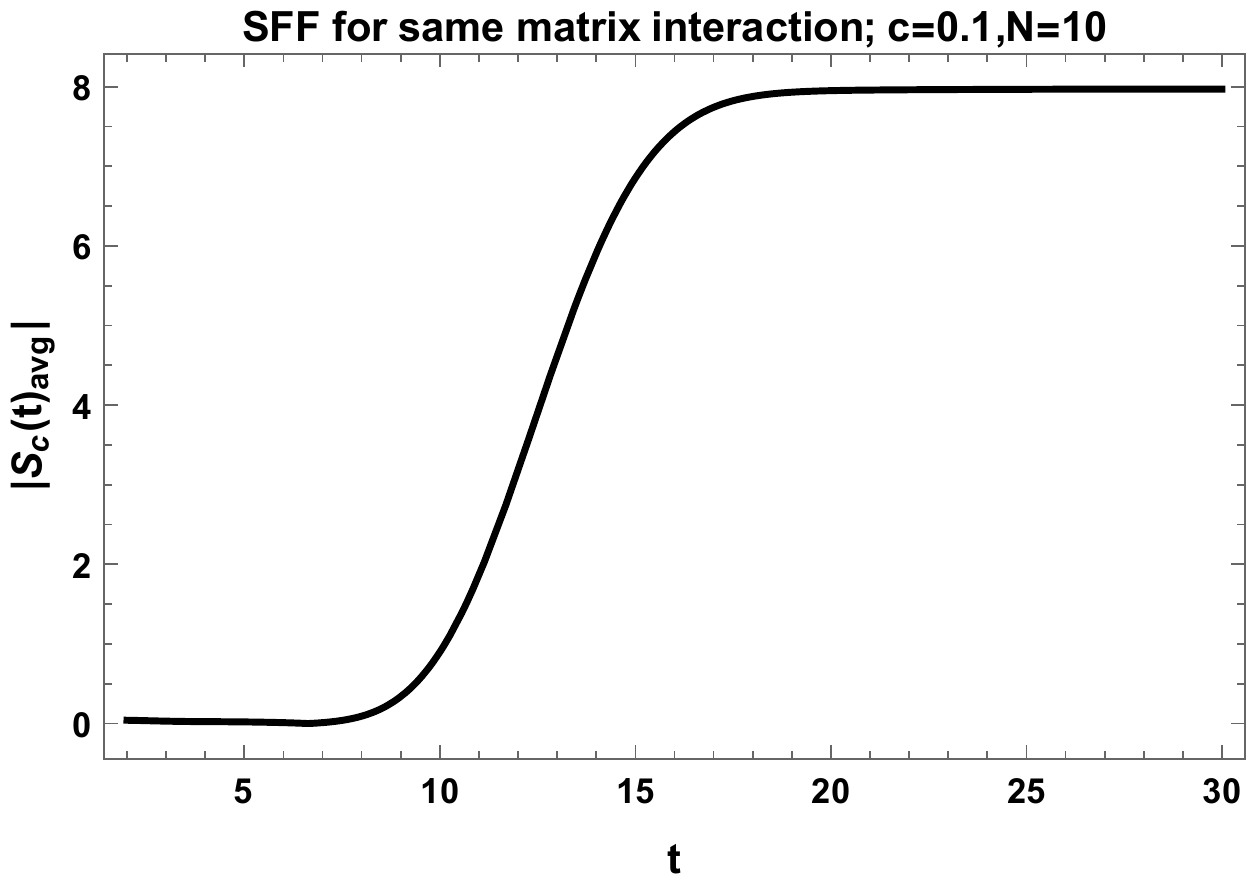}
     \label{}
            }  
             \caption[Optional caption for list of figures]{Spectral Form Factor for same matrix interaction w.r.t $t$ for different $c=e^{-t}$ and dimension of Matrix(N)} 
                                                                    \label{g1112}
                                                                    \end{figure} 
From Fig:-\ref{g1112} SFF for same matrix interaction ($M_{1}-M_{1}$ interaction) have $c$ dependence. When we go towards $c\rightarrow0$  we get the exact behavior of SFF as of different matrix interaction. In other cases it changes its magnitude and period with values of $c$ as well as $N$.



\section{Instanton action for two matrix model}
In \cite{SFFRMT} the two point function for two matrix model was calculated and given in the following form
\bea\label{brezinhikami}
\begin{array}{lll}
\displaystyle
\rho(\tau)=\int d \omega e^{i \omega \tau}\left(-\frac{1-c^{2}}{16 \pi^{2} N^{2}} \frac{1}{\cos \varphi} \frac{1}{\left[\frac{(1-c)^{2}}{4 c^{2}}+\sin ^{2} \varphi\right]^{2}} f \right)
\end{array}
\eea
where $f$ is 
\be
f=\left[-2 \cos \varphi \cos [N h(\varphi)-\varphi]+\left(1+c^{2}\right) \cos N h(\varphi)\right]^{2}
\ee
and $\og=\frac{2}{\sqrt{1-c^{2}}}\sin \varphi$. We have evaluated this same equation in Eq:-\ref{bessel}.
Now the function has poles at  $\omega= -\frac{i\sqrt{c^2-1}}{c}$ and $\og=\frac{i\sqrt{c^2-1}}{c}$. Evaluating the contour integral for the pole at  $\frac{i\sqrt{c^2-1}}{c}$gives functional dependence of $\rho(\tau)$ \be
\rho(\tau) \simeq \tau e^{-\frac{\sqrt{1-c}}{\epsilon} \tau}
\ee
The non-perturbative term (instanton) comes from the pole of Eq:-\ref{brezinhikami}.
\\
There is interesting identity between spectral form factor and Laguerre polynomials discussed in \cite{SFFRMT} for $c=0$,
 in \cite{forrester2020differential}.Through this identity, we could discuss the instanton effect. For time dependent case, $c\neq 0$, we follow the analysis of 
\ref{brezinhikami} from \cite{SFFRMT}, where the two point function is expressed as Fourier transform. As shown in \ref{brezinhikami} or from Eq:-\ref{bessel}, one of this pole contributes to the rounding behavior around Heisenberg time.
\\
For one matrix model, the effect of instanton has been previously studied in \cite{Okuyama:2018yep,Okuyama:2018gfr}.
 Firstly we separate out two eigenvalues, one from each matrices, and then compute the rest of the terms exactly for large $N$ to get an effective potential. The instanton action is derived from values of effective potential at the saddle points and one reference point along the support of eigen value density. For two matrix model density of states is distributed along the interval $[-\frac{2}{\sqrt{c^2-1}},\frac{2}{\sqrt{c^2-1}}]$.
\\
Now we know the eigenvalue distribution /density of states has the form
\[
\displaystyle
\rho(\lb)=\frac{\sqrt{1-c^2}}{2\pi}\sqrt{4-(1-c^2)\lb^{2}}
\]
 and normalized over $[-\frac{2}{\sqrt{c^2-1}},\frac{2}{\sqrt{c^2-1}}]$. We have considered the behavior of instanton action in plateau regime, $\tau>1$. First we consider the one point function and evaluated the instanton action from it. Then we have repeated the same method for two-point function. We have computed them explicitly and compared their nature. 
\\
One-point function defined in \ref{f1} can be rewritten as
 \bea
\begin{array}{lll}
\displaystyle
U(\tau)=\langle \frac{1}{N} \mathrm{Tr}(e^{2iN\tau M_{1}}\rangle_{N}
\\
\displaystyle
~~~~~=\int \mathrm{Tr}(e^{2iN\tau M_{1}})e^{-\frac{1}{2}\mathrm{Tr} M_{1}^{2}-\frac{1}{2}\mathrm{Tr} M_{2}^{2}+c \mathrm{Tr} M_{1} M_{2}}dM_{1}dM_{2}
\end{array}
\eea
For the term 
\bea
\begin{array}{lll}
\displaystyle
\mathrm{Tr}(e^{2iN\tau M_{1}})=\left\{\sum_{\ag=1}^{N-1} e^{2i N\tau r_{i}}\right\} +e^{2iN\tau x}
\end{array}
\eea
we can ignore the $N-1$ eigenvalues and only consider the $e^{2iN\tau x}$ term. Here x is considered to be the only eigenvalue with non-zero coupling.  
\[
\begin{array}{lll}
\displaystyle
U(\tau)=\int e^{2iN\tau x}e^{-\frac{N}{2}\sum_{i} r_{i}^{2}-\frac{N}{2}\sum_{i}\xi_{i}^{2}+cN\sum_{i}r_{i}\xi_{i}} \frac{\Del^2(r)\Del^2(\xi)\prod_{i}dr_{i}\prod_{i}d\xi_{i}}{\Del(r)\Del(\xi)}
\\
\displaystyle
=\int dx dy~ e^{2iN\tau x-\frac{N}{2}x^2-\frac{N}{2}y^{2}+cN xy } \int \prod_{i=1}^{N-1}d r_{i}\prod_{j=1}^{N-1}d \xi_{j}\prod_{i< j}^{N-1}(r_{i}-r_{j})\prod_{i=1}^{N-1}(x-r_{i})
\\
\displaystyle
~~~~~~~~~~~~~\prod_{i< j}^{N-1}(\xi_{i}-\xi_{j})\prod_{i=1}^{N-1}(y-\xi_{i})~e^{2iN\tau x-\frac{N}{2}\sum_{i=1}^{N-1}r_{i}^{2}-\frac{N}{2}\sum_{j=1}^{N-1}\xi_{j}^{2}+cN\sum_{i=1}^{N-1}r_{i}\xi_{i}}
\end{array}
\]
Now we define the second integral over $N-1$ eigenvalues as a matrix interval. We define two matrices composed from $N-1$ eigenvalue of the previous one. $\bar{M}_{1}|_{N-1\times N-1}$ has $N-1$ eigenvalues and in large $N$ limit the distribution of this $N-1$ eigenvalues follows the same density of states and distributed along the  interval $[-\frac{2}{\sqrt{c^2-1}},\frac{2}{\sqrt{c^2-1}}]$. Same is true for $\bar{M}_{2}$. So in matrix notation this can be represented as
\[
\begin{array}{lll}
\displaystyle
U(\tau)=\int dx dy e^{2iN\tau x-\frac{N}{2}x^2-\frac{N}{2}y^{2}+cN xy }
\\
\displaystyle
~~~~~~~~~~~~~~~\times \int d\bar{M}_{1}d\bar{M}_{2} Det(x-\bar{M}_{1})Det(y-\bar{M}_{2}) e^{-\frac{1}{2}\mathrm{Tr} \bar{M}_{1}^{2}-\frac{1}{2}\mathrm{Tr} \bar{M}_{2}^{2}+c \mathrm{Tr} \bar{M}_{1} \bar{M}_{2}}
\\
\displaystyle
~~~~~~~~=\int dx dy e^{2iN\tau x-\frac{N}{2}x^2-\frac{N}{2}y^{2}+cN xy } \langle Det(x-\bar{M}_{1})Det(y-\bar{M}_{2})\rangle_{N}
\end{array}
\]
In large $N$ limit we can use the factorization property of determinant
\[
\langle Det(A) Det (B)\rangle_{N}=\langle Det(A)\rangle_{N}\langle Det (B)\rangle_{N}
\] and $\mathrm{Exp}[\mathrm{Tr}(\mathrm{Log}[A])]=Det(A)$ which simplifies the integral to
\[
\begin{array}{lll}
\displaystyle
U(\tau)=\int dx dy e^{2iN\tau x-\frac{N}{2}x^2-\frac{N}{2}y^{2}+cN xy } \langle Det(x-\bar{M}_{1})\rangle_{N}\langle Det(y-\bar{M}_{2})\rangle_{N}
\\

\displaystyle
~~~~~~=\int dx dy e^{2iN\tau x-\frac{N}{2}x^2-\frac{N}{2}y^{2}+cN xy +\langle\mathrm{Tr}(\mathrm{Log}(x-\bar{M}_{1}))\rangle_{N}+\langle\mathrm{Tr}(\mathrm{Log}(y-\bar{M}_{2}))\rangle_{N}}
\\
\displaystyle
~~~~~~=\int dx dy e^{-N A_{eff}(x,y)}
\end{array}
\]
Therefore
\[
\begin{array}{lll}
\displaystyle
A_{eff}(x,y)=-2i\tau x+\frac{1}{2}x^2+\frac{1}{2}y^{2}-c xy -\langle\mathrm{Tr}(\mathrm{Log}(x-\bar{M}_{1}))\rangle_{N}-\langle\mathrm{Tr}(\mathrm{Log}(y-\bar{M}_{2}))\rangle_{N}~~~
\end{array}
\]
To evaluate the saddle point we solve the following equations
\bea
\begin{array}{lll}
\displaystyle
\frac{\partial A_{eff}(x,y)}{\partial x}=0~~~~~~~~~\frac{\partial A_{eff}(x,y)}{\partial y}=0
\\
\displaystyle
-2i\tau+x-cy-\langle\mathrm{Tr}(\frac{1}{x-\bar{M}_{1}})\rangle_{N}=0
\\
\displaystyle
y-cx -\langle\mathrm{Tr}(\frac{1}{y-\bar{M}_{2}})\rangle_{N}=0
\end{array}
\eea
\bea
\begin{array}{lll}
\displaystyle
\langle\mathrm{Tr}(\frac{1}{x-\bar{M}_{1}})\rangle_{N}=\int_{-\frac{2}{\sqrt{1-c^2}}}^{\frac{2}{\sqrt{1-c^2}}} dz \frac{\rho(z)}{x-z}=\frac{x}{2}(1-c^2)-\frac{1}{2}\sqrt{(1-c^2)^{2}x^{2}-4}
\end{array}
\eea
So the saddle points are:-
\bea
\begin{array}{lll}
\displaystyle
x_{*}=-\frac{2 i \sqrt{\tau ^2-1}}{c^2-1},\frac{2 i \sqrt{\tau ^2-1}}{c^2-1} \\
\displaystyle
y_{*}=-i\left\{\frac{\left(c^2+1\right) \sqrt{\tau ^2 -1}- \left(1-c^2\right) \tau }{c \left(c^2-1\right)}\right\},\\
\displaystyle
~~~~~~~~~~~~~~~~~i\left\{\frac{\left(c^2+1\right) \sqrt{\tau ^2-1}- \left(c^2-1\right) \tau }{c \left(c^2-1\right)}\right\}
\end{array}
\eea
Here we will consider only $\tau>1$ part for Large $N$ limit. So the saddle point are conjugate in nature.\\
Now we evaluate $A_{eff}(x,y)$ by
\bea
\begin{array}{lll}
\displaystyle 
\int \frac{\partial A_{eff}}{\partial x} dx = B(x,y)+C(y) 
\\
\displaystyle
\frac{\partial B(x,y)}{\partial y}+\frac{d C(y)}{d y}=\frac{\partial A_{eff}}{\partial y}
\end{array}
\eea
comparing the coefficients in the last equation we evaluate the $C(y)$ term upto a constant. Now the final form of $A_{eff}(x,y)$ is
\bea
\begin{array}{lll}
\displaystyle
A_{eff}(x,y)=\frac{1}{4 \left(c^2-1\right)}\bigg[\left(c^2-1\right) (x (\sqrt{\left(c^2-1\right)^2 x^2-4}-4 c y-8 i \tau)
\\
\displaystyle
~~~~~~~~~~~~~+\left(c^2+1\right) x^2+y (\sqrt{\left(c^2-1\right)^2 y^2-4}+c^2 y+y))
\\
\displaystyle
~~~~~~~-4 \mathrm{Log}(\sqrt{(c^2-1)^2 x^2-4}+\left(c^2-1\right) x)-4 \mathrm{Log} (\sqrt{\left(c^2-1\right)^2 y^2-4}+(c^2-1) y)\bigg]
\end{array}
\eea
So,
\bea \label{instantonact}
\begin{array}{lll}
\displaystyle
A_{inst}(\tau)=S_{eff}(x_{*}^{1},y_{*}^{1})-S_{eff}(x_{*}^{2},y_{*}^{2})
\\
\displaystyle
A_{inst}(\tau)=-\frac{2}{\left(c^2-1\right)}\bigg[\left(\tau  \sqrt{\tau ^2-1}-\log \left(\sqrt{\tau ^2-1}+\tau \right)\right)\bigg]
\end{array}
\eea
\bea 
\begin{array}{lll}
\displaystyle
A_{eff}(x^{1}_{*},y^{1}_{*})=-\frac{2\tau  \sqrt{\tau ^2-1}}{c^2-1} +\frac{2}{c^2-1}\left(cosh^{-1}(\tau)\right)
\\
\displaystyle
A_{eff}(x^{2}_{*},y^{2}_{*})=\frac{2 \tau  \sqrt{\tau ^2-1}}{c^2-1}-\frac{cosh^{-1}(-\tau)+cosh^{-1}(\tau)}{c^2-1}
\end{array}
\eea
Now two point correlation can be defined as follows:-
\[
\displaystyle
U(\tau,-\tau)=\int \mathrm{Tr}(e^{2iN\tau M_{1}})\mathrm{Tr}(e^{-2iN\tau M_{2}})e^{-\frac{1}{2}\mathrm{Tr} M_{1}^{2}-\frac{1}{2}\mathrm{Tr} M_{2}^{2}+c \mathrm{Tr} M_{1} M_{2}}dM_{1}dM_{2}
\]
Now we have used same argument as of one-point function we have ignored $N-1$ eigenvalue and considered only $x$ and $y$ has non-zero coupling. Also in large $N$ limit factorization of determinant is used.
\bea \label{fe1}
\begin{array}{lll}
\displaystyle
U(\tau,-\tau)=\int e^{2iN\tau x-2iN\tau y}e^{-\frac{N}{2}\sum_{i} r_{i}^{2}-\frac{N}{2}\sum_{i}\xi_{i}^{2}+cN\sum_{i}r_{i}\xi_{i}} \frac{\Del^2(r)\Del^2(\xi)\prod_{i}dr_{i}\prod_{i}d\xi_{i}}{\Del(r)\Del(\xi)}
\\
\displaystyle
~~~~~~~~~~~=\int dx dy e^{2iN\tau x-2iN\tau y-\frac{N}{2}x^2-\frac{N}{2}y^{2}+cN xy } \langle Det(x-\bar{M}_{1})\rangle_{N}\langle Det(y-\bar{M}_{2})\rangle_{N}
\\
\displaystyle
~~~~~~~~~~~=\int dx dy e^{2iN\tau x-2iN\tau y-\frac{N}{2}x^2-\frac{N}{2}y^{2}+cN xy +\langle\mathrm{Tr}(\mathrm{Log}(x-\bar{M}_{1}))\rangle_{N}+\langle\mathrm{Tr}(\mathrm{Log}(y-\bar{M}_{2}))\rangle_{N}}
\\
\displaystyle
~~~~~~~~~~~=\int dx dy e^{-N A'_{eff}(x,y)}
\end{array}
\eea
Therefore
\[
\begin{array}{lll}
\displaystyle
A'_{eff}(x,y)=-2i\tau x+2i\tau y+\frac{1}{2}x^2+\frac{1}{2}y^{2}-c xy -\langle\mathrm{Tr}(\mathrm{Log}(x-\bar{M}_{1}))\rangle_{N}-\langle\mathrm{Tr}(\mathrm{Log}(y-\bar{M}_{2}))\rangle_{N}
\end{array}
\]

\bea
\begin{array}{lll}
\displaystyle
\frac{\partial A'_{eff}(x,y)}{\partial x}=0~~~~~~~~~\frac{\partial A'_{eff}(x,y)}{\partial y}=0
\\
\displaystyle
-2i\tau+x-cy-\langle\mathrm{Tr}(\frac{1}{x-\bar{M}_{1}})\rangle_{N}=0
\\
\displaystyle
2i\tau+ y - cx -\langle\mathrm{Tr}(\frac{1}{y-\bar{M}_{2}})\rangle_{N}=0
\end{array}
\eea
Again we have used the density of states expression
\[
\begin{array}{lll}
\displaystyle
\langle\mathrm{Tr}(\frac{1}{x-\bar{M}_{1}})\rangle_{N}=\int_{-\frac{2}{\sqrt{1-c^2}}}^{\frac{2}{\sqrt{1-c^2}}} dz \frac{\rho(z)}{x-z}=\frac{x}{2}(1-c^2)-\frac{1}{2}\sqrt{(1-c^2)^{2}x^{2}-4}
\end{array}
\]
After evaluating the saddle points are
\bea
\begin{array}{lll}
\displaystyle
x^{1}_{*}=\frac{i \left(-\sqrt{(c-1)^2 \tau ^2+c}+c \tau +\tau \right)}{c (c+1)}~~,x^{2}_{*}=\frac{i \left(\sqrt{(c-1)^2 \tau ^2+c}+c \tau +\tau \right)}{c (c+1)}
\\
\displaystyle
y^{1}_{*}=\frac{i \left(-\left(c^2+c+1\right) \sqrt{(c-1)^2 \tau ^2+c}+(c+1) ((c-3) c+1) \tau \right)}{c^2 \left(c+1\right)}~~
\\
\displaystyle
~~~~~~~~~~~~~~~~~~~~~~,y^{2}_{*}=\frac{i \left(\left(c^2+c+1\right) \sqrt{(c-1)^2 \tau ^2+c}+(c+1) ((c-3) c+1) \tau \right)}{c^2 (c+1)}
\end{array}
\eea
,we get the instanton action as:-
\bea\label{instantonact2}
\begin{array}{lll}
\displaystyle
A'_{inst}(\tau)=A'_{eff}(x^{1}_{*},y^{1}_{*})-A'_{eff}(x^{2}_{*},y^{2}_{*})
\\
\displaystyle

\end{array}
\eea

Instanton effect for one matrix model has been studied in \cite{Okuyama:2018gfr}
for large N limit in $\tau>1$ plateau regime in context of Gaussian one matrix model. For eigenvalue instanton in $N\times N$ Gaussian hermitian matrix instanton action is given as-
\be\label{instantonact3}
 A_{inst}(\tau)=2[\tau\sqrt{\tau^2-1}-cosh^{-1}(\tau)]
 \ee
\begin{figure}[H]
       \centering
       \subfigure[Instanton action comparison for c=0.2 in Log-Log scale]{
      \includegraphics[width=7cm,height=6cm] {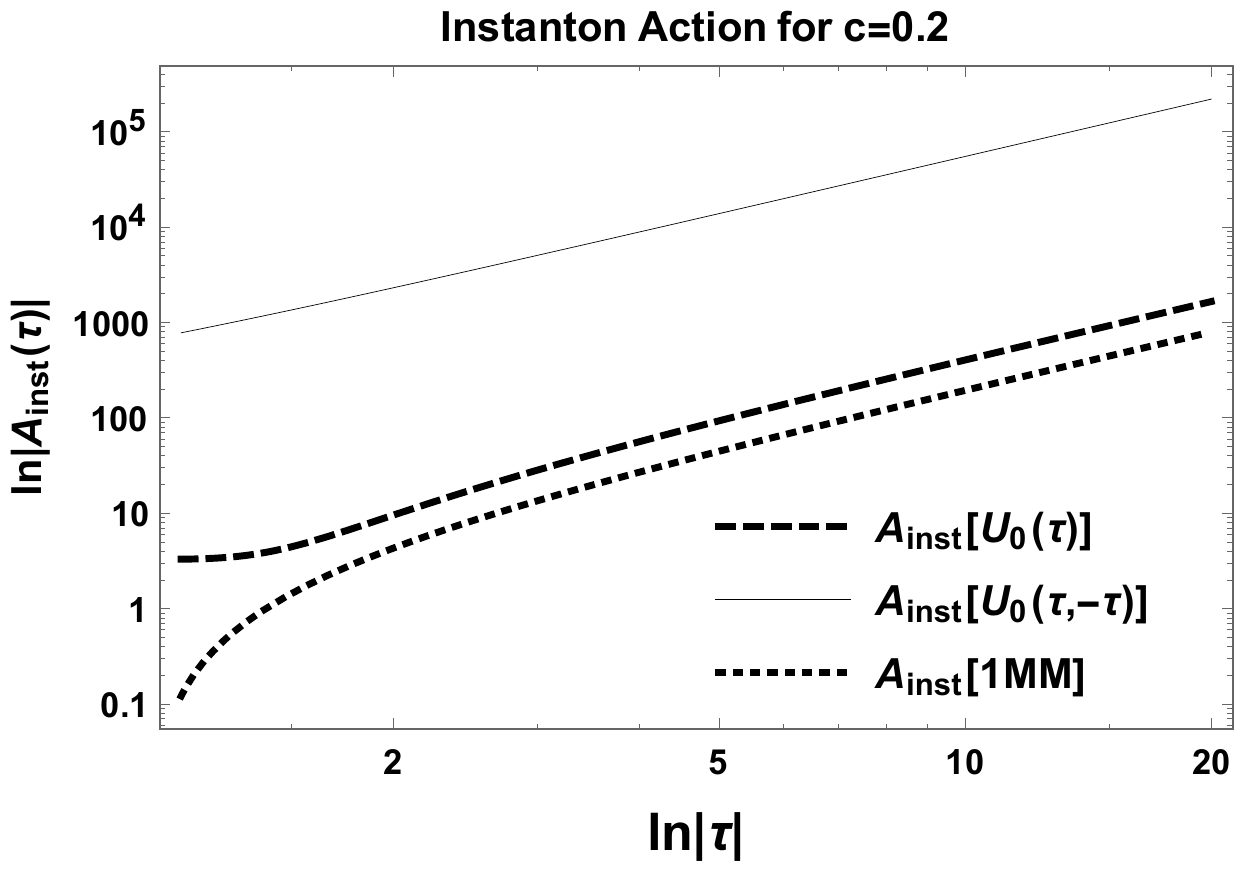}
     \label{}
         }
        \subfigure[Instanton action comparison for c=0.5 in Log-Log scale]{
      \includegraphics[width=7cm,height=6cm] {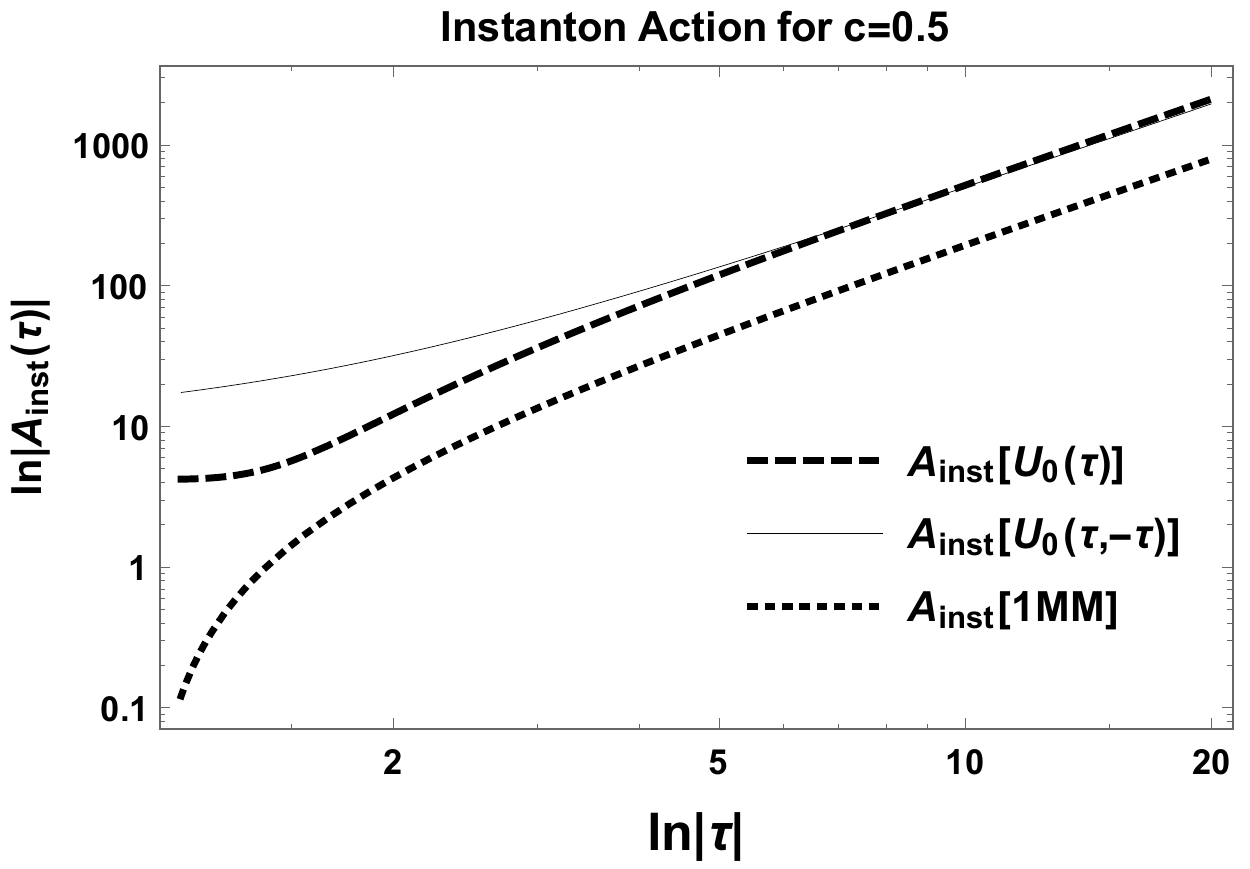}
     \label{}
         }
        \subfigure[Instanton action comparison for c=0.7]{
      \includegraphics[width=7cm,height=6cm] {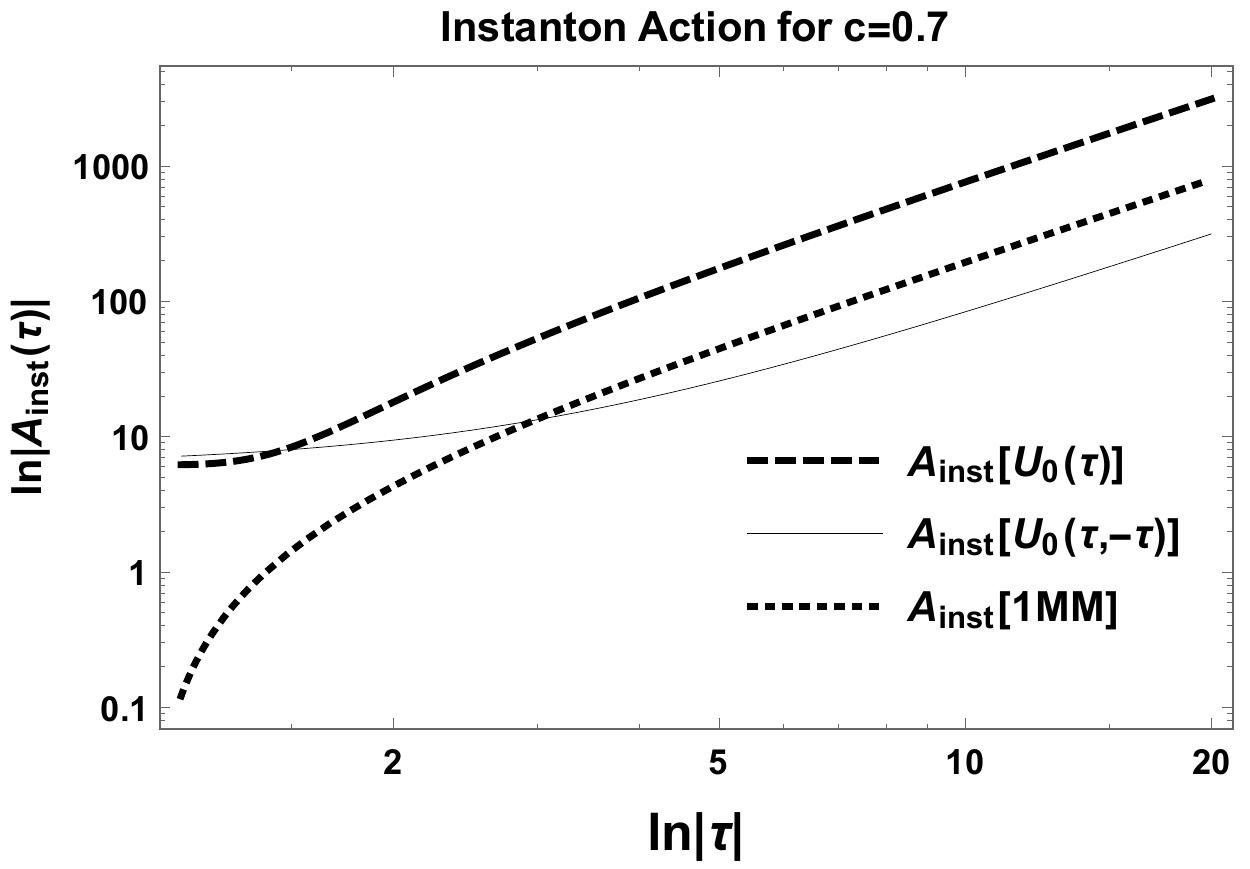}
     \label{}
         }
        \subfigure[Instanton action comparison for c=0.8]{
      \includegraphics[width=7cm,height=6cm] {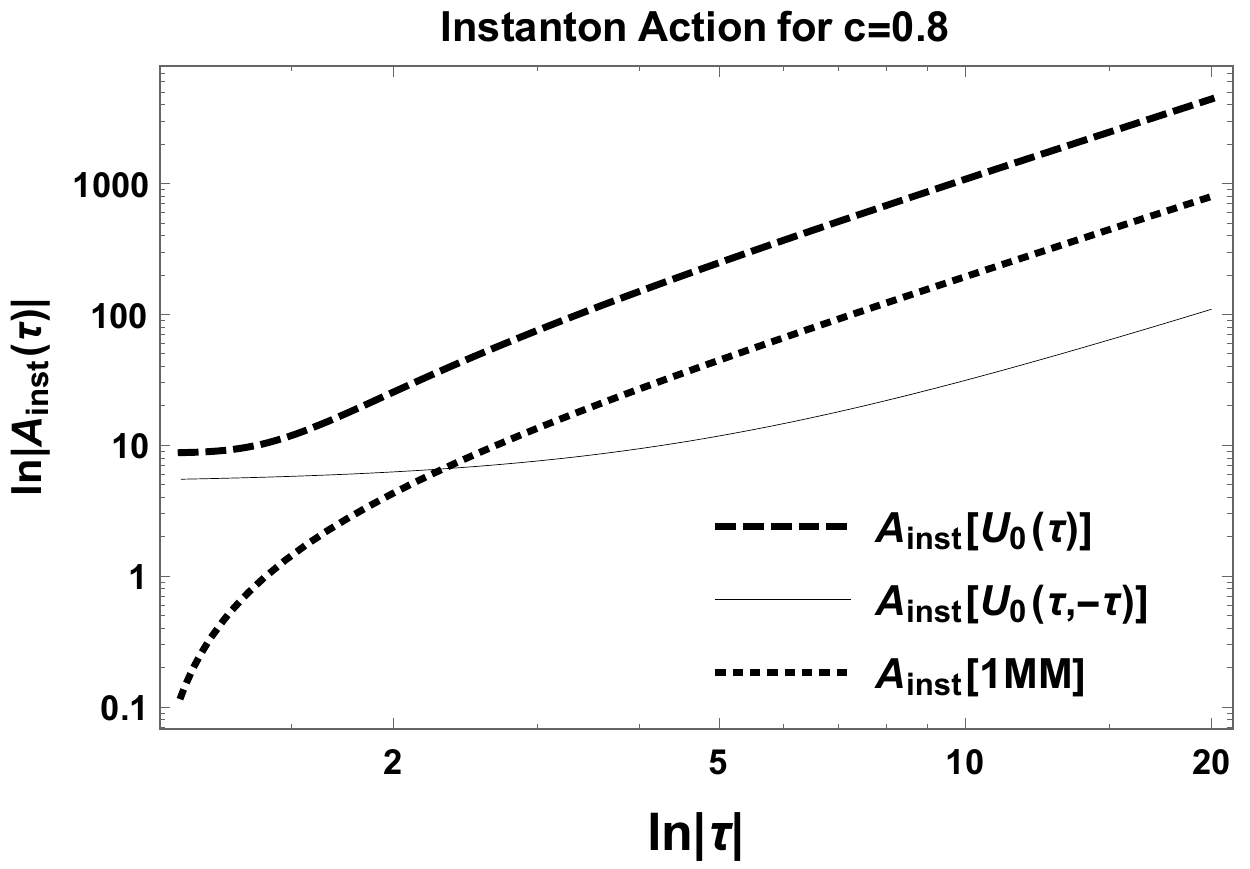}
     \label{}
         }
\caption[Optional caption for list of figures]{Instanton action from One-point function( Eq:-\ref{instantonact}), from two-point function( Eq:-\ref{instantonact2}) is compared with know eigenvalue instanton action for one matrix model (Eq:-\ref{instantonact3}). We have scaled them appropriately to show their same nature in large $\tau$ limit } 
\label{g98}
\end{figure}   
In \cite{Okuyama:2018gfr} it is claimed that $A'_{inst}(\tau)$ for two -point correlation function is proportional instanton action derived from one point function for one matrix model.To verify this we  have shown their behavior and found this proportionally to hold in large $\tau$ limit only.
 Now we can compare instanton action from one and two point function and check their proportional behavior.
 \begin{figure}[H]
       \centering
       \subfigure[For c=0.7 linearity in instanton action]{
      \includegraphics[width=7cm,height=5cm] {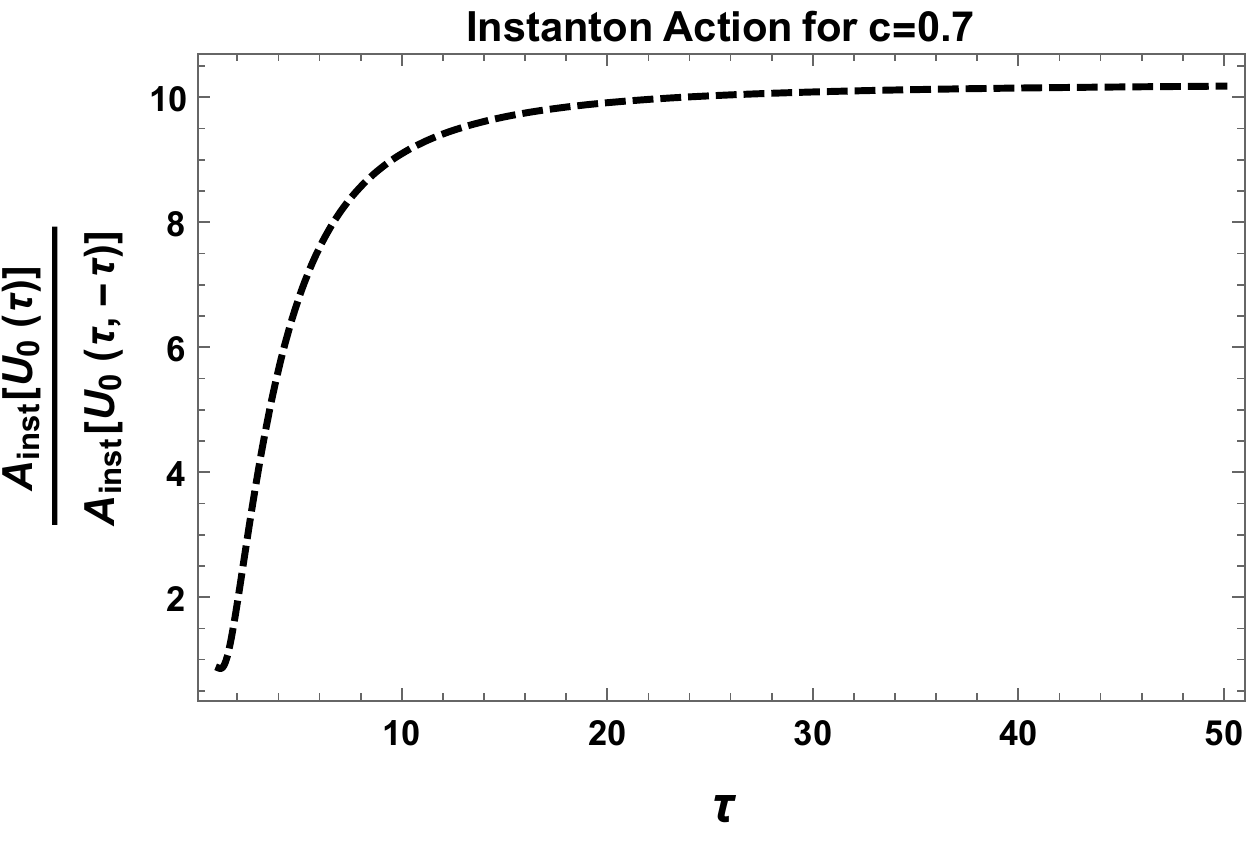}
     \label{}
         }
        \subfigure[For c=0.4 linearity in instanton action]{
      \includegraphics[width=7cm,height=5cm] {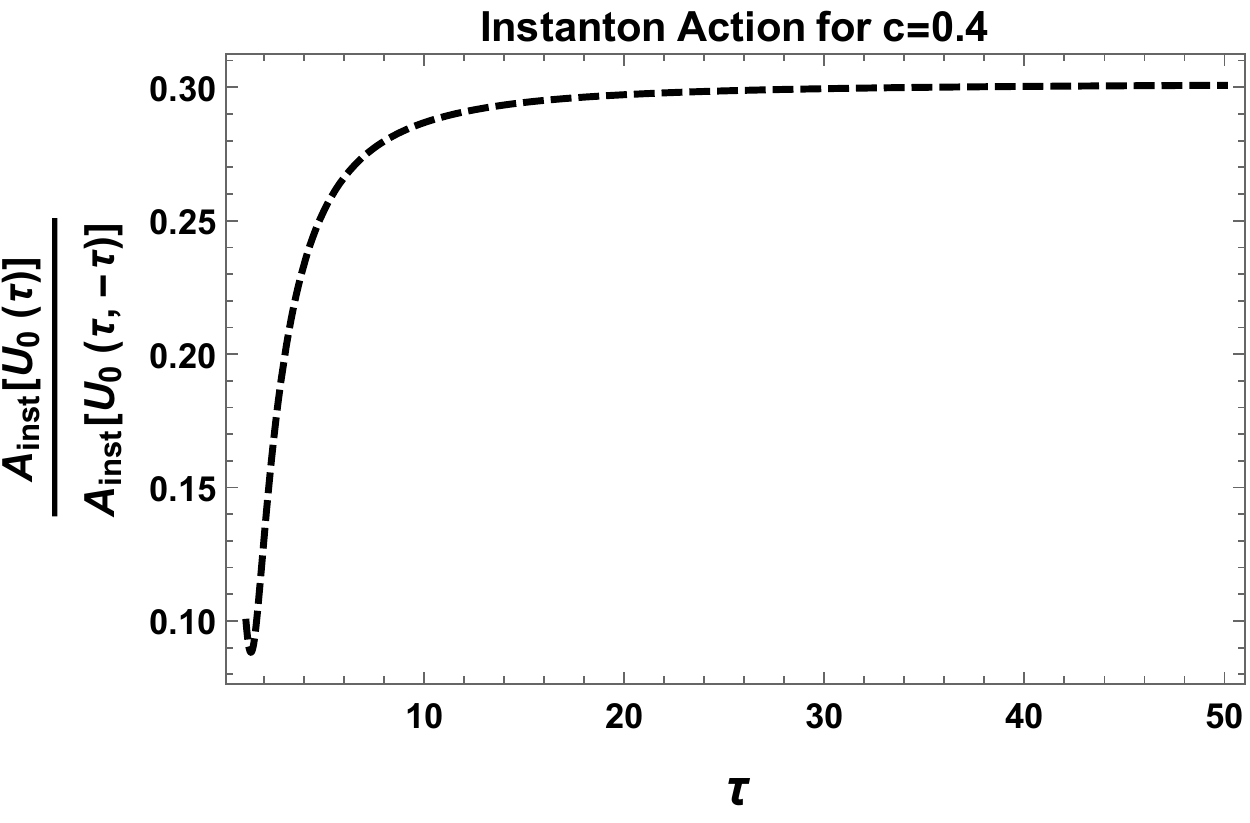}
     \label{}
         }
\caption[Optional caption for list of figures]{Ratio of Instanton action from One-point function( Eq:-\ref{instantonact}), from two-point function( Eq:-\ref{instantonact2}). Here $A_{inst}[U_{0}(\tau)]$ is the instanton action from one-point function( Eq:-\ref{instantonact}) and $A_{inst}[U_{0}(\tau,-\tau)]$ is from two-point function( Eq:-\ref{instantonact2}). In large $\tau$ limit the ratio become constant which supports the predicted proportional nature. } 
                                                                    \label{g98}
                                                                    \end{figure}   

\section{Duality relation for two matrix model}
Correlation function for characteristic polynomial of two matrix model has been studied in \cite{Brezin:2008bv}. We here review the duality formula found in \cite{Brezin:2008bv} for the later use of it in discussion of the logarithmic potential.
\be \label{firstfn}
J=\langle \prod_{\ag=1}^{k_1} \det(\lb_{\ag}-M_{1}) \prod_{\bg=1}^{k_2} \det(\mu_{\bg}-M_{2})\rangle
\ee
$M_{1}$ and $M_{2}$ are N$\times$N Hermitian matrix as seen in Eq:-(\ref{corr1}), Eq:-(\ref{corr2})
The average is over distribution for two matrix model [Eq:-(\ref{corr1})]
\be
\displaystyle
P(M_{1},M_{2})=\frac{1}{Z} e^{-\frac{1}{2}trM_{1}^{2}-\frac{1}{2}trM_{2}^{2}-c trM_{1}M_{2}-trM_{1}A}
\ee
The duality formula found in \cite{Brezin:2008bv} are expressed as
\bea
\begin{array}{lll}
\displaystyle
J=\int dB_{1}dB_{2} dD^{\dagger}dD e^{-\frac{N}{2}\mathrm{Tr}(B_{1}^{2}+B^{2}_{2}+2D^{\dagger}D)}\prod_{i=1}^{N} \det (X_{i})
\end{array}
\eea
\[
X_{i}=\begin{array}{lll}
\begin{bmatrix}
\displaystyle
\{(\lb_{\ag}-\frac{a_{i}}{1-c^2})\del_{\ag,\ag'}+\frac{i}{\sqrt{1-c^2}}B_{1}\} &\sqrt{\frac{c}{1-c^{2}}}D \\
\displaystyle
\sqrt{\frac{c}{1-c^{2}}}D^{\dagger} & \{(\mu_{\bg}-\frac{ca_{i}}{1-c^2})\del_{\bg,\bg'}+\frac{i}{\sqrt{1-c^2}}\}B_{2}\}
\end{bmatrix}
\end{array}
\]
where $B_{1}$ and $B_{2}$ are hermitian square matrices and $D$ is complex rectangular  matrix.
Now we use a transformation 
\[
B'_{1}\rightarrow B_{1}+i\sqrt{1-c^{2}}\lb_{\ag,\ag'}\del_{\ag,\ag'}~~~~~~
\\
B'_{2}\rightarrow B_{2}+i\sqrt{1-c^{2}}\mu_{\bg,\bg'}\del_{\bg,\bg'}
\] 
This simplifies the integral :-
\bea\label{we11}
\begin{array}{lll}
	\displaystyle
J=C \int dB_{1}dB_{2} dD^{\dagger}dD \mathrm{Exp}\Bigg[-\frac{N}{2}\mathrm{Tr}(B_{1}^{2}+B^{2}_{2}+2D^{\dagger}D)-iN\sqrt{1-c^2}trB_{1}\Lambda_{1}
\\
\displaystyle
~~~~~~~~~~~~~~~~~~~~~~~~~~~~~~~~~~~~~-iN\sqrt{1-c^2}\mathrm{Tr} B_{2}\Lambda_{2}-\sum_{i=1}^{N} \mathrm{Tr}(\mathrm{Log}(1-K_{i}))\Bigg]
\end{array}
\eea
Here we have used $\det(A)=e^{\mathrm{Tr}(\rm{Log}(A))}$ and the matrix K is reduced from X
\\
$K_{i}=\begin{array}{lll}
\begin{bmatrix}
\displaystyle
i\sqrt{1-c^2}\frac{B_{1}}{a_{i}} &\sqrt{c(1-c^{2})}\frac{D}{a_{i}} \\
\displaystyle
\frac{\sqrt{c(1-c^{2})}}{c}\frac{D^{\dagger}}{a_{i}} & -\frac{i\sqrt{1-c^2}}{c}\frac{B_{2}}{a_{i}}
\end{bmatrix}
\end{array}$$ \rightarrow$$
K=\begin{array}{lll}
\begin{bmatrix}
\displaystyle
iB_{1} &\sqrt{c}D \\
\displaystyle
\frac{D^{\dagger}}{\sqrt{c}} & -\frac{iB_{2}}{c}\\
\end{bmatrix}
\end{array}$ \\
We set A=aI with constraint $a=\sqrt{1-c^2}$ .
Now we can expand  \rm{Log}(1-K) in Taylor series upto 3rd term,
\be\label{err}
\rm{Log}(1-K)=-K-\frac{K^{2}}{2}-\frac{K^{3}}{3}-\frac{K^{4}}{4}
\ee
Considering upto $K^{3}$ gives the term in power of exponential [Eq:-(\ref{we11})] as:-
\bea\label{intersect}
\begin{array}{lll}
\displaystyle
J=\int  dB_{1}dB_{2}dD^{\dagger}dD~~\mathrm{Exp}\bigg[\mathrm{Tr}\bigg\{-N\big[i B_{1}(1-\sqrt{1-c^{2}}\lambda_{1})-\frac{i}{c}B_{2}(1-c\sqrt{1-c^{2}}\lambda_{2})~~~~~~~
\\
\displaystyle
~~~~~~~~~~~~~~~~~~~~~~~~~~~~~~~~~~~~~~~~~~~-\frac{1}{2}(1-\frac{1}{c^{2}})B_{2}^{2}+\frac{i}{3}B_{1}^{3}-\frac{i}{3c^{3}}B_{2}^{3}+\frac{2i}{3}DD^{\dagger}B_{1}
\\
\displaystyle
~~~~~~~~~~~~~~~~~~~~~~~~~~~~~~~~~~~~~~~~~~~-\frac{2i}{3c}B_{2}D^{\dagger}D+\frac{i}{3c}DB_{2}D^{\dagger}-\frac{i}{3}D^{\dagger}B_{1}D\big]\bigg\}\bigg]
\end{array}
\eea

\bea
\begin{array}{lll}
\displaystyle
J=\int dB_{1}dB_{2}dD^{\dagger}dD \mathrm{Exp}\bigg\{-iN \mathrm{Tr}(B_{1}\Lambda_{1})-iN\mathrm{Tr}(B_{2}\Lambda_{2})+\frac{i}{3}N\mathrm{Tr}(B^{3}_{1})
\\
\displaystyle
~~~~~~~~~-\frac{N}{2}(1-\frac{1}{c^2})\mathrm{Tr}B_{2}^{2}+iN\mathrm{Tr}(DD^{\dagger}B_{1}-1)-\frac{iN}{c}B_{2}(-1+DD^{\dagger})\bigg\}
\end{array}
\eea
Now at the edge of the spectrum for the matrix $M_{1}$ edge scaling limit at large N gives:- 
\be \label{scale12}
B_{1}\sim O(N^{-\frac{1}{3}})~~~~B_{2}\sim O(N^{-\frac{1}{2}})~~~~~D\sim O(N^{-\frac{1}{3}})
\ee
Dropping the negligible terms $(B_{2}DD^{\dagger}\sim O(N^{-\frac{7}{3}}))$ 
\bea
\begin{array}{lll}
\displaystyle
Z=\int dB_{1}dB_{2}dD^{\dagger}dD e^{-iN\mathrm{Tr}(B_{1}\Lambda_{1})-iN\mathrm{Tr}(B_{2}\Lambda_{2})+\frac{i}{3}N\mathrm{Tr}(B^{3}_{1})-\frac{N}{2}(1-\frac{1}{c^2})\mathrm{Tr}B_{2}^{2}+iN\mathrm{Tr}(DD^{\dagger}B_{1})}
\\
\displaystyle
Z=Q\int dB_{1}dD^{\dagger}dD e^{-i\mathrm{Tr}B_{1}\Lambda_{1}+\frac{i}{3}\mathrm{Tr}B_{1}^{3}+i\mathrm{Tr}DD^{\dagger}B_{1}}
\end{array}
\eea
Q is the decoupled part generated after integration over $B_{2}$
Integrating out $D^{\dagger}$ and D gives logarithmic term:-
\bea\label{ads}
\begin{array}{lll}
\displaystyle
Z=\int dB_{1} e^{\frac{i}{3}\mathrm{Tr}B_{1}^{3}-k_{2}\mathrm{Tr}\mathrm{Log}(B_{1})-i\mathrm{Tr}(B_{1}\Lambda_{1})}
\end{array}
\eea
This has been related to Airy Matrix model coupled with a  logarithmic potential (Kontsevich - Penner model ) in \cite{Brezin:2011ka} 

\subsection*{\textbf{Derivation for $B_{1}^{4}$ term}}
$K=\begin{array}{lll}
\begin{bmatrix}
\displaystyle
iB_{1} &\sqrt{c}D \\
\displaystyle
\frac{D^{\dagger}}{\sqrt{c}} & -\frac{iB_{2}}{c}\\
\end{bmatrix}
\end{array}$
Expanding upto 4th term
\be
\rm{Log}(1-K)=-K-\frac{K^{2}}{2}-\frac{K^{3}}{3}-\frac{K^{4}}{4}
\ee
So, \rm{Tr}(\rm{Log}(1-K)) has terms from four contribution, as trace is there we can consider only the diagonal terms in each of $\mathrm{Tr}[K^{n}]$.
So for  $\mathrm{Tr}[\frac{1}{3}K^{3}]$ term $\rightarrow$ 
\bea
\begin{array}{lll}
\displaystyle
\mathrm{Tr}\bigg(-\frac{i}{3} B_{1}^{3}+\frac{1}{3} D D^{\dagger} B_{1}+\frac{i}{3} B_{1} D D^{\dagger}+ \frac{i}{3c}D B_{2} D^{\dagger}+\frac{i}{3} D^{\dagger} B_{1} D-\frac{i}{3 \sqrt{c}} B_{2} D^{\dagger}D
\\
\displaystyle
~~~~~~~~~~~~~~~~~-\frac{i}{3 \sqrt{C}} D D^{\dagger} B_{2}+\frac{i}{3c^{3}} B_{2}^{3}\bigg)
\end{array}
\eea
 $\mathrm{Tr}[\frac{1}{4}K^{4}]$ term $\rightarrow$ 
\bea
\begin{array}{lll}
\displaystyle
\mathrm{Tr}\bigg(\frac{1}{4} B_{1}^{4}+\frac{i}{4} D D^{\dagger} B_{1}^{2}-\frac{1}{4} B_{1} D D^{\dagger} B_{1}+\frac{1}{4 c} D B_{2} D^{\dagger} B_{1}-\frac{1}{4} B_{1} D B_{1} D^{\dagger}+\frac{1}{4} D D^{\dagger} D D^{\dagger}\\
\displaystyle
~~~~~~-\frac{1}{4 c^{2}} D B_{2}^{2} D^{\dagger}-\frac{1}{4} D^{\dagger} B_{1}^{2} D+\frac{1}{4 \sqrt{c}} B_{2} D^{\dagger} B_{1} D+\frac{1}{4} D D^{\dagger} D^{\dagger} D-\frac{1}{4c^{2}} B^{2} D^{\dagger} D\\
\displaystyle
~+\frac{1}{4c} B_{1} D B_{2} D^{\dagger}+\frac{1}{4c} B_{1} D B_{2} D^{\dagger}+\frac{i}{4 c^{3} \sqrt{c}} B_{2}^{4}-\frac{1}{4c\sqrt{c}} B_{2} D^{\dagger} D B_{2}-\frac{1}{4c\sqrt{c}} D D^{\dagger} B_{2} B_{2}\bigg)
\end{array}
\eea

If we consider upto $K^{4}$ term of Eq:-(\ref{err})
This integral is solved in similar way. Now with existing edge scaling Eq:-(\ref{scale12}), after integral over $B_{2}$ and $D,D^{\dagger}$
\bea \label{b4}
\begin{array}{lll}
\displaystyle
Z=\int dB_{1} \mathrm{Exp}\left[-i \mathrm{Tr}(B_{1}\Lambda_{1})-\frac{i}{c^3}\mathrm{Tr}(B_{1}^{3})-\frac{1}{4}\mathrm{Tr}(B_{1}^{4})-k_{3}\mathrm{Tr}(\rm{Log}[B_{1}])\right]
\end{array}
\eea
Although $\mathrm{Tr}( B^4)$ term is absent in the edge scaling, this term can be derived as \cite{brezin1998level,Brezin:1998zz
}. Two converging saddle points gives rise to fold singularity as in the $B_{1}^{3}$ expression. This is related to Airy kernel 
 For extended Airy Kernel Eq:-(\ref{b4}) cubic singularity becomes quartic term. This is expressed in terms of Pearcey function and showed in \cite{brezin1998level,Brezin:1998zz} on the level spacing distribution for hermitian random matrices with an  external field. If $H$=$H_{0}$+$V$ where $H_{0}$ is a fixed matrix and $V$ is an $N\times N$ random GUE matrix. $H_{0}$ has eigenvalues $\pm a$ each with multiplicity $\frac{N}{2}$. Spectrum of $H_{0}$ is such that there is a gap in the average density of eigenvalues of $H$ which is thus split into two pieces. With $N\rightarrow \infty$ density of eigenvalues supported on single or double interval depending on size of a. At the closing of gap the limiting eigenvalue distribution has Pearcey kernel structure. When the spectrum of $H_{0}$ is tuned so that the gap closes limiting eigenvalue distribution have the same structure as Pearcey kernel.
\\
\subsection*{Connecting the Two point correlation function with Open partition function}
At first consider the equation Eq:- (\ref{intersect}) with $B_{2}\rightarrow i B_{2}c, ~~ B_{1} \rightarrow -i B_{1}$ and  $\Lambda_{1}\rightarrow (1-\sqrt{1-c^{2}}\Lambda_{1})$ and $\Lambda_{2}\rightarrow (1+c\sqrt{1-c^{2}}\Lambda_{2})$
\bea
\begin{array}{lll}
\displaystyle
J=\int  dB_{1}dB_{2}dD^{\dagger}dD\mathrm{Exp}\bigg[-N \mathrm{Tr}\big\{B_{1}\Lambda_{1}+B_{2}\Lambda_{2}-\frac{1}{2}(c^{2}-1) B_{2}^{2}-\frac{1}{3}B_{1}^{3}-\frac{1}{3}B_{2}^{3}
\\
\displaystyle
~~~~~~~~~~~~~~~~~~~~~~~~~~~~~~~~+\frac{2}{3}DD^{\dagger}B_{1}+\frac{2}{3}B_{2}D^{\dagger}D-\frac{1}{3}DB_{2}D^{\dagger}-\frac{1}{3}D^{\dagger}B_{1}D\big\}\bigg]~~

\end{array}
\eea
Then we  integrate over $dD$ and $dD^{\dagger}$ and made the transformation $B_{1}\rightarrow B_{1}+\sqrt{\Lambda_{1}}$.We rewrite the equation in $H$ and $Z$ replacing $B_{1}$ and $B_{2}$
\bea
\begin{array}{lll}
\displaystyle
\displaystyle
J=\int\limits_{H_{k_{1}}\times Z_{k_{2}}} \mathrm{Exp}\Bigg[\frac{2N}{3}\mathrm{Tr}(\Lambda_{1}^{\frac{3}{2}})-\frac{N}{3}\mathrm{Tr}(H^{3})-N \mathrm{Tr}(H^{2}\sqrt{\Lambda_{1}})+Ntr(Z\Lambda_{2})-\frac{N}{3}Z^{3}\\
\displaystyle
~~~~~~~~~~~+N(\frac{c^{2}-1}{2})Z^{2}\Bigg]\times \mathrm{Exp}\left[\big\{\frac{N}{3} \mathrm{Tr}(\mathrm{Log}(H+\sqrt{\Lambda_{1}}))+\frac{N}{3}\mathrm{Tr}(\mathrm{Log}(Z)) \big\}\right]dH dZ
\end{array}
\eea
Matrix integral representation of two point correlation function of two matrix model:-
\[
\begin{array}{lll}
\displaystyle
J=\int_{H_{k_{1}}\times Z_{k_{2}}} dH dZ ~\mathrm{Exp} \big[\frac{2N}{3}\mathrm{Tr}(\Lambda_{1}^{\frac{3}{2}})-\frac{N}{3}\mathrm{Tr}(H^{3})-N \mathrm{Tr}(H^{2}\sqrt{\Lambda_{1}})+N\mathrm{Tr}(Z\Lambda_{2})\\
\displaystyle
~~~~~~~~~~~~~~~~+N(\frac{c^{2}-1}{2})Z^{2}-\frac{N}{3}Z^{3}+\big\{N k \mathrm{Tr}(\mathrm{Log}(H+\sqrt{\Lambda_{1}}))+N k' \mathrm{Tr}(\mathrm{Log}(Z)) \big\}\big]
\end{array}
\]

\[
\begin{array}{lll}
\displaystyle
J=\mathrm{Exp}\big[\frac{\tilde{N}}{3}\mathrm{Tr}(\Lambda_{1}^{\frac{3}{2}})\big]\int_{H_{k_{1}}\times Z_{k_{2}}} \mathrm{Exp} \Bigg[ \frac{\tilde{N}}{2}(\frac{c^{2}-1}{2})Z\tilde{Z}^{T}- \frac{\tilde{N}}{2} \mathrm{Tr}(H^{2}\sqrt{\Lambda_{1}})
\\
\displaystyle
~~~~~~~~~~~~~~~~~~~~~~~~~~~~~~~~~~- \frac{\tilde{N}}{6}\mathrm{Tr}(H^{3})- \frac{\tilde{N}}{6}\mathrm{Tr}Z^{3}+ \frac{\tilde{N}}{2}\mathrm{Tr}(Z\Lambda_{2})+ \frac{\tilde{N}}{2}\mathrm{Tr}(Z\Lambda_{2})
\\
\displaystyle
~~~~~~~~~~~~~~~~~~~~~~~~~~~~~~~~~~~~~+\frac{\tilde{N}}{6}\big\{ \mathrm{Tr}(\mathrm{Log}(H+\sqrt{\Lambda_{1}}))+ \mathrm{Tr}(\mathrm{Log}(Z)) \Bigg] dH dZ
\end{array}
\]

Now from Eq:-(\ref{firstfn})
\[
\Lambda_{1}|_{\alpha}=(1-\sqrt{1-c^{2}})\lambda_{\alpha} ~~~~~ and~~~~~~~~~~~ \Lambda_{2}|_{\beta}=(1-\sqrt{1-c^{2}})\mu_{\beta} ~~~ N=\frac{\tilde{N}}{2}
\]
and $Z_{k_{2}\times k_{2}}$ is hermitian matrix so $Z\tilde{Z}^{T}=Z^{2}$
\bea \label{twcr}
\begin{array}{lll}
\displaystyle
J=K\int\limits_{H_{k_{1}}\times Z_{k_{2}}} \mathrm{Exp} \Bigg[ \frac{\tilde{N}}{2}(\frac{c^{2}-1}{2})Z\tilde{Z}^{T}- \frac{\tilde{N}}{2} \mathrm{Tr}(H^{2}\sqrt{\Lambda_{1}})- \frac{\tilde{N}}{6}\mathrm{Tr}(H^{3})- \frac{\tilde{N}}{6}trZ^{3}
\\
\displaystyle
~~~~~~~~~~~~~~~~~~~~~~~~+ \frac{\tilde{N}}{2}\mathrm{Tr}(Z\Lambda_{2})\Bigg]
~ \big[\det(H+\sqrt{\Lambda_{1}})\big]^{\frac{\tilde{N}}{6}}\times \big[\det(Z)\big]^{\frac{\tilde{N}}{6}}dH dZ
\end{array}
\eea
Now we look at the very refined open partition function as derived in \cite{Alexandrov:2017ysm}. They have provided the matrix model for very refined open partition function as matrix integrals in the given form:-
\bea \label{vrf}
\begin{array}{lll}
\displaystyle
\left.\tilde{\tau}^{o}\right|_{t_{i}=t_{i}(\Lambda)}= \frac{c_{\Lambda, M}}{(2 \pi)^{N^{2}}} \int\limits_{\mathcal{H}_{M} \times \operatorname{M}_{N, N}} \operatorname{\det} \frac{\Lambda \otimes \mathrm{I}_{N}+\sqrt{\Lambda^{2} \otimes \mathrm{I}_{N}-\mathrm{I}_{M} \otimes \bar{Z}^{t}}-H \otimes \mathrm{I}_{N}+\mathrm{I}_{M} \otimes Z}{\Lambda \otimes \mathrm{I}_{N}+\sqrt{\Lambda^{2} \otimes \mathrm{I}_{N}-\mathrm{I}_{M} \otimes Z^{t}}-H \otimes \mathrm{I}_{N}-\mathrm{I}_{M} \otimes Z} 
\\
\displaystyle
 ~~~~~~~~~~~~~\mathrm{Exp}\Bigg[-\frac{1}{2} \operatorname{\mathrm{Tr}} H^{2} \Lambda-\frac{1}{2} \operatorname{\mathrm{Tr}} Z \bar{Z}^{t}+\frac{1}{6} \operatorname{\mathrm{Tr}} H^{3}+\frac{1}{6} \operatorname{\mathrm{Tr}} Z^{3}+\frac{1}{2} \operatorname{\mathrm{Tr}} \bar{Z}^{t} \Theta \Bigg] 
d H d Z
\end{array}
\eea
For $N \geq 1$ the space of Hermitian matrices is denoted by $\mathcal{H}_{M}$ and the space of complex  $N \times N$ matrices by  M$_{N\times N}(\mathbb{C})$ . Volume $d~Z$ is denoted by 
\[
d Z:=\prod_{1 \leq i, j \leq N} d\left(\operatorname{Re} z_{i, j}\right) d\left(\operatorname{Im} z_{i, j}\right)
\]
 and Gaussian probability measure on space of complex matrices is given by 
\[
\frac{1}{(2 \pi)^{N^{2}}} e^{-\frac{1}{2} \operatorname{\mathrm{Tr}} Z \bar{Z}^{T}} d Z
\]
$\theta_{i, j}, 1 \leq i, j \leq N,$ are considered as an extra set of complex variables:-
\bea
\begin{array}{lll}
\displaystyle
\Theta :=\left(\theta_{i, j}\right)_{1 \leq i, j \leq N} \in \operatorname{M}_{N, N}(\mathbb{C}) 
\\
\displaystyle
q_{m}(\Theta) &:=\operatorname{\mathrm{Tr}} \Theta^{m}, \quad m \geq 0
\end{array}
\eea
And,
\begin{equation}
c_{\Lambda, M}:=(2 \pi)^{\frac{M^{2}}{2}} \prod_{i=1}^{M} \sqrt{\lambda_{i}} \prod_{1 \leq i<j \leq M}\left(\lambda_{i}+\lambda_{j}\right)
\end{equation}

Now comparing Eq:-(\ref{twcr}) and Eq:-(\ref{vrf}) two matrix model two point correlation function and very refined open partition function are similar with $\sqrt{\Lambda_{1}}=\Lambda$ and $\Lambda_{2}=\Theta$
and $K=\mathrm{Exp}\big[\frac{\tilde{N}}{3}\mathrm{Tr}(\Lambda_{1}^{\frac{3}{2}})\big]$ is the extra constant term multiplied in front.
More detailed discussion on open partition function and refined open partition function can be found in \cite{
Alexandrov:2017ysm
}.
\\
In \cite{Brezin:2008bv} two matrix model correlation function has been related to Kontsevich-Penner Matrix model near Heisenberg time. Using Replica method they have studied the intersection number discussion in this context. In our previous calculation we have obtained a rounding off behavior near Heisenberg time. The universal behavior of SFF ramp region Dyson sine kernel is now changed. It suggests that some new kind of description is needed in this region. Kontsevich\cite{kontsevich1992} and Penner\cite{penner1988} Matrix models gives the edge behavior and open boundaries for the punctured open Riemann surfaces. This has been explained in\cite{Brezin:2008bv,Brezin:2007iv,Brezin:2015dza}. Universal Dyson sine kernel gives one important feature of underlying Gaussian Unitary Ensemble , its stationary nature under Dyson Brownian motion. But now universality of sine kernel are no more available. To explain the rounding off behavior we need to consider Brownian motion near edges. This Brownian motion effect is related to time dependence of the model, which involves higher singularities.

\section{Discussion}
Authors of \cite{SFFRMT} converted time dependent matrix model of Eq:-\ref{td11} into two matrix model and formulated two point correlation functions in the integral form. We revisited this approach, specially for the spectral form factor, from the point of view of the universal signature of the quantum chaos. We confirm by the numerical works the behavior of the large $N$ limit due to the exact $N$ expression by Hermite polynomials, and made a detailed comparison to saddle point results.
This time dependent model has interesting interpretation as open
intersection numbers, which is derived from the logarithmic potential representing the boundaries \cite{Brezin:2011ka}.
The rounding behavior around Heisenberg time, which we have confirmed 
in this paper, is shown to be related to such boundary problems
. We have considered two type of correlation function and also the next order contribution of $1/N$ expansion, for saddle point integral. SFF for different matrix correlation has been shown to have a rounding off near Heisenberg time $\tau=\tau_{c}$, a crossover in this point.
\\
This two matrix model may be related to wormhole between different CFT states and to black hole statistics \cite{Chakravarty:2020wdm,Cotler:2020hgz}.
For our same matrix correlation function and SFF it gives a decaying average spectral form factor which is consistent with GUE behavior of SFF. Second term contribution calculated here from the $\frac{1}{N}$ expansion of saddle point integral gives rounding off behavior and appear as correction to the first order solution. Change in Heisenberg time for this correction are computed explicitly. The second term of saddle point contribution controls the shift in saturation value for different $N$. And the calculation for Instanton action for two matrix model appears to have same eigenvalue instanton equation with scaling $\frac{1}{c^2-1}$. Previously it has been predicted that instanton action from two point function is proportional to instanton equation of one-point function. In two matrix case they are not same/proportional to one-point function instanton action. But in large $\tau$ limit this solution has proportional structure.

\section*{Acknowledgment}
A.M. thanks OIST for a visiting internship during this work and S.H. thanks JSPS KAKENHI 19H01813 for the support.



\providecommand{\href}[2]{#2}\begingroup\raggedright\endgroup

\begin{appendices}
\section{Two matrix model density of states}
Density of state $\rho(\lambda)$ derived by Fourier transform of $U_{A}(z)$
\begin{equation}
U_{A}(z)=\big\langle\frac{1}{N}\mathrm{Tr} e^{izM_{1}}\big\rangle
\end{equation}
We have considered an external matrix $\rm{A}$ coupled to matrix $M_{1}$ acting as a source. At last step we will put it zero to get our desired result. The eigenvalue of $M_1$ and $M_2$ are denoted by $r_i$ and $\xi_i$. We follow the formulation of \cite{SFFRMT}.
\bea \label{f1}
\begin{array}{lll}
\displaystyle
U_{A}(z)=\frac{1}{Z_{A}N}\int \mathrm{Tr} e^{izM_{1}} \times e^{-\frac{1}{2}\mathrm{Tr} M_{1}^{2}-\frac{1}{2}\mathrm{Tr} M_{2}^{2}+c \mathrm{Tr} M_{1} M_{2}-\mathrm{Tr} AM_{1}}dM_{1}dM_{2} \\
\displaystyle
=\frac{1}{NZ}\sum_{\ag=1}^{N}\int e^{izr_{\ag}}e^{-\frac{N}{2}\sum_{i} r_{i}^{2}-\frac{N}{2}\sum_{i}\xi_{i}^{2}+cN\sum_{i}r_{i}\xi_{i}-N\sum_{i}a_{i}r_{i}} \frac{\Del^2(r)\Del^2(\xi)\prod_{i}dr_{i}\prod_{i}d\xi_{i}}{\Del(r)\Del(\xi)\Del(A)\Del(r)}~~~~
\end{array}
\eea
In \cite{SFFRMT}, HarishChandra-Itzykson-Zuber formula is used to change the measure from integration over matrix to integration over eigenvalues of the matrix. $\Del(r)=\prod_{i<j}(r_{i}-r_{j})$ is the Vandermonde determinant.
\be
\int e^{\mathrm{Tr} M_{1}M_{2}-\mathrm{Tr} A M_1}dM_{1}dM_{2}=\int e^{N(\sum r_{i}\xi_{i}-\sum a_{i}r_{i})}\frac{\Del^2(r)\Del^2(\xi)\prod_{i}d r_{i}\prod_{j}d \xi_{j}}{\Del(r)\Del(\xi)\Del{A}\Del{r}}
\ee
Now using the above expression in Eq:-(\ref{f1}) we first do the Gaussian integral over $\prod dr_{i}$ and get the form as:- 
\bea
\begin{array}{lll}
\displaystyle
U_{A}(z)=\frac{1}{NZ_{A}\Del(A)}\sum_{\ag=1}^{N}\int \prod d\xi_{j}\Del(\xi)e^{-\frac{N}{2}\sum \xi_{j}^{2}} \int \prod dr_{i}  e^{iz r_{\ag}}e^{-\frac{N}{2}\sum r_{i}^{2}-N\sum a_{i}r_{i}} e^{cN\sum \xi_{j} r_{j}} \\
\displaystyle
~~~~~~=\frac{1}{NZ_{A}\Del(A)}\sum_{\ag=1}^{N}\prod_{i<j}\bigg\{ \frac{i c z}{N(1-c^2)}\{\del_{i,\ag}-\del_{j,\ag}\}-c(a_{j}-a_{i})\bigg\}e^{-\frac{z^2}{2N(1-c^2)}-\frac{i z a_{\ag}}{\sqrt{1-c^2}}}\\
\displaystyle
~~~~~~=\frac{c}{NZ_{A}}\sum_{\ag=1}^{N}\prod_{\gamma\neq \ag}\frac{\left(a_{\ag}-a_{\gamma}-\frac{i z}{N(1-c^2)}\right)}{(a_{\ag}-a_{\gamma})}e^{-\frac{z^2}{2N(1-c^2)}-\frac{iz a_{\ag}}{\sqrt{1-c^2}}}\\
\displaystyle
~~~~~~=-\frac{c\sqrt{1-c^2}}{iz}\oint\frac{du}{2\pi i}\prod_{\gamma=1}^{N}\bigg(\frac{u-a_{\gamma}-\frac{i z}{N(1-c^2)}}{u-a_{\gamma}}\bigg)^{N}e^{-\frac{z^2}{2N(1-c^2)}-\frac{iz a_{\ag}}{\sqrt{1-c^2}}}
\end{array}
\eea
If we take the external source term to zero ( $a_{i}\rightarrow0$)
\be
U_{0}(z)=-\frac{c\sqrt{1-c^2}}{iz}\oint\frac{du}{2\pi i}\bigg(1-\frac{iz}{Nu\sqrt{1-c^2}}\bigg)^{N}e^{-\frac{izu}{\sqrt{1-c^2}}-\frac{z^2}{2N(1-c^2)}}
\ee
Density of states is defined as the Fourier transform of this function:-
\be
\rho(\lambda)=-\frac{c\sqrt{1-c^2}}{2\pi i}\int \frac{dz}{z}e^{-iz\lambda}\oint\frac{du}{2\pi i}\bigg(1-\frac{iz}{Nu\sqrt{1-c^2}}\bigg)^{N}e^{-\frac{izu}{\sqrt{1-c^2}}-\frac{z^2}{2N(1-c^2)}}
\ee
 The density of state $\rho(\lambda)$ becomes in the large $N$ limit,
 \be
 \rho(\lambda) = \frac{\sqrt{1 - c^2}}{2\pi} \sqrt{4-(1- c^2) \lambda^2}
 \ee
The kernel $K_N(\lambda,\mu)$ is written by Hermite polynomial $H_l(t)$ in (\ref{Hermite}),
\be
K_N(\lambda,\mu) = \frac{1}{\sqrt{2 N \pi}} \sum_{l=0}^{N-1} \frac{1}{l!} H_l(\sqrt{N} \lambda) H_l(\sqrt{N} \mu) e^{-\frac{N}{2}\lambda^2}
\ee
The density of state $\rho(\lambda)$ is
\be\label{density}
\rho(\lambda)= K_N(\lambda,\lambda)
\ee
It is normalized as 
\be
\int_{-\infty}^\infty d\lambda \rho(\lambda) = 1
\ee
In the large N limit, (\ref{density})  approaches to the semi-circle law, $\rho(\lambda) = \frac{1}{2 \pi}\sqrt{4- \lambda^2}$.

The Fourier transform of the product of the density of state is
\bea
&&\int_{-\infty}^\infty dE \rho(\lambda + E)\rho(\lambda - E) e^{i t E}
= \frac{1}{(2\pi)^2}\int_{-2}^2 (4 - E^2)  {\rm cos} E t\nonumber\\
&&= \frac{1}{(2\pi)^2}( - \frac{8}{t^2}{\rm cos}(2 t) + \frac{4}{t^3}{\rm sin}( 2 t) )
\eea
where we put $\lambda=0$.
This term gives the dip (decay) region for the specral form factor of order one ($\frac{1}{N^0}$).

The two point function $\rho^{(2)}(\lambda,\mu)$ is \cite{SFFOLD}
\bea\label{rho2}
\rho^{(2)}(\lambda,\mu) &=& <\frac{1}{N}{\rm tr}\delta(\lambda- M)\frac{1}{N}{\rm tr}\delta(\mu- m)>\nonumber\\
&=&\frac{1}{N}\delta(\lambda- \mu)\rho(\lambda) - \frac{1}{N^2} K_N(\lambda,\mu)K_N(\mu,\lambda)
\eea
We followed the derivation of sine kernel in \cite{SFFOLD} with the integral representation of $K_N(\lambda,\mu)$.
\be
K_N(\lambda,\mu) = N \int\frac{dt}{2\pi}\oint \frac{du}{2i\pi}\frac{1}{it} (1- \frac{it}{Nu})^N e^{-\frac{t^2}{2N}- i u t - i t\lambda + N u (\lambda-\mu)}
\ee

By the change $u = i t u$,  and  the exponentiating $(1-\frac{1}{Nu})^N = e^{-\frac{1}{u}}$, and neglecting $t^2/(2N)$ and$1/N$ term  in the large $N$ limit, it becomes after the Gaussian integration of $t$,
\be
K_N(\lambda,\mu) = - i N \sqrt{\pi} \oint \frac{du}{2i \pi} \frac{1}{\sqrt{u}} e^{-\frac{1}{u} (1-\frac{\lambda^2}{4}) + \frac{1}{4} y^2 u - \frac{1}{2} \lambda y}
\ee
with $y = N(\lambda- \mu)$.
Using $a= - y^2(1- \lambda^2/4)/4$, 
\be
\int_0^\infty dt e^{- t - \frac{a}{t}} = 2 a^{1/2} K_{-1/2}(2\sqrt{a})
\ee
where $K_{-1/2}(x)= K_{1/2}(x)= \sqrt{\pi/(2x)} e^{-x}$ is a modified Bessel function.
We obtain in the large $N$ limit,
\be
K_N(\lambda,\mu) =  e^{-\frac{\lambda^2}{2}}\frac{{\rm sin}[\pi y \rho(\lambda)]}{\pi y}
\ee
with $\rho(\lambda) = \frac{1}{2\pi}\sqrt{4- \lambda^2}$. Another $K_N(\mu,\lambda)$ is obtained similarly and their product becomes
\be
\rho_c^{(2)}(\lambda,\mu) = e^{- \frac{\lambda-\mu}{2}y} \frac{1}{\pi^2 y^2} {\rm sin}[\pi y \rho(\lambda)] {\rm sin}[\pi y \rho(\mu)]
\ee
Since we take the large $N$ limit with a fixed $y=N(\lambda-\mu)$,
\be
K_N(\lambda,\mu) K_N(\mu,\lambda) = \frac{1}{\pi^2 N^2 (\lambda-\mu)^2} {\rm sin}[\pi N (\lambda-\mu) \rho(\lambda)]^2
\ee
This term gives a ramp for the spectral form factor of order $\frac{1}{N^2}$, and the first term of (\ref{rho2}) gives a plateau term of the spectral form factor of order $\frac{1}{N}$.
\end{appendices}
\end{document}